\documentclass[%
 aip,
 amsmath,amssymb,
 reprint,%
]{revtex4-1}

\usepackage{graphicx}% Include figure files
\usepackage{dcolumn}% Align table columns on decimal point
\usepackage{bm}% bold math
\usepackage[utf8]{inputenc}
\usepackage[T1]{fontenc}
\usepackage{mathptmx}
\usepackage{etoolbox}
\usepackage{subfigure}
\usepackage{subcaption}
\usepackage{multirow}
\usepackage{xcolor}
\newtheorem{t1}{Theorem}

\newtheorem{d1}{Definition}
\newtheorem{r1}{Remark}

\newtheorem{proof}{Proof}

\makeatletter
\def\@email#1#2{%
 \endgroup
 \patchcmd{\titleblock@produce}
  {\frontmatter@RRAPformat}
  {\frontmatter@RRAPformat{\produce@RRAP{*#1\href{mailto:#2}{#2}}}\frontmatter@RRAPformat}
  {}{}
}%
\makeatother
\begin{document}

\preprint{AIP/123-QED}

\title[ ]{{A Compounded Burr Probability Distribution for Fitting Heavy-Tailed Data with Applications to Biological Networks}}
% Force line breaks with \\
\author{Tanujit Chakraborty}
\affiliation{ 
SAFIR, Sorbonne University Abu Dhabi, UAE.}
\affiliation{ 
Sorbonne Center for Artificial Intelligence, Sorbonne University, Paris, France.}
\author{Swarup Chattopadhyay}%
\affiliation{ 
Department of Computer Science \& Engineering, XIM University, Bhubaneswar, India.%\\This line break forced with \textbackslash\textbackslash
}%

\author{Suchismita Das}
 % \homepage{http://www.Second.institution.edu/~Charlie.Author.}
\affiliation{%
Department of Data Science, SP Jain School of Global Management, Mumbai, India.%\\This line break forced% with \\
}%
\author{Shraddha M. Naik}
 % \homepage{http://www.Second.institution.edu/~Charlie.Author.}
\affiliation{Department of Computer Science, Khalifa University, UAE.%
%\\This line break forced% with \\
}%
\author{Chittaranjan Hens}
% \homepage{http://www.Second.institution.edu/~Charlie.Author.}
\email{chittaranjan.hens@iiit.ac.in}
\affiliation{%
Center for Computational Natural Science and Bioinformatics, International Institute of Information Technology, Hyderabad, India%\\This line break forced% with \\
}%

\date{\today}% It is always \today, today,

% \title{{A Compounded Burr Probability Distribution for Fitting Heavy-Tailed Data with Applications to Biological Networks}}
% \author{Tanujit Chakraborty$^a$$^b$}\email{tanujit.chakraborty@sorbonne.ae}
% \author{Swarup Chattopadhyay$^c$}\email{swarupc@xim.edu.in}
% \author{Suchismita Das$^d$}\email{suchismita.das@spjain.org}
% \author{Shraddha M. Naik$^a$}\email{shraddha.naik@ku.ac.ae}
% \author{Chittaranjan Hens$^e$}\email{chittaranjan.hens@iiit.ac.in}
% \affiliation{$^a$SAFIR, Sorbonne University Abu Dhabi, UAE.\\
% 	$^b$Sorbonne Center for Artificial Intelligence, Sorbonne University, Paris, France.\\
%     $^c$Department of Computer Science \& Engineering, XIM University, Bhubaneswar, India.\\
%     $^d$Department of Data Science, SP Jain School of Global Management, Mumbai, India.\\
%     $^e$Center for Computational Natural Science and Bioinformatics, International Institute of Information Technology, Hyderabad, India.}
% % 
% % \email{}
% \date{\today}

\begin{abstract}
Complex biological networks, encompassing {metabolic pathways, gene regulatory systems, and protein–protein interaction networks}, often exhibit scale-free {structures characterized by heavy-tailed degree distributions}. However, empirical studies reveal significant deviations from ideal power-law behavior, underscoring {the need for more flexible and accurate probabilistic models.} In this work, we propose the Compounded Burr (CBurr) distribution, a novel four-parameter family derived by compounding the Burr distribution with a discrete mixing process. This model is specifically designed to capture both the body and tail behavior of real-world network degree distributions with applications to biological networks. We rigorously derive its statistical properties, including moments, hazard and risk functions, and tail behavior, and develop an efficient maximum likelihood estimation framework. The CBurr model demonstrates broad applicability to networks with complex connectivity patterns, particularly in biological, social, and technological domains. Extensive experiments on large-scale biological network datasets show that CBurr consistently outperforms classical power-law, log-normal, and other heavy-tailed models across the full degree spectrum. By providing a statistically grounded and interpretable framework, the CBurr model enhances our ability to characterize the structural heterogeneity of biological networks.
\end{abstract}

\maketitle

\begin{quotation}
The traditional Erdős-Rényi model, known for assuming random networks with a Poisson degree distribution, falls short in capturing the complex connectivity patterns exhibited by real-world complex networks. This limitation led to the introduction of scale-free networks, which are characterized by a power-law degree distribution and the presence of highly connected hubs alongside nodes with lower degrees. However, it is important to acknowledge that fitting the entire degree distribution to a pure power-law model may not always accurately represent the intricacies of biological network data, necessitating the exploration of alternative distribution models. Motivated by the above discussion, this study explores the degree distribution of biological networks and proposes a compounded probability model combining Burr distribution with the Poison model. This study proposes a new probability distribution designed to fit the node degree of the biological network that can provide a more accurate representation, which may enhance our ability to interpret biological phenomena. Biological networks also play crucial roles in understanding biological processes, disease mechanisms, and evolutionary dynamics. Therefore, it is also important to accurately model their degree distributions in order to find key nodes (genes, proteins, and metabolites) that are highly connected and have a big impact on how the network works. Hence, a new distribution that better fits biological networks may improve our ability to predict how perturbations or mutations affect network behavior and biological outcomes. 
\end{quotation}

\section{Introduction} \label{Introduction}
Biological networks often exhibit a scale-free topology, where a small number of nodes act as highly connected hubs, while the majority possess {relatively few connections} \cite{barabasi1999emergence, broido2019scale, chattopadhyay2021uncovering}. This {heterogeneous structure provides a crucial lens into the organizational principles, efficiency, and resilience} of biological systems \cite{wuchty2001scale, arita2005scale, khanin2006scale, wuchty2006architecture}. Driven by this, considerable efforts have been made to model real-world biological networks such as metabolic networks \cite{jeong2000large, fell2000small, ma2003connectivity}, protein-protein interaction networks \cite{jeong2000large, uetz2000comprehensive, schwikowski2000network, maslov2002specificity, rual2005towards}, protein domain networks \cite{rzhetsky2001birth, wuchty2001scale}, gene interaction maps \cite{tong2004global}, and genetic regulatory networks \cite{lee2002transcriptional, farkas2003topology, luscombe2004genomic}. These networks represent real-world complex systems and have been widely analyzed through the lens of graph theory \cite{haggarty2003chemical, dorogovtsev2003evolution}. Biological networks serve as a {valuable framework} for characterizing the {intricate structure} of biological systems \cite{lima2009powerful, miranda2016theoretical, saucan2021simple}. In such networks, nodes represent {biological entities (e.g., genes, proteins, metabolites)}, and edges capture {interactions or functional relationships} between them \cite{newman2010networks, barabasi2013network, albert2002statistical}. A node’s connectivity—its degree—is categorized into in-degree (incoming edges) and out-degree (outgoing edges), and {understanding the statistical distribution of degrees. This is critical for identifying central nodes and capturing global structural properties} \cite{artime2024robustness}.

The degree distribution plays a pivotal role in distinguishing between homogeneous and heterogeneous topologies, guiding how networks {evolve, adapt, and reorganize}. Analytical tools based on graph theory and degree distribution {allow researchers to decode the structural and functional characteristics} of complex biological systems \cite{barabasi1999emergence, newman2003structure, clote2020rna}. The advent of deep learning has further {augmented our ability to extract hidden patterns from these systems}, enhancing biological interpretation and predictive modeling \cite{muzio2021biological, jin2021application, pan2022dwppi, yang2022deep, ma2023predicting}. Although classical probability models like power-law, exponential, log-normal, and Poisson distributions have been extensively applied to model degree distributions, these approaches {often fall short in accurately capturing the nuances and heterogeneity} inherent in heavy-tailed biological networks. For instance, Khanin and Wit \cite{khanin2006scale} found that many biological networks, such as gene interaction networks defined by synthetic lethal interactions \cite{tong2004global}, and metabolic and protein interaction networks \cite{ito2000toward, schwikowski2000network}, {deviate from pure power-law behavior}. This has been echoed by Clauset et al. \cite{clauset2009power}, who showed that while some datasets conform to a power-law, others do not, based on {rigorous goodness-of-fit tests and maximum likelihood estimations}.

Similar critiques have emerged in subsequent works \cite{goldstein2004problems, salem2018biological}, highlighting the need for {distributions that better capture complex real-world network topologies}. Broido and Clauset \cite{broido2019scale} {underscored the rarity of strictly scale-free networks} and introduced nuanced classifications such as weak or strong scale-free types. Clote \cite{clote2020rna} {provided an efficient algorithm for estimating the degree distribution in RNA secondary structure networks}, revealing that these structures {do not conform to power-law models}. These limitations in existing distributions {necessitate the development of new models capable of accurately capturing the entire degree spectrum} of biological networks. Voitalov et al. \cite{voitalov2019scale} also demonstrated that {Pareto type-II (Lomax) distributions may better approximate scale-free behavior}. Recent efforts have introduced more flexible alternatives such as the {modified Lomax \cite{chattopadhyay2021modified}, Burr \cite{chakraborty2022searching}}, and other {heavy-tailed models \cite{chakraborty2022new}}. Still, even these enhanced models {sometimes fail to represent the full range of connectivity}, particularly among low-degree nodes for heavy-tailed complex networks \cite{stumpf2005subnets}. The Burr distribution \cite{burr1942cumulative}, known for its {versatility and capacity to model heavy-tailed behaviors}, offers a {promising foundation}. It can flexibly accommodate diverse topologies and levels of heterogeneity observed in biological networks. Theoretical extensions such as the exponentiated Burr \cite{ahmed2021new} and its nonlinear variants \cite{chakraborty2022new}, and Marshall-Olkin generalized Burr (Burr-MO) \cite{jayakumar2008generalization} have demonstrated superior performance across a variety of complex network settings. However, as shown in \cite{chakraborty2022searching}, even these models {sometimes fall short of fully capturing the distributional complexities of large-scale heavy-tailed biological networks.}

Motivated by the above discussion, this study proposes a novel Compounded Burr (CBurr) distribution to model the degree distributions of biological networks more accurately. By compounding the Burr distribution with a Poisson structure (introducing a Poison-shifted parameter), the proposed distribution enhances flexibility and better reflects empirical network characteristics — especially in cases where networks exhibit {heavy truncated tails}. The CBurr model {introduces a new parameterization that captures both the heavy-tailed nature and heterogeneity observed in real data}. Our empirical evaluation demonstrates that CBurr consistently outperforms classical distributions across various biological network datasets, without the need to {discard low-degree nodes}. Furthermore, we establish its theoretical properties, including estimation via maximum likelihood, making it practically useful.

Real-world networks such as those in {biology, sociology, and technology} often reflect {multiscale behaviors that defy strict adherence to classical distributional assumptions} \cite{wang2021arbitrary}. Hence, the CBurr distribution, by {blending multiple distributional behaviors}, offers a {robust alternative} for modeling the complex and diverse architectures of real-world networks. {Unlike social or technological networks, where preferential attachment may continue relatively unchecked, biological systems are governed by physical, functional, and evolutionary constraints. These constraints motivate building Poisson-Shifted Burr (we named it CBurr) model, which incorporates an initial heavy tail followed by a softened decay, aligning well with biological realism. Moreover, the study of risk and residual life functions in this context can directly inform biological interpretations, such as estimating the likelihood of further molecular interactions or identifying saturation effects in complex signaling pathways. While the model is broadly applicable, starting with biological networks ensures a tight linkage between statistical properties and biological meaning, and paves the way for future generalizations to other domains.}

The structure of the remaining paper is as follows: Section \ref{motivation} discusses a motivating example from biological networks, and Section \ref{model} presents the genesis of the proposed family of distributions and its statistical properties to provide a better understanding of the proposed model. Section \ref{proposed_method} introduces the Compounded Burr (CBurr) distribution, including its statistical properties such as parameter estimation. In Section \ref{experiments}, we present experimental results and analyze them in the context of various real-world complex networks. Finally, Section \ref{conclusion} concludes the paper with a concise summary and discussion.

\section{Motivating Example}\label{motivation}

\begin{figure*}
     \centering
     % \begin{subfigure}%[b]{0.47\textwidth}
     %     \centering
     \includegraphics[width=8.1cm, height=7.0cm]{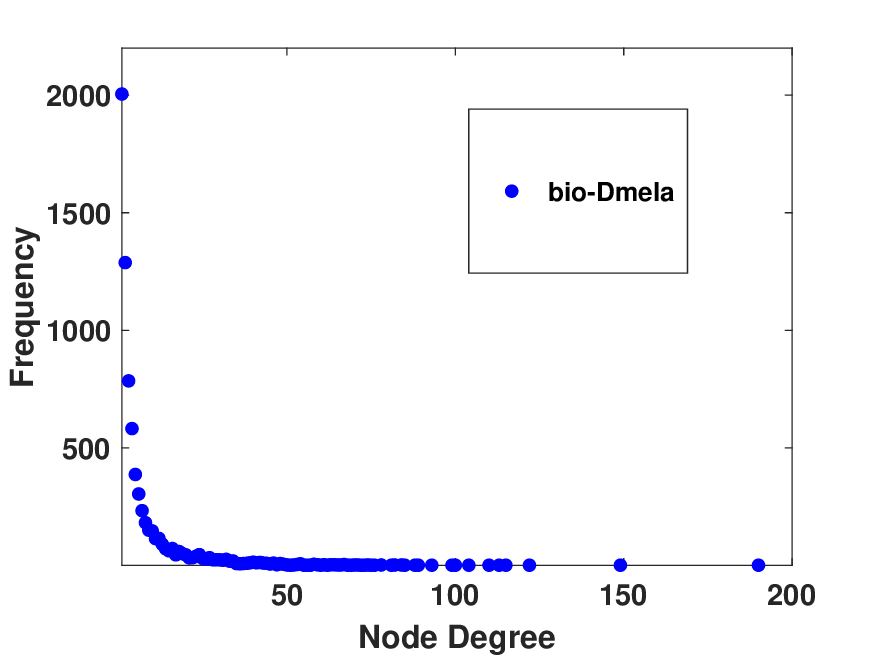}
      \includegraphics[width=8.1cm, height=7.0cm]{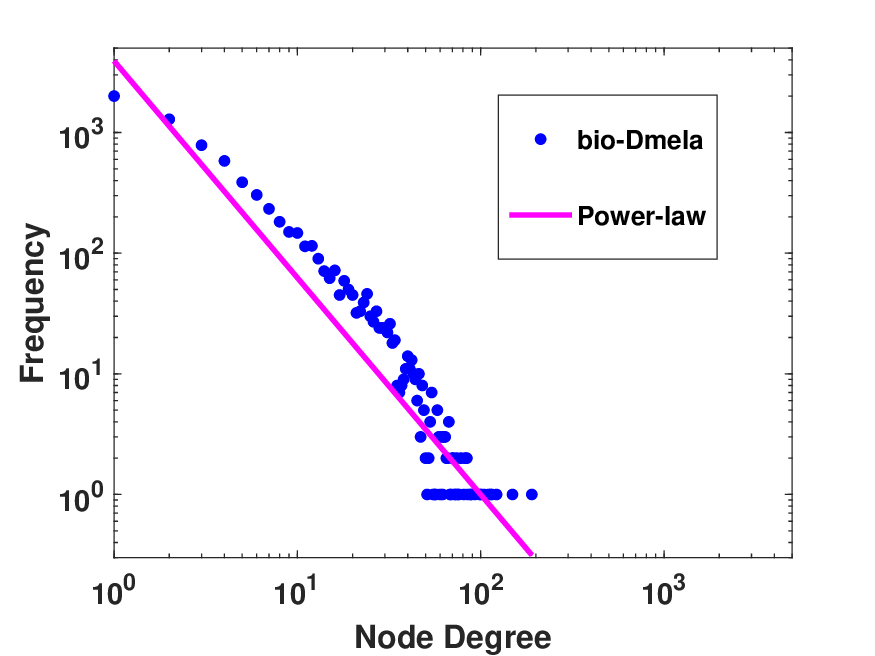}
    \caption{{Plots of the node degree distribution of bio-Dmela network in the Original scale (left) and in the Log-Log scale (right). It is clear from the plot (right) that a power-law distribution does not provide a `good' fitting for the bio-Dmela data, and there is a need for a more flexible heavy-tailed probability distribution to provide a `good' nonlinear fitting.}}
    \label{original_loglog_bio_dmela_dd}
\end{figure*}

One of the key structural features in analyzing real-world complex networks is the degree distribution. Degree distribution of nodes is one of the most significant structural characteristics of social, information, and biological networks. Previous empirical studies on real-world network patterns suggest that their degree distributions follow a single power-law, often called a scale-free network \cite{newman2003structure}. The main reason behind the emergence of this phenomenon is the mechanism of preferential attachment, which states that in a growing network, a node with a higher degree is more likely to receive new links. In 1999, Barabási \cite{barabasi1999emergence} modeled the degree distribution of the World Wide Web (WWW) using a single power-law. Since then, this structural pattern has been extensively studied across various real-world networks, including metabolic, gene regulatory, collaboration, communication, and social networks \cite{newman2010networks}. These networks are inherently dynamic, as their vertices and/or edges continue to grow over time. This dynamic nature can be effectively captured by studying the underlying pattern of node degree distribution. However, upon closer examination of the entire network data, this single power-law distribution often fails to accurately represent the data due to the presence of large fluctuations and sparsity in the upper and lower tails of the degree distribution. If the degree distribution followed a single power-law, then a single straight line would be enough to fit all the data points completely when viewed on the log-log scale. 

{For illustration, we consider bio-dmela (Drosophila melanogaster) PPI (Protein-Protein Interaction) networks, which are characterized by their heavy-tailed behaviour \cite{zhang2019protein}. These networks are typically highly skewed, with many low-degree nodes and few hubs (high-degree nodes). In particular, bio-dmela are known to follow sub-exponential or power-law–like distributions, but often deviate from a pure power-law due to biological constraints or network truncations \cite{broido2019scale}.}
We analyze a bio-Dmela heavy-tailed network \cite{singh2008-isorank-multi} and visualize the node degree distribution on both the original scale and the log-log scale (see Fig. \ref{original_loglog_bio_dmela_dd}) as a motivating example. When attempting to fit a single power-law model to the entire bio-Dmela network data, we observed that the pink straight line (see Fig. \ref{original_loglog_bio_dmela_dd} (right)) representing the power-law fit failed to align with many data points, particularly those corresponding to lower-degree nodes, indicating a poor fit in those regions. This suggests that when modeling with a single power-law, greater emphasis is placed on higher-degree nodes, while many lower-degree nodes, potentially containing valuable information about the network, are disregarded. Finally, our observation from Fig. \ref{original_loglog_bio_dmela_dd} (right) reveals that the node degree distribution of the bio-Dmela network, when plotted on a log-log scale, exhibits nonlinearity and heavy-tailed nature upon closer inspection. This behavior differs from the pattern observed when a single power-law model is applied and has also been noted in many other biological networks (discussed in Sec \ref{experiments}). Therefore, when fitting the degree distribution, it is crucial to consider all nodes, including both lower-degree and higher-degree nodes, to accurately capture the true pattern of the network's degree distribution. This motivates the need to find a model that can accurately describe the main features of heavy-tailed and long-tailed behavior in the entire degree distribution of real-world complex networks. This article contributes to that effort by introducing a new heavy-tailed compounded Burr distribution. In the proposed model, the Poisson shift parameter ($\lambda$) gives compounded Burr distribution an advantage over static models like log-normal or standard power-law.

{The proposed Poisson-shifted generalization and resulting CBurr model are not limited only to biological networks — they are applicable to a wide range of real-world complex networks (e.g., technological and social networks) that exhibit heavy-tailed degree distributions with finite-size effects or saturation behavior. The Poisson-shifting mechanism introduces a flexible exponential tail, allowing the model to capture non-monotonic behaviors and cutoff effects that arise naturally in complex real-world heavy-tailed networks. We focused on biological networks because they offer a biologically meaningful setting where both heavy-tailed degree distributions and saturation effects are prominently observed.}

\section{Model Formulation}\label{model}
{The proposed Compounded Burr (CBurr) model is motivated by the following intuition that node degree results from a random number of latent interactions, each with heavy-tailed potential. For example, if we consider protein-protein interactions (PPI), the degree of a node (i.e., the number of interactions) may result from a random number of underlying biological processes (e.g., gene expression levels, functional complexes, evolutionary duplications), which the Poisson component captures. The Burr distribution is popularly used in statistics and survival analysis for modeling heavy-tailed data effectively. This makes it suitable for the long right tails seen in biological network degrees. The proposed compounded Burr model uses a Poisson-shifting that adds flexible truncation or tempering effect. By taking the minimum of a Poisson-distributed number of Burr variables, the resulting distribution gains heavier or tempered tails, enabling it to mimic real-world degree distributions more closely than Burr alone. This compounded nature allows the model to capture both global variability (from Poisson) and local heavy-tail scaling (from Burr) — ideal for biological networks where some hubs dominate while others have moderate or sparse connectivity.

The rationale behind proposing a CBurr model, where $N-1 \sim \operatorname{Poisson}(\lambda)$ with $\lambda \in (0,\infty)$ and $X_1, X_2, \cdots \sim$ Burr - lies in capturing both heavy-tail behavior and variability in the number of competing mechanisms in biological networks.
Let $Y =\min\{X_1, X_2,\cdots, X_N\}$ be the effective or observable node (protein in PPI network example) degree, which gives the minimum of multiple potential contributions. Each $X_i$ is the degree a node would have under a specific interaction-generating mechanism (e.g., a biological process, experimental condition, structural domain, or evolutionary module), and each follows a Burr distribution. Here, $N$ is the number of such mechanisms the protein is exposed to, or the number of interaction opportunities or modules a protein is involved in, modeled as a Poisson random variable. For example, a protein might participate in a random number of functional domains, biological pathways, or interaction modules. Each of these modules offers an opportunity for the protein to gain connections, so the total number of such opportunities (interactions) is modeled as a Poisson variable plus 1 (since the model requires at least one event), i.e., $N-1 \sim \operatorname{Poisson}(\lambda)$. Since $X_i$ represents the potential interaction strength or connectivity level contributed by a single latent mechanism, therefore, Burr distribution is used because it can capture heavy-tailed behavior, representing that some mechanisms may yield many interactions (hubs) while most yield few. So, if a protein is part of $N$ such modules, the degree of the node is modeled as the minimum of these Burr-distributed random variables, mimicking competition or limiting influence between mechanisms. Therefore, this compounded framework helps account for over-dispersion, truncation, and tail behavior observed in real-world biological networks. Now, we present this new compounded family of distributions for the general case.}

Let $X_1, X_2, \ldots$ be i.i.d. (independent and identically distributed) random variables with CDF $F(x)$ and let
$
Y=\min \left\{X_1, \ldots, X_N\right\}, \quad \text { where } N-1 \sim \operatorname{Poisson}(\lambda) \; \text{with} \; \lambda \in (0,\infty),
$ 
where $\lambda$ controls the expected number of potential interaction trials. So, we have a random number of i.i.d. shocks, but there is always at least one shock with $N \in\{1,2,3, \ldots\}$, and $$\mathbb{P}(N=n)=\frac{e^{-\lambda} \lambda^{n-1}}{(n-1)!}, n=1,2, \ldots$$
This reflects a realistic biological setting where every node (e.g., a protein) has at least one potential interaction. But additional connections or roles (beyond the first) are random, modeled via a Poisson distribution. Then the survival function for $Y$ becomes:
\begin{align} \label{e2}
    \bar{G}(y) & = P\left[X_1 \geq y, X_2 \geq y, \ldots, X_n \geq y \right] := P(Y \geqslant y) \nonumber \\ 
    & = \sum_{n = 1}^{\infty} P\left[Y \geq y | N-1 = n-1\right] P\left[N-1 = n-1\right] \nonumber \\ 
    & = \displaystyle\sum_{n=1}^{\infty}\left[\bar{F}(y)\right]^{n} \frac{e^{-\lambda}\lambda^{n-1}}{(n-1)!} =\frac{\bar{F}(y)}{\exp\{\lambda F(y)\}}, 
\end{align}
{where $\bar{F}(y)$ and $\bar{G}(y)$ represent the survival function of $X$ and $Y$. Mathematically, $\bar{F}=1-F$, $\bar{G}=1-G$, and $F(x)=P[X \leq x]$ is the distribution function of $X$. For non-negative random variables, the survival function is more meaningful with a more convenient form than the better-known cumulative distribution function. $G(y)$ is a valid, smooth, unimodal distribution with a tunable tail via the Poison shift parameter $\lambda$. For modeling biological networks, where data has heavy-tailed behavior as discussed in Sec. \ref{motivation}, we choose $\bar{F}$ as the survival function of the Burr random variables. Thus, the survival function in \ref{e2} is not a member of the original Marshall-Olkin (MO) family \cite{marshall1997new}, but could be seen as a Poisson-shifted Generalized MO distribution.

\begin{r1}
\begin{enumerate}
    \item {\bf Role of $\lambda$:} Suppose $F$ is a Burr random variable, then lower value of $\lambda \; (\sim 0.1-1)$ implies that the model behaves close to Burr; tail is heavy but less suppressed. Very high value of $\lambda \; (>5)$ suggests that the tail becomes more sharply truncated $\rightarrow$ (mimics power-law cutoff). The intermediate value of $\lambda$ introduces subtle curvature in the body of the distribution, modeling non-monotonic or multi-slope behavior.
    \
    \item {\bf Interpretation of $\lambda$ in biological networks:} For practical use, $\lambda$ can be tuned to reflect biological heterogeneity, for e.g., networks with more alternative binding modes or greater stochasticity in forming connections will require a larger value of $\lambda$. In the PPI network example or other biological networks, a higher value of $\lambda$ indicates more variability and more trials before observing the minimal interaction event and a lower value of $\lambda$ suggests fewer competing trials, so the minimum may be less extreme. Overall, it helps in mimicking realistic saturation or resource constraints, when the probability of forming new connections diminishes as the degree grows.
\end{enumerate}
\end{r1}}

\noindent It is worth noting that the ratio ${\bar{F}(y)}/{\bar{G}(y)}$ is non-decreasing for $y \geqslant 0$, shown as
\begin{equation}\label{e1}
    \frac{\bar{F}(y)}{\bar{G}(y)} =\exp\{\lambda F(y)\}.
\end{equation}
The associated PDF of $Y$ is expressed as:
\begin{equation}\label{MMO1}
    g(y;\lambda)=\frac{f(y)\left[1+\lambda\bar{F}(y)\right]}{\exp\{\lambda
F(y)\}},~~~~-\infty<y<\infty,~0\leqslant \lambda <\infty. 
\end{equation}
We also investigate the distribution of the maximum, i.e., $Z =\max\{X_1,X_2,\cdots,X_N\}.$
This will be useful when the focus is on rare, highly connected nodes, and tail behavior dominates the network's structure. We write the distribution function of $Z$ as follows:
\begin{align*}
    {G}(z):= P(Z \leqslant z) &=\displaystyle\sum_{n=1}^{\infty}\left[F(z)\right]^{n} \frac{e^{-\lambda}\lambda^{n-1}}{(n-1)!} =\frac{F(z)}{\exp\{\lambda \bar{F}(z)\}},
\end{align*}
where $\bar{F}(z)$ represents the survival function of $X$. The associated probability density function (PDF) of $Z$, denoted as $g(z;\lambda)$, is given by:
\begin{align}
    g(z;\lambda)=\frac{f(z)\left[1+\lambda F(z)\right]}{\exp\{\lambda\bar{F}(z)\}},~~~~-\infty<z<\infty,~0\leqslant \lambda <\infty.
\label{MMO2}
\end{align}
{Therefore, $G(\cdot)$ is a generalization of the base distribution $F(\cdot)$ and $G$ satisfies all the properties of a cumulative distribution function when $X$ is a continuous random variable with the cumulative distribution function $F(\cdot)$. However, this study focuses on the distribution function defined in (\ref{e2}) using a Poisson-distributed number of Burr shocks and taking the minimum that is more relevant for biological networks.

A skewing mechanism is a method for introducing additional asymmetry or tail variation into a distribution by applying a probabilistic transformation.} This compounding method acts as a skewing mechanism, as illustrated by the following theorem:
\begin{t1}\label{th1}
If the original distribution $F$ of $X$ is symmetric, then the distribution of $-Y$ is equivalent to the distribution of $Z$.
\end{t1}
\begin{proof}
    \begin{align}
    g(-y;\lambda)=\frac{f(y)\left[1+\lambda {F}(y)\right]}{\exp\{\lambda\bar{F}(y)\}},~~~~-\infty<y<\infty,~0\leq \lambda <\infty.
\end{align}
It is evident that the PDFs in (\ref{MMO1}) and (\ref{MMO2}) of $Y$ and $Z$ can be expressed as weighted versions of any random variable using the following weight functions:
$$ w(y; \lambda) = \frac{1+\lambda \bar{F}(y)}{\exp\{\lambda
F(y)\}} \; \; \text{and} \; \; w(z; \lambda) = \frac{1+\lambda
F(z)}{\exp\{\lambda \bar{F}(z)\}}.$$
The weight functions $w(y;\lambda)$ and $w(z;\lambda)$ exhibit different monotonicity properties. Specifically, $w(y;\lambda)$ is a non-increasing function, while $w(z;\lambda)$ is a non-decreasing function, for $\lambda \geq 0$. Consequently, if the base random variable (RV) $X$ has a decreasing PDF, then $Y$ also has a decreasing PDF. Additionally, in the case where $X$ follows an unimodal PDF, it can be observed that the mode of $Y$ is less than the mode of $X$ for $\lambda > 0$.
\end{proof}
The following theorem demonstrates that the proposed method results in a generalization of the base distribution, preserving the same moment existence properties as the original distribution for any general $F$.  This theorem establishes that the newly derived distribution retains the moment characteristics of the underlying distribution, ensuring consistency in terms of moment existence.

\begin{t1}\label{th2}
The moments of $G$ exist in the same order as in the original distribution $F$.
\end{t1}
\begin{proof}
\begin{align}
    g(y;\lambda)= w(y;\lambda)f(y),
\end{align}
where $w(y;\lambda)$ takes values between $(1/e,1)$ when $0< \lambda <1$ and $(0,1)$ if $\lambda > 1$. The result follows. 
\end{proof}
\begin{r1}
Also, note that if the method is applied twice, another new family turns up, viz.,
$$\bar{H}(y) = \frac{\bar{F}(y)}{\exp\{\lambda [F(y)+G(y)]\}}.$$
\end{r1}

\subsection{Hazard rates and risk functions}
In the context of network data analysis, we introduced a transformation to the distribution $G(y: \lambda)$, which incorporates a new parameter $\lambda \geq 0$ into any existing family of continuous distributions. 
%This transformation is defined based on the survival function of the distribution given:
% \begin{align*}
%     \bar{G}(y) &=\frac{\bar{F}(y)}{\exp\{\lambda F(y)\}}.
% \end{align*}
% By assuming the continuity of $F$ throughout, we can derive the corresponding density function as follows:
% \begin{align*}
%     g(y;\lambda)&=\frac{f(y)\left[1+\lambda\bar{F}(y)\right]}{\exp\{\lambda F(y)\}},~~~~-\infty<y<\infty,~0\leq \lambda <\infty. 
% \end{align*}
The parameter $\lambda$ can also be interpreted in terms of the ratio of the hazard rates between $G$ and $F$, as expressed by the following equation:
\begin{equation}\label{haz}
    \frac{r_{G}(y)}{r_{F}(y)} = 1+\lambda \bar{F}(y). 
\end{equation}
Notably, $G$ encompasses a broader class of distributions compared to the original distribution $F$; we name it \textit{compounded family of distributions}. The hazard ratio, denoted as $[1+\lambda \bar{F}(y)]$, indicates that the model follows a proportional hazard-based representation similar to the Cox proportional hazard model \cite{cox1972regression}.
Note that $$\displaystyle \lim_{y \to - \infty}  r_{G}(y; \lambda)=\displaystyle \lim_{y \to - \infty} (1+\lambda) r_{F}(y) \; \text{and} \;  \displaystyle \lim_{y \to  \infty}  r_{G}(y; \lambda)=\displaystyle \lim_{y \to \infty}  r_{F}(y).$$ {The hazard rate of the compounded model can be interpreted as follows: (a) If $\lambda>0$, then the hazard is inflated in the early range; (b) If $\lambda<0$ (theoretically impossible but, in practice, happens due to poor initialization), then the hazard is deflated, possibly leading to non-monotonic shapes; (c) When $\lambda=0$, the model reduces to the baseline distribution. In the PPI network example, at lower degrees, the hazard rate is typically low, indicating that proteins with few interactions have a reduced propensity to form new links. As node degree increases, the hazard rate tends to rise, reflecting a phase of rapid interaction accrual commonly associated with hub-like or multifunctional proteins. This non-monotonic hazard behavior supports the relevance of the CBurr formulation in capturing nuanced biological constraints observed in empirical PPI networks and other biological networks.}

The following theorem establishes limits on the hazard rate function and the survival function.
\begin{t1}
For all $\lambda >0$, 
\begin{enumerate}
\item [(i)] $r_{F}(y) \leqslant r_{G}(y;\lambda) \leqslant \left(1+\lambda\right) r_{F}(y);$
\item [(ii)]$\bar{F}^{1+\lambda}(y) \leqslant \bar{G}(y;\lambda) \leqslant \bar{F}(y).$
\end{enumerate}
\end{t1}
\begin{proof}
From (\ref{haz}), we can easily verify that for $\lambda>0$, $ r_{G}(y;\lambda)\geqslant r_{F}(y)$. Again, one can see that for $\lambda>0$,
\begin{align*}
    r_{G}(y;\lambda)- \left(1+\lambda\right)r_{F}(y)&= \left(1+\lambda \bar{F}(y)\right)r_{F}(y)-\left(1+\lambda\right)r_{F}(y)\\
&=-\lambda F(y)r_{F}(y)\leqslant 0.
\end{align*}
Therefore, the result (i) is deduced, and by utilizing the result (i), we can establish the validity of the result (ii).
\end{proof}
The subsequent theorem provides valuable insights into the concepts of excess risk and relative risk, which will be of interest in epidemiology and medical science \cite{panja2023epicasting}. {Excess risk in biological networks quantifies how much more likely a node (protein) is to gain further interactions beyond a baseline, helping to identify emerging hubs. Relative risk compares the likelihood of connectivity between two nodes, offering insights into functional disparity. Together, they aid in pinpointing critical proteins and understanding interaction dynamics in health and disease contexts.}
\begin{t1}
For all $\lambda >0$, 
$$0\leqslant G(y)-F(y) \leqslant e^{\lambda}-1~~ \; \mbox{and} \; ~~\frac{\bar{G}(y)}{\bar{F}(y)} \; \mbox{is decreasing in}~~y.$$
\end{t1}
\begin{proof}
From Eq. (\ref{e1}), it follows that for $\lambda >0$, $\bar{G}(y) \leqslant \bar{F}(y)$. Furthermore, it is evident that for values of $\lambda$ greater than zero,
\begin{align*}
    G(y)-F(y)&\leqslant \left[\exp\{\lambda F(y)\}-1\right]\bar{F}(y) \leqslant e^{\lambda}-1.
\end{align*}
$ \mbox{Hence, for} \; \lambda> 0, \; 0\leqslant G(y)-F(y) \leqslant e^{\lambda}-1.$ \\
Moreover, \noindent $\frac{\bar{G}(y)}{\bar{F}(y)}=\exp\{-\lambda F(y)\}$ is decreasing in $y$ for $\lambda> 0$. 
\end{proof}
Thus, relative risk provides an increase or decrease in the likelihood of an event based on some exposure and is an important consideration in biological and medical science.
\subsection{Moments} \label{statistics_prop}
In probability and statistics, moments are popularly used to describe the characteristics of a distribution. For the compounded family of distributions, we can find closed-form expressions of moments derived below.
The density function of $Y$ given in (\ref{MMO1}) can be written as
\begin{align}\label{e4}
    g(y;\lambda)=\left[1+\lambda\bar{F}(y)\right]e^{-\lambda F(y)}f(y), \; -\infty<y<\infty, \; 0\leq \lambda <\infty. 
\end{align}
Now, for $\lambda>0$, Eq. (\ref{e4}) can be expressed as 
\begin{align}\label{e5}
    g(y;\lambda)=&f(y)\left[1+\lambda \bar{F}(y)\right]\displaystyle\sum_{j=0}^{\infty}\frac{\left(\lambda \bar{F}(y)-\lambda\right)^{j}}{j!}\nonumber \\ \nonumber 
    =&f(y) \left[ \displaystyle\sum_{j=0}^{\infty}\displaystyle\sum_{k=0}^{j}\frac{1}{j!}\binom jk (-1)^{j-k}\lambda^{j}\bar{F}^{k}(y) \; + \right.\\ \nonumber & \left. \quad \quad \displaystyle\sum_{j=0}^{\infty}\displaystyle\sum_{k=0}^{j}\frac{1}{j!}\binom jk (-1)^{j-k}\lambda^{j+1}\bar{F}^{k+1}(y) \right]\nonumber   \\ 
    =&  f(y)\left[\displaystyle\sum_{j=0}^{\infty}\displaystyle\sum_{k=0}^{j}a_{j,k}\bar{F}^{k}(y)+\displaystyle\sum_{j=0}^{\infty}\displaystyle\sum_{k=0}^{j}\lambda a_{j,k}\bar{F}^{k+1}(y)\right],
 \end{align}
\noindent where $a_{j,k}=\frac{1}{j!}\binom jk (-1)^{j-k}\lambda^{j}$. Probability weighted moments (PWMs), proposed by Greenwood et al. \cite{greenwood1979probability}, are statistical measures that represent expectations of certain functions of a random variable when its mean exists. If we express the PWMs in terms of the density function $f(y)$, then the corresponding equation is given by
\begin{align*}
    M_{p,r,s}&=\displaystyle\int_{-\infty}^{\infty}x^{p}(F(x))^{r}\left(1-F(x)\right)^{s}f(x)dx,
\end{align*}
where $p, r$, and $s$ are positive integers. Note that the moments $M_{p,0,0}$ are the noncentral conventional moments. The PWMs for the baseline survival function $\bar{F}(y)$ are defined by
\begin{align*}
    M_{p,0,s}&=\displaystyle\int_{-\infty}^{\infty}x^{p}\left(\bar{F}(x)\right)^{s}f(x)dx. 
\end{align*}
Hence, from Eq. (\ref{e5}), the $r$-{th} moment of $Y$ for $\lambda>0$ can be written as
\begin{align}\label{e11}
E\left(Y^{r}\right)&=\displaystyle\int_{-\infty}^{\infty}y^{r}g(y)dy\nonumber\\
&=\displaystyle\int_{-\infty}^{\infty}\displaystyle\sum_{j=0}^{\infty}\displaystyle\sum_{k=0}^{j}a_{j,k}y^{r}\bar{F}^{k}(y)f(y)dy \; + \nonumber \\
& \quad \displaystyle\int_{-\infty}^{\infty}\displaystyle\sum_{j=0}^{\infty}\displaystyle\sum_{k=0}^{j}\lambda a_{j,k}y^{r}\bar{F}^{k+1}(y)f(y)dy\nonumber\\
&=\displaystyle\sum_{j=0}^{\infty}\displaystyle\sum_{k=0}^{j}a_{j,k}\displaystyle\int_{-\infty}^{\infty}y^{r}\bar{F}^{k}(y)f(y)dy \; + \nonumber\\
% & \quad \displaystyle\sum_{j=0}^{\infty}\displaystyle\sum_{k=0}^{j}\lambda a_{j,k}\displaystyle\int_{-\infty}^{\infty}y^{r}\bar{F}^{k+1}(y)f(y)dy\nonumber\\
&=\displaystyle\sum_{j=0}^{\infty}\displaystyle\sum_{k=0}^{j}a_{j,k}M_{r,0,k}+\lambda \displaystyle\sum_{j=0}^{\infty}\displaystyle\sum_{k=0}^{j} a_{j,k}M_{r,0,k+1}.
\end{align}
Therefore, we can see a {semi-closed-form for the moments of the proposed family of distributions (when the baseline distribution is Burr/Pareto distributions). In biological network modeling, even if closed-form moments are unavailable, empirical moments and simulation-based estimates (bootstrap, MCMC) remain powerful tools to quantify mean connectivity, variability, and skewed risk of interaction across different protein classes.}
The following result shows that the mean residual life (MRL) function of the random variable $Y$ can be represented by PWMs of the residual random variable $X_{t}=\left(X-t|X>t\right)$.
\begin{t1}
The mean residual, given the survival to $t$ until the protein-protein interaction network experiences its first damaging event of the compounded family of distribution, can be obtained as follows, following \cite{marshall1997new}:
\begin{align*}
    \mu_{G}(t)= \; & E\left(Y-t|Y>t\right)\\
     = \; & e^{\lambda F(t)}\displaystyle\sum_{j=0}^{\infty}\displaystyle\sum_{k=0}^{j}a_{j,k}\left[M_{1,0,k}(t)+\lambda M_{1,0,k+1}(t)\right].
\end{align*}
\end{t1}
\begin{proof}
Suppose that $Y_{t}=\left(Y-t \; | \; Y>t\right)$ be the residual life of the system at age $t$ with density function $g(y+t)/\bar{G}(t)$, for $\bar{G}(t)>0$. Then for $\lambda>0$, the mean residual lifetime of the random variable $Y$ is given by
\begin{align*}
    \mu_{G}(t)&=\frac{1}{\bar{G}(t)}\displaystyle\int_{0}^{\infty}y g(y+t)dy\\
    &=\frac{1}{\bar{G}(t)}\displaystyle\int_{0}^{\infty}y f(y+t)  \left[\displaystyle\sum_{j=0}^{\infty}\displaystyle\sum_{k=0}^{j}a_{j,k}(\bar{F}(y+t))^{k} \; + \right. \nonumber \\  & \quad \quad\quad\quad\quad\quad\quad\quad\quad \left. \displaystyle\sum_{j=0}^{\infty}\displaystyle\sum_{k=0}^{j}\lambda a_{j,k}(\bar{F}(y+t))^{k+1}\right]dy\\
    &=\frac{e^{\lambda F(t)}}{\bar{F}(t)}\displaystyle\sum_{j=0}^{\infty}\displaystyle\sum_{k=0}^{j}  \biggl[\displaystyle\int_{0}^{\infty} y (\bar{F}(y+t))^{k}f(y+t)dy \; + \nonumber \\  & \quad \quad\quad\quad\quad\quad\quad\lambda \displaystyle\int_{0}^{\infty} y (\bar{F}(y+t))^{k+1}f(y+t)dy\biggr]\\
    &= e^{\lambda F(t)}\displaystyle\sum_{j=0}^{\infty}\displaystyle\sum_{k=0}^{j}\left[M_{1,0,k}(t)+\lambda M_{1,0,k+1}(t)\right],
\end{align*}
where, $M_{1,0,k}(t)=\frac{1}{\bar{F}(t)}\displaystyle\int_{0}^{\infty} y (\bar{F}(y+t))^{k}f(y+t)dy$.
\end{proof}
The MRL function $\mu_{G}(t)$ at age $t$ is defined to be the expected remaining life given survival to age $t$; it is a function of interest in actuarial studies, survival analysis, and reliability. {The MRL function plays an important role in understanding network growth dynamics in biological systems like PPI networks. In these networks, nodes represent proteins, and edges represent the interactions between them. The MRL quantifies the expected number of additional interactions a protein may gain, given it already has a certain number of connections.}

\section{Proposed CBurr distribution}\label{proposed_method}
Irving W. Burr introduced $12$ distinct cumulative distribution functions, which have been valuable for data analysis \cite{burr1942cumulative}. Among these functions, the Burr $XII$ distribution (commonly known as the Burr distribution) has garnered significant attention in the past decade due to its ability to generate a broad range of skewness and kurtosis values \cite{tadikamalla1980look}. The Burr distribution is particularly well-suited for analyzing heavy-tailed network data, surpassing distributions such as Exponential, Gamma, and Weibull \cite{chakraborty2022new}. 

\subsection{Definition and Statistical Properties}
The cumulative distribution function (CDF) and PDF of the Burr distribution are mathematically formulated as given below \cite{johnson1995continuous}.
\begin{d1} 
Let $X$ be a random variable following the Burr distribution with parameters $\alpha$, $\gamma$, and $c$. The cumulative distribution function (CDF) of $X$ can be expressed as:
\begin{equation} \label{e13}
    F(y;\alpha, \gamma, c)= 1-\left[1+\left(\frac{y}{\gamma}\right)^{c}\right]^{-\alpha}, y>0, \alpha, \gamma, c>0,
\end{equation}
\noindent where, $\gamma$ represents the scale parameter, and $\alpha$ and $c$ are shape parameters. The probability density function (PDF) of the Burr distribution is given by:
\begin{equation} \label{e14}
    f(y;\alpha, \gamma, c)= c\alpha \gamma^{-c}y^{c-1}\left[1+\left(\frac{y}{\gamma}\right)^{c}\right]^{-\alpha-1}, y>0, \alpha, \gamma, c>0.
\end{equation}
\end{d1}
The survival function corresponding to the Burr distribution can be defined as follows:
\begin{equation}\label{e15}
\bar{F}(y;\alpha, \gamma, c)= \left[1+\left(\frac{y}{\gamma}\right)^{c}\right]^{-\alpha}, y>0, \alpha, \gamma, c>0.
\end{equation}
The hazard rate function associated with the Burr distribution can be expressed as:
\begin{equation}\label{e16}
h(y;\alpha, \gamma, c)=c\alpha\gamma^{-c}y^{c-1}\left[1+\left(\frac{y}{\gamma}\right)^{c}\right]^{-1}, y>0, \alpha, \gamma, c>0.
\end{equation}
In the specific case where the random variables $X_i$ follow the Burr distribution with the distribution function given in Eq. (\ref{e13}), the proposed compounded family of distribution in Eq. (\ref{e2}) can be written as:
\begin{equation}\label{cburrcdf}
\bar{G}(y)=e^{-\lambda}\left[1+\left(\frac{y}{\gamma}\right)^{c}\right]^{-\alpha}e^{\lambda\left(1+\left(\frac{y}{\gamma}\right)^{c}\right)^{-\alpha}}.
\end{equation}
The associated PDF for CBurr distribution is given by:
\begin{equation}\label{e12}
    g(y)=\frac{c\alpha y^{c-1} e^{-\lambda}e^{\lambda\left(1+\left(\frac{y}{\gamma}\right)^{c}\right)^{-\alpha}}}{\gamma^{c}\left[1+\left(\frac{y}{\gamma}\right)^{c}\right]^{\alpha+1}}\left[1+\frac{\lambda}{\left(1+\left(\frac{y}{\gamma}\right)^{c}\right)^{\alpha}}\right],
\end{equation}
\noindent and the corresponding hazard rate function is given by:
\begin{equation*}
r_{G}(y)=\frac{c\alpha \gamma^{-c}y^{c-1}}{1+\left(\frac{y}{\gamma}\right)^{c}}\left[1+\frac{\lambda}{\left(1+\left(\frac{y}{\gamma}\right)^{c}\right)^{\alpha}}\right].
\end{equation*}
{
The hazard starts small for small $y$ and grows non-monotonically depending on $\gamma, \alpha, c, \lambda$. When $\lambda$ is large, the exponential term dampens the tail, reducing hazard at high $y$. When $\lambda<0$ (theoretically implausible but in practice sometimes observed while fitting), hazard can spike unnaturally, indicating modeling instability or data underfit. The hazard rate in the CBurr Model captures the instantaneous risk of observing a higher connectivity (degree) in a biological network, given that the current degree is $y$. In the context of PPI networks, this reflects how likely a node (protein) is to acquire additional interactions as its degree increases. A non-monotonic hazard shape—rising initially and then flattening or declining—can indicate that proteins with moderate degrees are more likely to gain new interactions, while high-degree nodes may saturate due to biological constraints like spatial limitations or functional specificity. This aligns with empirical observations in real-world biological network data, where hubs exist but are biologically bounded.}

By utilizing Eq. (\ref{e5}), we can express the density function as follows:
% \scriptsize{
\begin{align*}
    g(y;\lambda)= & c \alpha \gamma^{-c} y^{c-1} \left(1+\left(\frac{y}{\gamma}\right)^{c}\right)^{-\alpha-1} \\ & \Biggl[\displaystyle\sum_{j=0}^{\infty}\displaystyle\sum_{k=0}^{j}a_{j,k}\left(1+\left(\frac{y}{\gamma}\right)^{c}\right)^{-k\alpha} \; + \\
    & \; \lambda\displaystyle\sum_{j=0}^{\infty}\displaystyle\sum_{k=0}^{j}a_{j,k} \left(1+\left(\frac{y}{\gamma}\right)^{c}\right)^{-(k+1)\alpha}\biggr]
\end{align*}
where $a_{j,k}=\frac{1}{j!}\binom jk (-1)^{j-k}\lambda^{j}$. By examining Eq. (\ref{e11}), it is evident that
\begin{equation*}
E\left(Y^{r}\right) =  \displaystyle\sum_{j=0}^{\infty}\displaystyle\sum_{k=0}^{j}a_{j,k}M_{r,0,k}+\lambda \displaystyle\sum_{j=0}^{\infty}\displaystyle\sum_{k=0}^{j} a_{j,k}M_{r,0,k+1},
\end{equation*}
where 
$$
M_{r,0,k}=\displaystyle\int_{-\infty}^{\infty}y^{r}\bar{F}^{k}(y)f(y)dy 
% &=\displaystyle\int_{0}^{\infty}y^{r}\left(1+\left(\frac{y}{\gamma}\right)^{c}\right)^{-k\alpha}c\alpha \gamma^{-c}y^{c-1}\left(1+\left(\frac{y}{\gamma}\right)^{c}\right)^{-\alpha-1}dy\\
% &=\frac{c\alpha}{\gamma^{c}}\displaystyle\int_{0}^{\infty}y^{r+c-1} \left(1+\left(\frac{y}{\gamma}\right)^{c}\right)^{-\alpha(k+1)-1}dy\\
=\alpha \gamma^{r}\frac{\Gamma\left(\alpha(k+1)-\frac{r}{c}\right)\Gamma\left(\frac{r}{c}+1\right)}{\Gamma\left(\alpha(k+1)+1\right)},$$ and
$$M_{r,0,k+1}=\alpha \gamma^{r}\frac{\Gamma\left(\alpha(k+2)-\frac{r}{c}\right)\Gamma\left(\frac{r}{c}+1\right)}{\Gamma\left(\alpha(k+2)+1\right)}.$$
Therefore, the $r^{th}$ moment of the random variable $Y$ can be expressed as
\begin{align*}
    E(Y^{r})=\alpha\gamma^{r}\Gamma\left(\frac{r}{c}+1\right)\Biggl[\displaystyle\sum_{j=0}^{\infty}\displaystyle\sum_{k=0}^{j}a_{j,k}\frac{\Gamma\left(\alpha(k+1)-\frac{r}{c}\right)}{\Gamma\left(\alpha(k+1)+1\right)} \\     +\lambda\displaystyle\sum_{j=0}^{\infty}\displaystyle\sum_{k=0}^{j}a_{j,k}\frac{\Gamma\left(\alpha(k+2)-\frac{r}{c}\right)}{\Gamma\left(\alpha(k+2)+1\right)}\Biggr].
\end{align*}
In particular, we have these two equations for the mean calculation and $2^{\text{nd}}$ order moment (relevant for the variance 
of the proposed CBurr distribution.
\begin{align*}
    E(Y)=\alpha\gamma\Gamma\left(\frac{1}{c}+1\right)\Biggl[\displaystyle\sum_{j=0}^{\infty}\displaystyle\sum_{k=0}^{j}a_{j,k}\frac{\Gamma\left(\alpha(k+1)-\frac{1}{c}\right)}{\Gamma\left(\alpha(k+1)+1\right)}\\  +\lambda\displaystyle\sum_{j=0}^{\infty}\displaystyle\sum_{k=0}^{j}a_{j,k}\frac{\Gamma\left(\alpha(k+2)-\frac{1}{c}\right)}{\Gamma\left(\alpha(k+2)+1\right)}\Biggr].
\end{align*}
\begin{align*}
    E(Y^{2})= &\alpha\gamma^{2}\Gamma\left(\frac{2}{c}+1\right)\Biggl[\displaystyle\sum_{j=0}^{\infty}\displaystyle\sum_{k=0}^{j}a_{j,k}\frac{\Gamma\left(\alpha(k+1)-\frac{2}{c}\right)}{\Gamma\left(\alpha(k+1)+1\right)}  + \\
    & \quad  \quad  \quad  \quad  \lambda\displaystyle\sum_{j=0}^{\infty}\displaystyle\sum_{k=0}^{j}a_{j,k}\frac{\Gamma\left(\alpha(k+2)-\frac{2}{c}\right)}{\Gamma\left(\alpha(k+2)+1\right)}\Biggr].
\end{align*}
{For small integer orders (e.g., first and second-order moment), we derived a semi-closed-form using known identities for Burr distribution; however, numerical approximation will be used in practice.}

In the case where the baseline distribution function follows the Burr distribution, the mean residual life function of the random variable $Y$ can be represented as
\begin{align*}
\mu_{G}(t)&=E\left(Y-t|Y>t\right)\\
 &=e^{\lambda F(t)}\displaystyle\sum_{j=0}^{\infty}\displaystyle\sum_{k=0}^{j}a_{j,k}\left[M_{1,0,k}(t)+\lambda M_{1,0,k+1}(t)\right].
\end{align*}
Now,
\begin{align*}
    M_{1,0,k}(t)&=\frac{1}{\bar{F}(t)}\displaystyle\int_{0}^{\infty}y\bar{F}^{k}(y+t)f(y+t)dy\\
    &=\frac{1}{\left[1+\left(\frac{t}{\gamma}\right)^{c}\right]^{-\alpha}} \displaystyle\int_{0}^{\infty}y \left[1+\left(\frac{y+t}{\gamma}\right)^{c}\right]^{-k\alpha} \\ & \quad \frac{c\alpha}{\gamma^{c}}(y+t)^{c-1}\left[1+\left(\frac{y+t}{\gamma}\right)^{c}\right]^{-\alpha-1}dy\\
    % &=\frac{c \alpha}{\gamma^{c}}\left[1+\left(\frac{t}{\gamma}\right)^{c}\right]^{\alpha} \\ & \displaystyle\int_{0}^{\infty}(y+t-t)(y+t)^{c-1}\left[1+\left(\frac{y+t}{\gamma}\right)^{c}\right]^{-k \alpha-\alpha-1}dy\\
    &=\alpha\left[1+\left(\frac{t}{\gamma}\right)^{c}\right]^{\alpha}
    \left(\frac{\gamma \Gamma\left(\alpha(k+1)-\frac{1}{c}\right)\Gamma\left(\frac{1}{c}+1\right)}{\Gamma\left(\alpha(k+1)+1\right)} - \right. \\ &\left. \quad \quad \quad \quad \quad \frac{t}{\alpha(k+1)}\right).
\end{align*}
Therefore, the MRL function can be expressed as
\begin{align*}
\mu_{G}(t)&=e^{\lambda F(t)}\displaystyle\sum_{j=0}^{\infty}\displaystyle\sum_{k=0}^{j}a_{j,k}\left[M_{1,0,k}(t)+\lambda M_{1,0,k+1}(t)\right]\\
&=\left[1+\left(\frac{t}{\gamma}\right)^{c}\right]^{\alpha}\Gamma\left(\frac{1}{c}+1\right)\\ & \; \exp\left[\lambda\left(1-\left(1+\left(\frac{t}{\gamma}\right)^{c}\right)^{-\alpha}\right)\right]  \\ & \; \displaystyle\sum_{j=0}^{\infty}\displaystyle\sum_{k=0}^{j}a_{j,k}\Biggl[\alpha\gamma\frac{\Gamma\left(\alpha(k+1)-\frac{1}{c}\right)}{\Gamma\left(\alpha(k+1)+1\right)} +\\
&\lambda\alpha\gamma\frac{\Gamma\left(\alpha(k+2)-\frac{1}{c}\right)}{\Gamma\left(\alpha(k+2)+1\right)}-\frac{t}{k+1}-\frac{\lambda t}{k+2}\Biggr]
\end{align*}
{For the CBurr Model, the MRL captures not only the heavy-tailed behavior (from the baseline Burr distribution) but also incorporates a decaying interaction probability through the Poisson-shift mechanism. This allows the model to realistically represent biological constraints, such as limited binding capacities or competitive inhibition, which naturally cap the number of possible future interactions.}

% The CBurr distribution has four parameters and is flexible enough to analyze highly skewed and heavy-tailed datasets. As explained in the next subsection, estimating the parameters is not difficult. It can also be naturally extended for a multivariate setup, which is a great advantage of the CBurr distribution in practical usage. 

\begin{figure*}
     \centering
     \begin{subfigure}%[b]{0.47\textwidth}
         \centering
         \includegraphics[width=6.0cm, height=4cm]{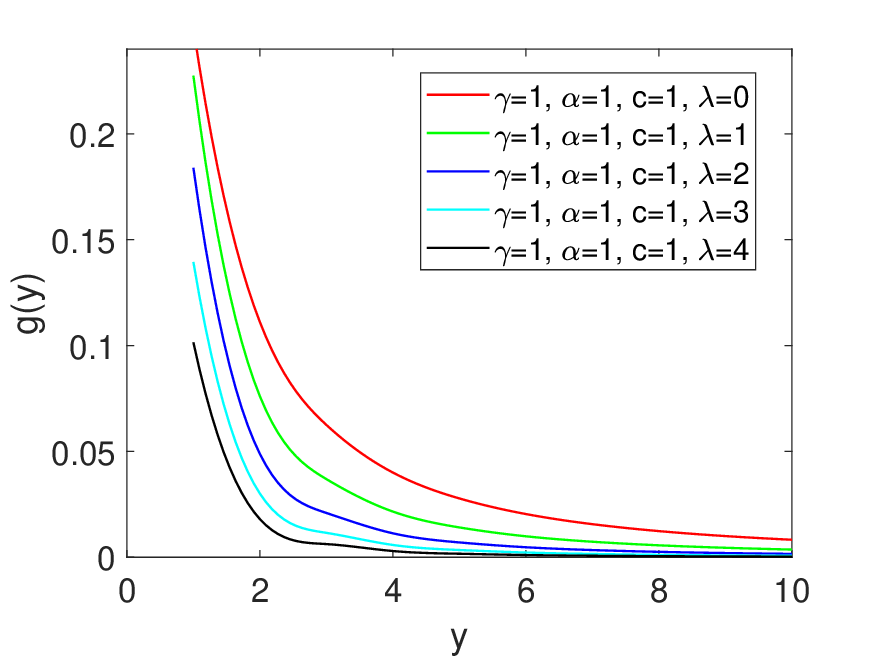}
         % \caption{bio-dmela}
     \end{subfigure}
     \begin{subfigure}%[b]{0.47\textwidth}
         \centering
         \includegraphics[width=6.0cm, height=4cm]{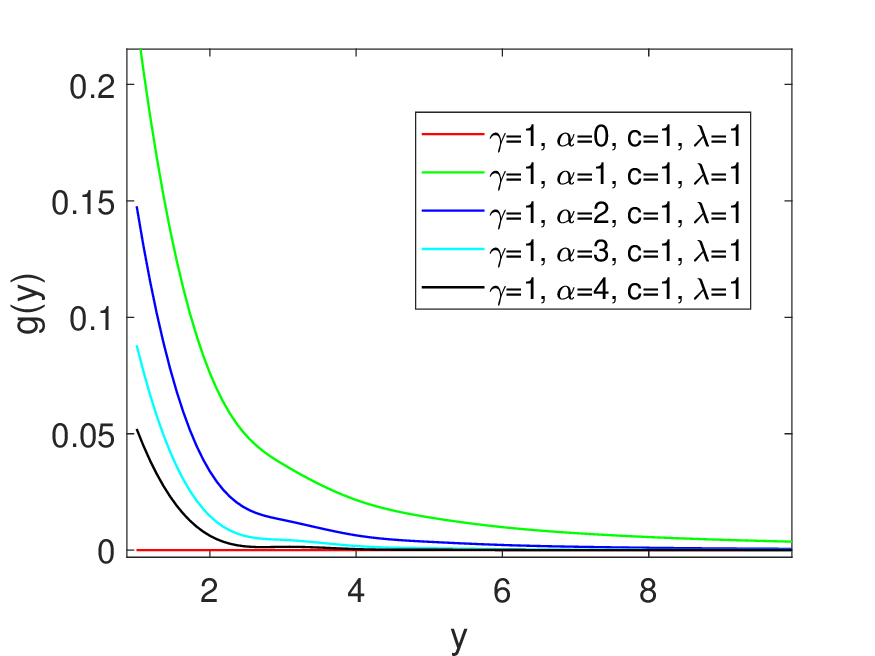}
         % \caption{bio-mouse-gene}
     \end{subfigure} \medskip 
     \begin{subfigure}%[b]{0.47\textwidth}
         \centering
         \includegraphics[width=6.0cm, height=4cm]{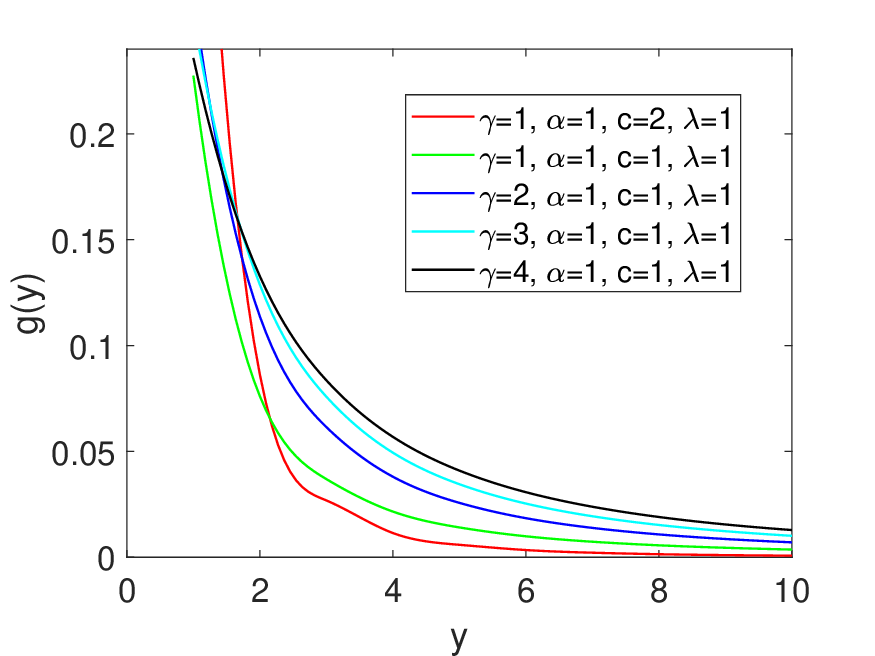}
         % \caption{deseasome}
     \end{subfigure}
     \begin{subfigure}%[b]{0.47\textwidth}
         \centering
         \includegraphics[width=6.0cm, height=4cm]{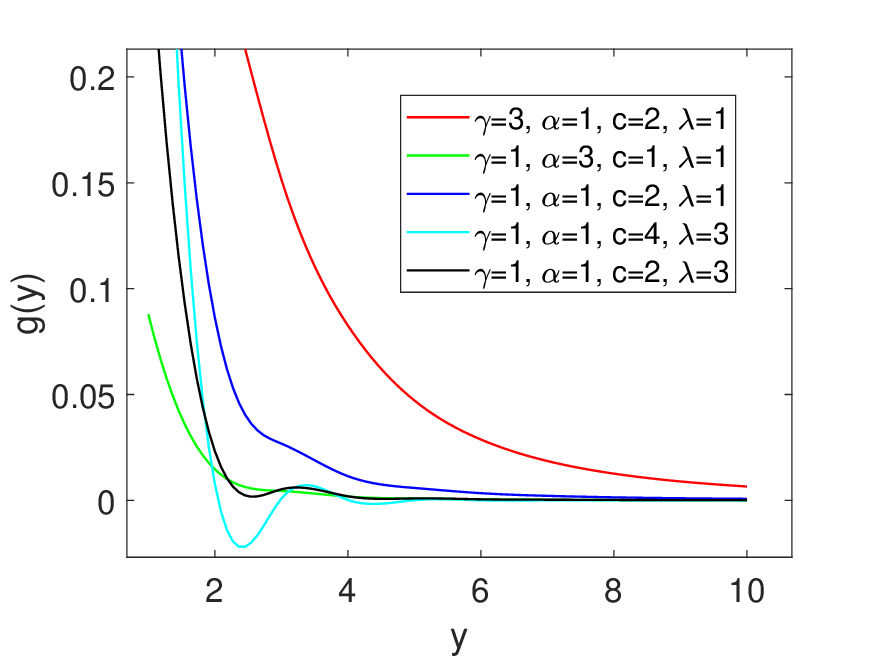}
         % \caption{deseasome}
     \end{subfigure}
     \caption{Plots corresponding to PDFs of CBurr distribution of varying values of parameters.}
     \label{fig_mlm_2}
\end{figure*}

{
\subsection{Interpretation of CBurr model parameters}
 The CBurr distribution exhibits different shapes due to the changing values of parameters, as depicted in Figs. \ref{fig_mlm_2}. 
 %and \ref{fig_mlm_3}. 
 {Fig. \ref{fig_mlm_2} shows that CBurr distribution captures non-monotonic, heavy-tailed, and skewed behaviors — all within a four-parameter space (one scale parameter, two shape parameters, and one Poisson shift parameter.} In the context of the CBurr model, the emergence of \textit{non-monotonic shapes} in the degree distribution is closely tied to the choice of parameters. Each parameter governs a specific aspect of the distribution’s behavior: (I) \textbf{The scale parameter ($\gamma$)} determines how stretched the distribution is along the $x$-axis; keeping $\gamma$ small (e.g., $0.5 \leq \gamma \leq 2$) ensures a sharp early rise in the distribution; (II) \textbf{The shape parameter ($c$)}, when set to moderate or high values ($3 \leq c \leq 7$), creates sufficient curvature in the mid-range degrees, allowing the model to reflect an initial increase in frequency followed by a decline; (III) \textbf{The tail sharpness parameter ($\alpha$)}, within a range like $2 \leq \alpha \leq 6$, ensures the distribution doesn’t decay too fast, enabling a flexible transition between the body and the tail; (IV) \textbf{The Poisson-shift parameter ($\lambda$)} introduces a skewing effect; moderate values ($1 \leq \lambda \leq 5$) allow the peak to form naturally without flattening or overpowering the body structure. Excessively high $\lambda$ values can suppress non-monotonicity by overemphasizing the tail decay.

These parameter settings are especially relevant when modeling biological networks, where degree distributions often show a peak followed by a heavy tail, reflecting a few highly connected proteins (hubs) alongside many proteins with moderate interactions. The non-monotonic pattern in such networks arises due to functional modularity, evolutionary pressures, and biophysical constraints.}

\subsection{Parameter Estimations: Maximum Likelihood Estimation}
Let ${y_{1}, y_{2},\cdots,y_{n}}$ denote a random sample of size $n$ from the new family of distributions characterized by the density function given in Eq. (\ref{e12}). The parameter vector $\Theta=(\lambda, \beta^{T})^{T}$ represents the unknown parameters, where $\beta=(c, \alpha, \gamma)^T$ corresponds to the parameters of the baseline distribution. The log-likelihood function for the parameter vector $\Theta$ can be expressed as follows:
\begin{align*}
    \ell(\Theta)&=\displaystyle\sum_{i=1} ^{ n} \log\left[1+\lambda\bar{F}(y_{i},\beta)\right]-\lambda\displaystyle\sum_{i=1} ^{ n}F(y_{i},\beta)+\displaystyle\sum_{i=1} ^{ n}\log f(y_{i},\beta)\\
    &=n\log c\alpha-n\lambda+(c-1)\displaystyle\sum_{i=1}^{n}\log y_{i} -nc\log\gamma -n\lambda \\ 
    &-(\alpha+1)\displaystyle\sum_{i=1} ^{ n}\log\left[1+\left(\frac{y_{i}}{\gamma}\right)^{c}\right]+\\ & \quad\displaystyle\sum_{i=1} ^{ n}\log\left[1+\frac{\lambda}{\left(1+\left(\frac{y_{i}}{\gamma}\right)^{c}\right)^{\alpha}}\right]+  \lambda\displaystyle\sum_{i=1} ^{n}\left(1+\left(\frac{y_{i}}{\gamma}\right)^{c}\right)^{-\alpha}.
\end{align*}
By calculating the partial derivatives of the log-likelihood function with respect to $\lambda$ and $\beta$, we can derive the normal equations. Setting these derivatives to zero, we obtain:
\begin{align*}
\frac{\partial\ell}{\partial \lambda} =\displaystyle\sum_{i=1} ^{n}\frac{\bar{F}(y_{i},\beta)}{1+\lambda\bar{F}(y_{i},\beta)}-\displaystyle\sum_{i=1} ^{n}F(y_{i},\beta) \; \text{ and }
\end{align*}

\begin{align*}
\frac{\partial\ell}{\partial \beta}= & \displaystyle\sum_{i=1} ^{n}\frac{\lambda}{1+\lambda\bar{F}(y_{i},\beta)}\frac{\partial\bar{F}(y_{i},\beta)}{\partial \beta}-\lambda\displaystyle\sum_{i=1} ^{n}\frac{\partial F(y_{i},\beta)}{\partial \beta}+\\ & \; \displaystyle\sum_{i=1} ^{n}\frac{\partial\log f(y_{i},\beta)}{\partial \beta}.
\end{align*}
Upon equating the partial derivatives to zero, a system of equations is obtained. Solving this system simultaneously yields the maximum likelihood estimates (MLE) $\hat{\Theta}=(\hat{\lambda},\hat{\beta}^{T})^{T}$ for $\Theta=(\lambda, \beta^{T})^{T}$. Due to the absence of closed-form solutions for these equations, numerical solutions obtained through iterative algorithms are required to solve them.

These partial derivatives are set to zero and are solved simultaneously to obtain the MLEs of the parameters of the CBurr model. As the closed-form solutions are not available, we have used an iterative method to solve these equations numerically. We have used the quasi-Newton Broyden-Fletcher-Goldfarb-Shanno (BFGS) algorithm available in the R package `optimx' for solving the unconstrained nonlinear optimization problem. This helps in finding the MLE estimates of the CBurr model parameters.  

% \begin{figure*}
%      \centering
%      \begin{subfigure}%[b]{0.47\textwidth}
%          \centering
%          \includegraphics[width=8.0cm, height=6cm]{cburr3d_fig1.eps}
%          % \caption{bio-dmela}
%      \end{subfigure}
%      \begin{subfigure}%[b]{0.47\textwidth}
%          \centering
%          \includegraphics[width=8.0cm, height=6cm]{cburr3d_fig2.eps}
%          % \caption{bio-mouse-gene}
%      \end{subfigure} \medskip 
%      \begin{subfigure}%[b]{0.47\textwidth}
%          \centering
%          \includegraphics[width=8.0cm, height=6cm]{cburr3d_fig3.eps}
%          % \caption{deseasome}
%      \end{subfigure}
%      \begin{subfigure}%[b]{0.47\textwidth}
%          \centering
%          \includegraphics[width=8.0cm, height=6cm]{cburr3d_fig4.eps}
%          % \caption{deseasome}
%      \end{subfigure}
%      \caption{3D Plots corresponding to PDFs of CBurr distribution of varying values of parameters $\gamma$ and $\alpha$. (c and $\lambda$ are constant).}
%      \label{fig_mlm_3}
% \end{figure*}

\section{Experimental Study} \label{experiments}

\subsection{Datasets}
The degree distribution of biological networks is an essential factor in comprehending their structure and function. For example, in PPI networks, node degree distribution indicates the presence of densely packed hub proteins and the majority of proteins having a lower degree. In DDI networks, node degree distributions provide insight into the drug interaction pattern and its pharmacological implications. Additionally, metabolic networks display a scale-free distribution, emphasizing the highly connected metabolites essential for metabolic pathways. Finally, GRN networks use node degrees to comprehend the pattern of gene regulation and to identify important regulatory genes. Analyzing node degree distributions provides researchers with valuable insight into the organizational principles and regulatory dynamics of biological networks. The following datasets were used for experiments in this study:

\textit{Bio-Dmela} \cite{singh2008global}: PPI networks were constructed for five species, such as baker's yeast, fruit fly, nematode worm, mouse, and human. The networks were created by combining data from multiple databases. The PPI data coverage varied among the species, with different numbers of edges observed. Singleton nodes were added based on sequence data to include proteins not present in the PPI network. 

\textit{Bio-Mouse-Gene} \cite{liu2015regnetwork}: Gene regulatory networks \cite{barabasi1999emergence, albert2005scale, newman2003structure} were studied using a variety of network feature indices in the Bio-Mouse-Gene study. The study focused on regulatory relationships between TFs, miRNAs, and genes, including experimental, inferred, and predicted interactions. The study used the extensive RegNetwork database to provide information on both transcriptional and post-transcriptional regulatory interactions.

\textit{Bio-Diseasome} \cite{goh2007human,goh2012exploring}:
The bio-diseasome dataset describes the relationship between diseases and the genes that play a role in their pathophysiology. It uses network-based approaches to establish relationships through mutation links to gain insights into the genetics of diseases. The bio-diseaseasome dataset enables researchers to systematically investigate the genetic elements that underlie various disorders and the relationships between them.

\textit{Protein-protein interaction (PPI) network in budding yeast} \cite{bu2003topological}:
The PPI network in budding yeast (Saccharomyces cerevisiae) is a comprehensive resource that consists of 2361 vertices (proteins) and 7182 edges (interactions), including 536 loops. This network has been extensively studied and is available as an example dataset in the PIN software package developed by Bu et al. \cite{bu2003topological}. It provides valuable insights into the functional relationships and relevance of protein interactions. Specifically, interactions such as Bio-CE-CX, SC-HT, SC-LC, HS-CX, HS-LC, DM-CX, and DM-HT have been analyzed in other research studies. In addition, we will experiment with the bio-SC-HT, bio-SC-LC, bio-HS-CX, and bio-HS-LC interactions along with the Yeast-PPIN dataset.

\textit{Bio-GRID (bio-grid-fruitfly; bio-grid-human; bio-grid-worm; bio-grid-yeast)}\cite{stark2006biogrid,oughtred2019biogrid}:
BioGRID is an open-access database that curates protein, genetic, and chemical interaction data for various organisms. It contains over 1.6 million curated interactions from 55,000 publications, covering 71 species. BioGRID includes information on protein interactions, post-translational modifications, and chemical interactions. For the experiment, the following datasets were chosen: BioGRID-Fruitfly, BioGRID-Human, BioGRID-Worm, and BioGRID-Yeast. 
\begin{figure*}[htbp]
     \centering
     % \begin{subfigure}%[b]{0.3\textwidth}
     %     \centering
         \includegraphics[width=0.45\textwidth]{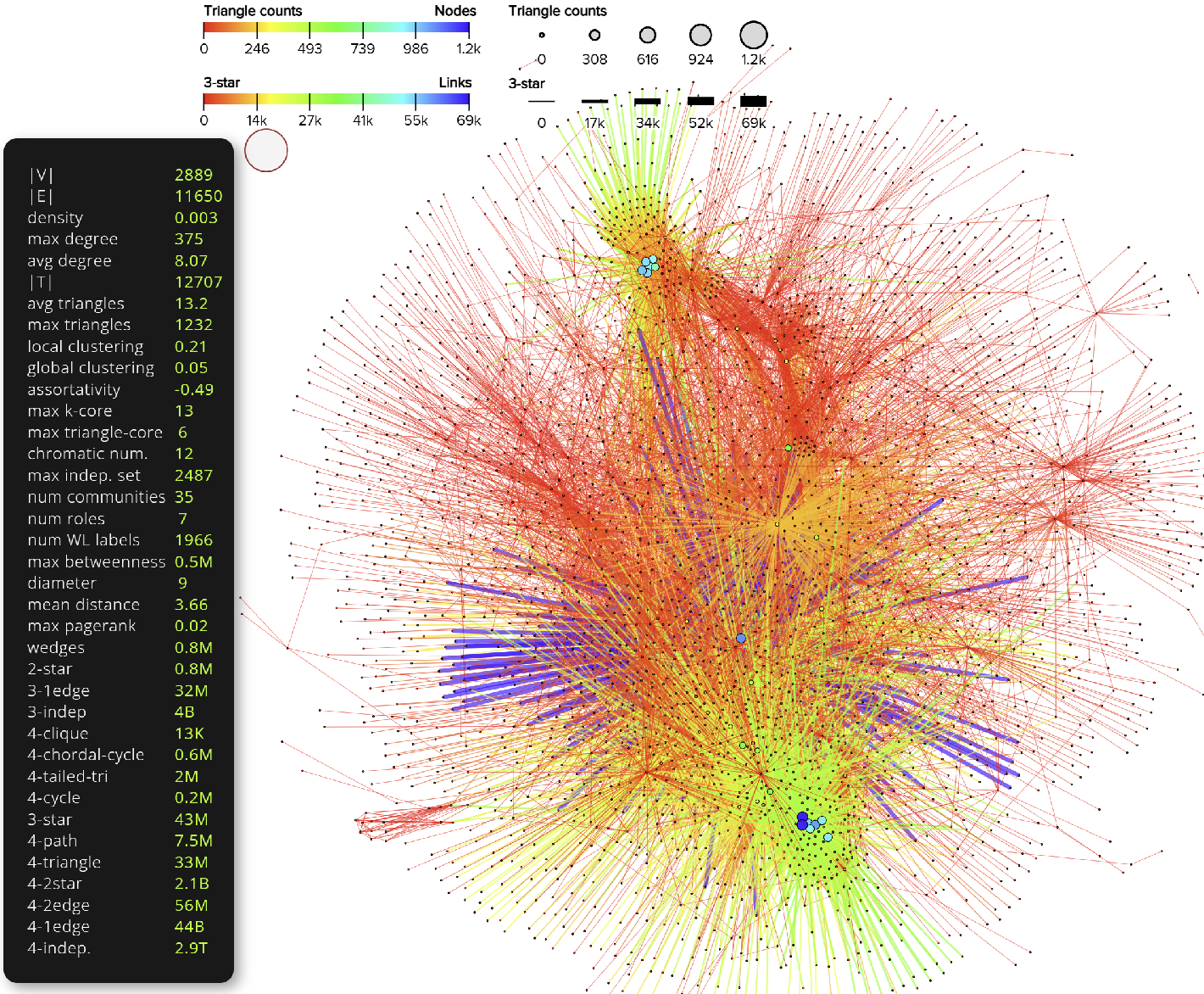}%[width=5.5cm, height=4.5cm]
         %\caption{bio-HS-CX data}
         %\label{fig1}
     % \end{subfigure}
     \hspace*{5em}
     % \begin{subfigure}%[b]{0.3\textwidth}%[b]{0.40\textwidth}
     %     \centering
         \includegraphics[width=0.45\textwidth]{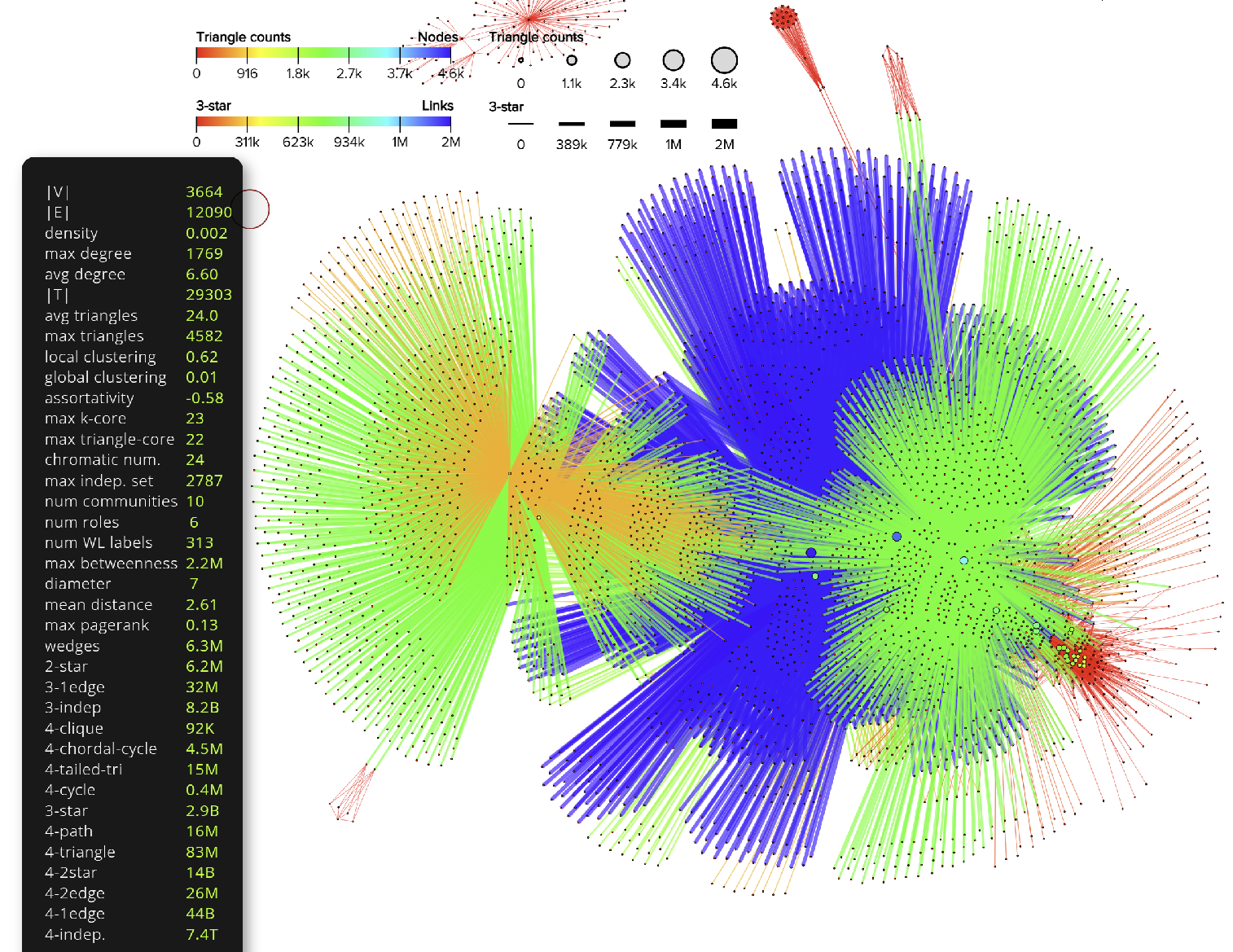}%[width=5.5cm, height=4.5cm]
         %\caption{bio-mouse-Gene}
         %\label{fig2}
     % \end{subfigure}
        \caption{A visual representation of the (left) bio-HS-CX data and (right) bio-mouse-Gene to provide insight into its structure, interactions, and functional relationships \cite{nr-aaai15}. {We have used the Interactive visualization platform, called ``GraphVis''\footnote{\url{https://networkrepository.com/graphvis.php}}, (Interactive Visual Graph Mining and Machine Learning) for visualizing the graph structures\footnote{\url{https://networkrepository.com/}}\cite{rossi2018interactive}}}
        \label{fig_visual}
\end{figure*}
Figure~\ref{fig_visual} depicts the visualization of the Bio-HS-CX and Bio-Mice-Gene network examples \cite{nr-aaai15}. The figure~\ref{fig_visual} also illustrates the ``list of parameters'' which comprises customizable variables allowing users to adapt their network dataset analysis and visualization to specific requirements. These parameters encompass node and edge attributes, interaction options, filtering settings, layout options, and statistical measures, empowering users to derive deeper insights and address specific research inquiries with effectiveness.

\subsection{Evaluation Criteria}
We use a variety of evaluation metrics to determine the goodness-of-fit for the proposed CBurr distribution for the degree distribution in a real-world complex network. Due to the fact that the actual distribution is discrete, we test the goodness-of-fit using the chi-square statistic. The chi-squared test is calculated by bootstrap rebalancing with 50K synthetic datasets. We also calculate additional statistical measures, such as Root Mean Square Error (RMSE) and KL-Difference (KLD), as well as Mean Absolute Error (MAE), to further compare the goodness-of-fit of the proposed CBurr distribution model to other considered distributions.

\subsection{Analysis of results}\label{analysis_results}
\begin{table*}[htbp]
%\footnotesize
\centering
\caption{The performance of the proposed CBurr Distribution model was assessed across different biological networks.}
\label{table:1estparameter}
\resizebox{14cm}{!}{
\begin{tabular}{|cl|c|c|ccc|cccc|c|}
\hline
\multicolumn{2}{|c|}{\multirow{2}{*}{Datasets}} & \begin{tabular}[c]{@{}c@{}}Node\\ Count\end{tabular} & \begin{tabular}[c]{@{}c@{}}Edge\\ Count\end{tabular} & \multicolumn{3}{c|}{Statistical Prop.} & \multicolumn{4}{c|}{Estimated Parameters} & \begin{tabular}[c]{@{}c@{}}Bootstrap\\ Chi-square \\ value\end{tabular} \\
\multicolumn{2}{|c|}{} & $|\operatorname{V}|$ & $|\operatorname{E}|$ & $s$ & $\mu$ & $\frac{s}{\mu}$ & $\widehat{\gamma}$ & $\widehat{\alpha}$ & $\widehat{c}$ & $\widehat{\lambda}$ & (p) \\ \hline
\multicolumn{1}{|c|}{\multirow{12}{*}{\begin{tabular}[c]{@{}c@{}}Biological \\ \\ Networks\end{tabular}}} & Bio-Dmela & 7393 & 25569 & \multicolumn{1}{c|}{10.782} & \multicolumn{1}{c|}{6.9170} & 1.5587 & \multicolumn{1}{c|}{50.877} & \multicolumn{1}{c|}{5.5685} & \multicolumn{1}{c|}{0.7234} & 0.6978 & 0.954 \\
\multicolumn{1}{|c|}{} & Bio-Mouse-Gene & 43101 & 14,506,199 & \multicolumn{1}{c|}{856.67} & \multicolumn{1}{c|}{643.27} & 1.3317 & \multicolumn{1}{c|}{5.9357} & \multicolumn{1}{c|}{0.0043} & \multicolumn{1}{c|}{0.5934} & 3.2029 & 0.960 \\
\multicolumn{1}{|c|}{} & Bio-Diseasome & 3,926 & 7,823 & \multicolumn{1}{c|}{9.1009} & \multicolumn{1}{c|}{5.5334} & 1.6447 & \multicolumn{1}{c|}{289.25} & \multicolumn{1}{c|}{14.161} & \multicolumn{1}{c|}{0.5125} & 0.4750 & 0.912 \\
\multicolumn{1}{|c|}{} & Bio-Yeast-PPIN & 2,361 & 7,182 & \multicolumn{1}{c|}{8.0800} & \multicolumn{1}{c|}{6.0838} & 1.3281 & \multicolumn{1}{c|}{176.39} & \multicolumn{1}{c|}{2.5393} & \multicolumn{1}{c|}{0.6576} & 5.7988 & 0.984 \\
\multicolumn{1}{|c|}{} & Bio-SC-HT & 2084 & 63028 & \multicolumn{1}{c|}{15.748} & \multicolumn{1}{c|}{7.7472} & 2.0328 & \multicolumn{1}{c|}{15.201} & \multicolumn{1}{c|}{1.4361} & \multicolumn{1}{c|}{0.5573} & -0.7228 & 0.826 \\
\multicolumn{1}{|c|}{} & Bio-SC-LC & 2004 & 20453 & \multicolumn{1}{c|}{30.655} & \multicolumn{1}{c|}{18.905} & 1.6214 & \multicolumn{1}{c|}{265.11} & \multicolumn{1}{c|}{9.9369} & \multicolumn{1}{c|}{0.7335} & -0.2575 & 0.698 \\
\multicolumn{1}{|c|}{} & Bio-HS-CX & 4413 & 108819 & \multicolumn{1}{c|}{30.425} & \multicolumn{1}{c|}{15.648} & 1.9442 & \multicolumn{1}{c|}{294.07} & \multicolumn{1}{c|}{5.4005} & \multicolumn{1}{c|}{0.6420} & -0.3727 & 0.931 \\
\multicolumn{1}{|c|}{} & Bio-HS-LC & 4227 & 39485 & \multicolumn{1}{c|}{76.324} & \multicolumn{1}{c|}{27.993} & 2.7265 & \multicolumn{1}{c|}{411.07} & \multicolumn{1}{c|}{0.3120} & \multicolumn{1}{c|}{0.5947} & 33.079 & 0.702 \\
\multicolumn{1}{|c|}{} & bio-grid-fruitfly & 7282 & 49788 & \multicolumn{1}{c|}{272.93} & \multicolumn{1}{c|}{81.730} & 3.3394 & \multicolumn{1}{c|}{115.93} & \multicolumn{1}{c|}{9.1965} & \multicolumn{1}{c|}{0.6324} & 0.4132 & 0.968 \\
\multicolumn{1}{|c|}{} & bio-grid-human & 9527 & 62364 & \multicolumn{1}{c|}{334.14} & \multicolumn{1}{c|}{82.771} & 4.0369 & \multicolumn{1}{c|}{18.449} & \multicolumn{1}{c|}{4.2740} & \multicolumn{1}{c|}{0.6279} & 0.0959 & 0.924 \\
\multicolumn{1}{|c|}{} & bio-grid-warm & 3507 & 13062 & \multicolumn{1}{c|}{256.78} & \multicolumn{1}{c|}{56.564} & 4.5397 & \multicolumn{1}{c|}{4.6275} & \multicolumn{1}{c|}{5.0673} & \multicolumn{1}{c|}{0.4156} & 1.2740 & 0.683 \\
\multicolumn{1}{|c|}{} & bio-grid-yeast & 6008 & 313890 & \multicolumn{1}{c|}{33.920} & \multicolumn{1}{c|}{16.782} & 2.0212 & \multicolumn{1}{c|}{2.3144} & \multicolumn{1}{c|}{1.1605} & \multicolumn{1}{c|}{0.6307} & -1.4910 & 0.945 \\ \hline
\end{tabular}}
\end{table*}

The proposed CBurr and other distributions were implemented using R Software version 4.2.2 on a Windows desktop, equipped with an Intel Core i7-8700 CPU at 3.20 GHz, 6 GB of RAM, and a 64-bit operating system. These models can be easily run and simulated on a personal computer across various platforms. In Table \ref{table:1estparameter}, we first provide the statistical properties of the different networks, including their node count, edge count, and descriptive statistics. In particular, the standard deviation (s), mean ($\mu$), and coefficient of variation (s/$\mu$). It is interesting to note that the coefficient of variation for all these datasets is higher than 1, indicating that the datasets have high variability {(highly skewed)}. Table \ref{table:1estparameter} also presents evidence of the proposed fitting using the Compounded Burr distribution for the node degree distribution in various networks. The parameters $(\gamma, \alpha, c, \lambda)$ of the Compounded Burr distribution were estimated numerically using the ``optim'' function with the L-BFGS-B algorithm in R software, initialized at $(\gamma, \alpha, c, \lambda) = (1,1,0.5,1.5)$. The advantage of using the ``optim'' function with the L-BFGS-B algorithm is its ability to efficiently find the optimal parameter values by iteratively optimizing the objective function based on the provided initial values and constraints. The estimated parameter values in Table \ref{table:1estparameter} satisfy the conditions $(\gamma > 0, \; \alpha > 0, \; c > 0, \; \text{and} \; \lambda > -2)$, which are necessary for the complete characterization of the CBurr distribution. 

{In our empirical analyses, we may observe the Poisson shift parameter $\lambda$ attaining negative values due to poor parameter initialization, model overflexibility, or numerical instability in the optimization process, especially when the data's tail behavior is already well captured by the baseline Burr distribution. From a statistical perspective, a negative $\lambda$ reverses the exponential skewing introduced via the Poisson-shifted minimum construction, effectively dampening the influence of the Poisson mechanism. This suggests that the data may not support—or even resist—additional compounding from the skewing layer, particularly when the baseline model (such as the Burr) already accommodates heavy-tailed characteristics observed in biological networks. Rather than signaling a misfit, a negative $\lambda$ indicates an adaptive response of the model, aligning with the underlying data by softening or inverting the shift intensity. 

To avoid negative values, one may enforce positivity constraints on 
$\lambda$ during parameter estimation. Yet, this can lead to under-identified models, where different parameter configurations yield similar likelihoods, making inference unstable. Constrained optimization provides a potential remedy, but at the cost of increased computational burden and sensitivity to starting values or tolerance thresholds. This can be considered as a future scope for further investigation.} {Notably, the estimated values of $\gamma$ tend to be higher than the other parameters $\alpha, c, \; \text{and} \; \lambda$, signifies that $\gamma$ compresses the distribution, shifting the mass towards smaller values of the degree of a node in the network (for a protein (node), it is the number of direct interaction partners (edges) it has). It suggests that a heavier penalty is imposed for higher degrees. For example, in a PPI network, a high $\gamma$ could suggest that proteins tend to have lower maximum interaction capacity, or that the interaction-generating mechanisms are more constrained in scale.} Furthermore, the estimated value of the parameter $\lambda$ ranges between $(-2, \infty)$, indicating a wide range of variability in the shape parameter's values, as observed from Table \ref{table:1estparameter}.

\begin{table*}[htbp]
\caption{The comparison metrics among competitive models across biological networks.}
%\tiny
\label{tab:performance}
\resizebox{16cm}{!}{\begin{tabular}{|ll|l|l|l|l|l|l|l|l|l|l|l|l|}%{|ll|l|l|l|l|l|l|l|l|l|l|l|l|}
\hline
\multicolumn{2}{|l|}{Datasets} & Dmela & Mouse-Gene & Diseasome & Yeast-PPIN & bio-SC-HT & bio-SC-LC & bio-HS-CX & bio-HS-LC & Grid-fruitfly & Grid-human & Grid-warm & Grid-yeast \\ \hline
\multicolumn{1}{|l|}{\multirow{3}{*}{\begin{tabular}[c]{@{}l@{}}Proposed \\ CBurr\end{tabular}}} & RMSE & \textbf{6.7306} & \textbf{5.0744} & \textbf{9.7518} & \textbf{4.4775} & \textbf{2.4666} & \textbf{4.7039} & \textbf{3.2407} & 3.8032 & \textbf{5.3814} & \textbf{5.5554} & 1.7573 & \textbf{2.0647} \\ %\cline{2-14} 
\multicolumn{1}{|l|}{} & KLD & \textbf{0.01314} & 0.10351 & 0.08767 & 0.0148 & \textbf{0.1806} & \textbf{0.09092} & 0.06979 & 0.07928 & 0.0117 & 0.0088 & \textbf{0.02471} & 0.05908 \\ %\cline{2-14} 
\multicolumn{1}{|l|}{} & MAE & \textbf{2.9237} & \textbf{1.5425} & 3.0616 & \textbf{2.4194} & \textbf{1.3657} & \textbf{2.48} & 1.7794 & 1.563 & \textbf{2.3064} & 1.4606 & 0.3877 & 0.4289 \\ \hline

\multicolumn{1}{|l|}{\multirow{3}{*}{Burr}} & RMSE & 14.9 & 11.661 & 10.45 & 4.7091 & 2.6404 & 4.8725 & 4.3199 & 3.7883 & 5.7591 & 4.301 & 1.895 & 2.5844 \\ %\cline{2-14} 
\multicolumn{1}{|l|}{} & KLD & \textbf{0.01314} & 0.13013 & 0.08788 & 0.0163 & 0.19102 & 0.0914 & 0.0651 & 0.07928 & \textbf{0.01143} & \textbf{0.00857} & 0.0248 & \textbf{0.05402} \\ %\cline{2-14} 
\multicolumn{1}{|l|}{} & MAE & 3.9705 & 1.9375 & 3.1013 & 2.7479 & 1.4803 & 2.5576 & 1.9034 & 1.5587 & 2.3222 & \textbf{1.2838} & 0.4068 & \textbf{0.4221} \\ \hline

\multicolumn{1}{|l|}{\multirow{3}{*}{Lomax}} & RMSE & 10.579 & 14.654 & 12.451 & 12.651 & 2.9554 & 6.1888 & 7.6341 & 8.2955 & 12.996 & 17.181 & 2.6553 & 3.6268 \\ %\cline{2-14} 
\multicolumn{1}{|l|}{} & KLD & 0.01759 & 0.19473 & 0.10202 & 0.02389 & 0.19526 & 0.09658 & 0.09352 & 0.08863 & 0.01862 & 0.01534 & 0.02594 & 0.06338 \\ %\cline{2-14} 
\multicolumn{1}{|l|}{} & MAE & 3.8219 & 2.3919 & 3.4575 & 5.4529 & 1.5541 & 2.8101 & 2.5272 & 2.2546 & 4.0163 & 3.253 & 0.5016 & 0.4463 \\ \hline

\multicolumn{1}{|l|}{\multirow{3}{*}{Power-law}} & RMSE & 143.71 & 41.371 & 26.006 & 75.325 & 23.882 & 43.613 & 52.398 & 52.729 & 140.2 & 130.76 & 33.719 & 30.195 \\ %\cline{2-14} 
\multicolumn{1}{|l|}{} & KLD & 0.1907 & 0.4566 & 0.2248 & 0.1999 & 0.51371 & 0.45476 & 0.46706 & 0.3048 & 0.1805 & 0.16483 & 0.17858 & 0.5661 \\ %\cline{2-14} 
\multicolumn{1}{|l|}{} & MAE & 21.415 & 3.9018 & 5.3567 & 19.013 & 3.3589 & 9.6831 & 7.6153 & 6.6692 & 21.353 & 14.954 & 2.841 & 2.1799 \\ \hline

\multicolumn{1}{|l|}{\multirow{3}{*}{Pareto}} & RMSE & 143.67 & 92.539 & 26.005 & 77.455 & 23.659 & 43.368 & 50.78 & 51.534 & 139.98 & 130.63 & 33.718 & 29.973 \\ %\cline{2-14} 
\multicolumn{1}{|l|}{} & KLD & 0.1907 & 0.5373 & 0.2248 & 0.1998 & 0.51371 & 0.45476 & 0.46706 & 0.3048 & 0.1805 & 0.16483 & 0.17858 & 0.5661 \\ %\cline{2-14} 
\multicolumn{1}{|l|}{} & MAE & 21.414 & 4.7342 & 5.3566 & 19.079 & 3.35 & 9.6659 & 7.5502 & 6.6351 & 21.349 & 14.952 & 2.841 & 2.1755 \\ \hline

\multicolumn{1}{|l|}{\multirow{3}{*}{Log-normal}} & RMSE & 46.271 & 17.199 & 23.282 & 29.928 & 4.2927 & 8.3206 & 8.5989 & 13.298 & 55.623 & 62.57 & 32.487 & 4.7906 \\ %\cline{2-14} 
\multicolumn{1}{|l|}{} & KLD & 0.0426 & 0.1878 & 0.1552 & 0.0496 & 0.21226 & 0.10934 & 0.10032 & 0.09745 & 0.04752 & 0.05111 & 0.15071 & 0.08533 \\ %\cline{2-14} 
\multicolumn{1}{|l|}{} & MAE & 9.2857 & 2.5372 & 4.8906 & 9.3869 & 1.716 & 3.259 & 2.8007 & 2.5799 & 10.203 & 8.1441 & 2.8605 & 0.5902 \\ \hline

\multicolumn{1}{|l|}{\multirow{3}{*}{Poisson}} & RMSE & 206.44 & 101.23 & 55.985 & 109.62 & 20.892 & 39.969 & 44.077 & 64.725 & 215.08 & 220.16 & 80.446 & 24.527 \\ %\cline{2-14} 
\multicolumn{1}{|l|}{} & KLD & 3.6221 & 15.318 & 3.0101 & 2.5149 & 28.416 & 9.8485 & 23.462 & 13.238 & 3.6734 & 4.0136 & 3.0504 & 15.194 \\ %\cline{2-14} 
\multicolumn{1}{|l|}{} & MAE & 45.991 & 10.254 & 12.001 & 39.181 & 7.4672 & 16.726 & 15.383 & 15.623 & 48.901 & 36.176 & 7.0054 & 3.7803 \\ \hline

\multicolumn{1}{|l|}{\multirow{3}{*}{\begin{tabular}[c]{@{}l@{}}Power-law \\ Cutoff\end{tabular}}} & RMSE & 24.091 & 9.277 & 9.3332 & 4.9595 & 3.1094 & 5.0699 & 3.2699 & 5.6953 & 13.45 & 11.285 & 2.4371 & 2.5613 \\ %\cline{2-14} 
\multicolumn{1}{|l|}{} & KLD & 0.0162 & \textbf{0.0943} & \textbf{0.0822} & 0.0175 & 0.1825 & 0.09335 & \textbf{0.05837} & 0.08727 & 0.01208 & 0.0122 & 0.02754 & 0.06309 \\ %\cline{2-14} 
\multicolumn{1}{|l|}{} & MAE & 5.0541 & 1.6036 & \textbf{2.8587} & 2.9178 & 1.462 & 2.6267 & \textbf{1.7022} & 1.8486 & 3.7699 & 2.2844 & 0.4695 & 0.4494 \\ \hline
\multicolumn{1}{|l|}{\multirow{3}{*}{\begin{tabular}[c]{@{}l@{}}Exponentiated \\  Burr\end{tabular}}} & RMSE & 13.179 & 7.7979 & 10.617 & 5.7997 & 2.6661 & 7.0909 & 6.3807 & \textbf{3.7625} & 9.3894 & 10.327 & 2.0389 & 6.025 \\ %\cline{2-14} 
\multicolumn{1}{|l|}{} & KLD & 0.01376 & 0.16554 & 0.09488 & \textbf{0.01417} & 0.18974 & 0.10873 & 0.09738 & \textbf{0.07917} & 0.01159 & 0.01183 & 0.02479 & 0.13751 \\ %\cline{2-14} 
\multicolumn{1}{|l|}{} & MAE & 3.7747 & 2.0698 & 3.2044 & 3.007 & 1.4784 & 3.3658 & 2.5721 & \textbf{1.5629} & 3.1225 & 2.2822 & 0.4246 & 0.8475 \\ \hline
\multicolumn{1}{|l|}{\multirow{3}{*}{Burr MO}} & RMSE & 11.057 & 6.1439 & 10.452 & 5.6466 & 2.7267 & 7.2417 & 4.3487 & 4.791 & 7.933 & 11.538 & \textbf{1.6707} & 2.578 \\ %\cline{2-14} 
\multicolumn{1}{|l|}{} & KLD & 0.0142 & 0.18673 & 0.08791 & 0.01681 & 0.18395 & 0.13325 & 0.07581 & 0.08159 & 0.0153 & 0.01226 & 0.02459 & 0.05789 \\ %\cline{2-14} 
\multicolumn{1}{|l|}{} & MAE & 3.4314 & 2.1186 & 3.0912 & 3.1456 & 1.4586 & 3.7697 & 2.022 & 1.7724 & 2.9544 & 2.4551 & \textbf{0.3723} & 0.4294 \\ 
\hline
\end{tabular}}
\end{table*}

\begin{figure*}
     \centering
         \includegraphics[width=8.1cm, height=7.0cm]{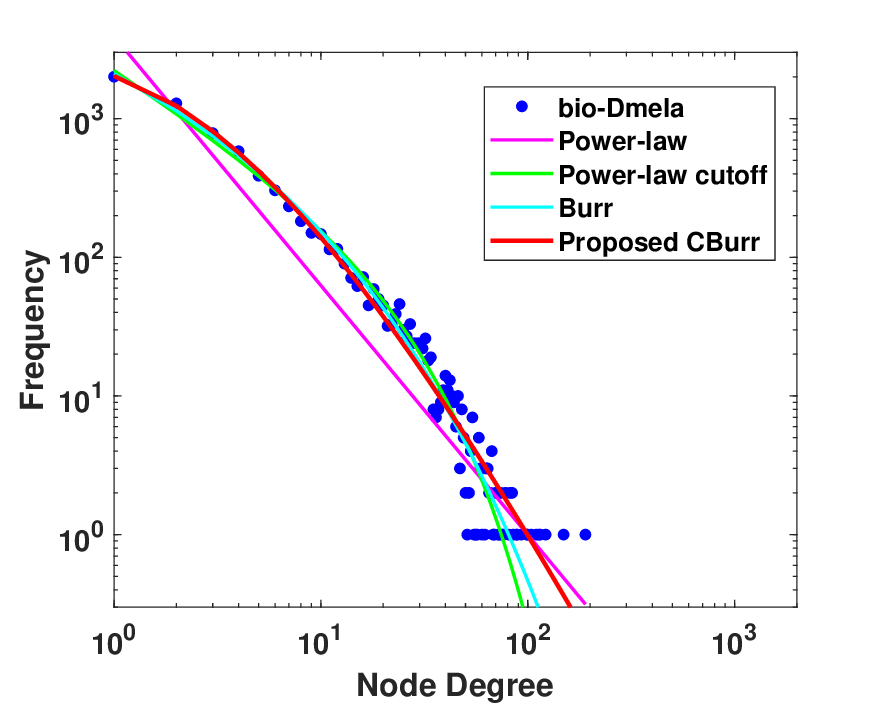}
         \includegraphics[width=8.1cm, height=7.0cm]{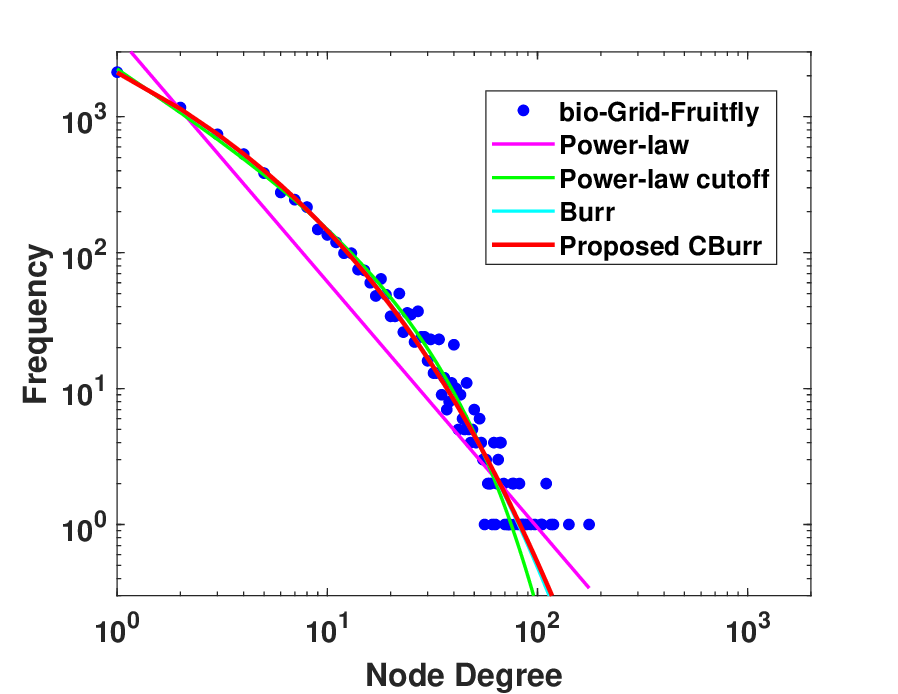}
         \includegraphics[width=8.1cm, height=7.0cm]{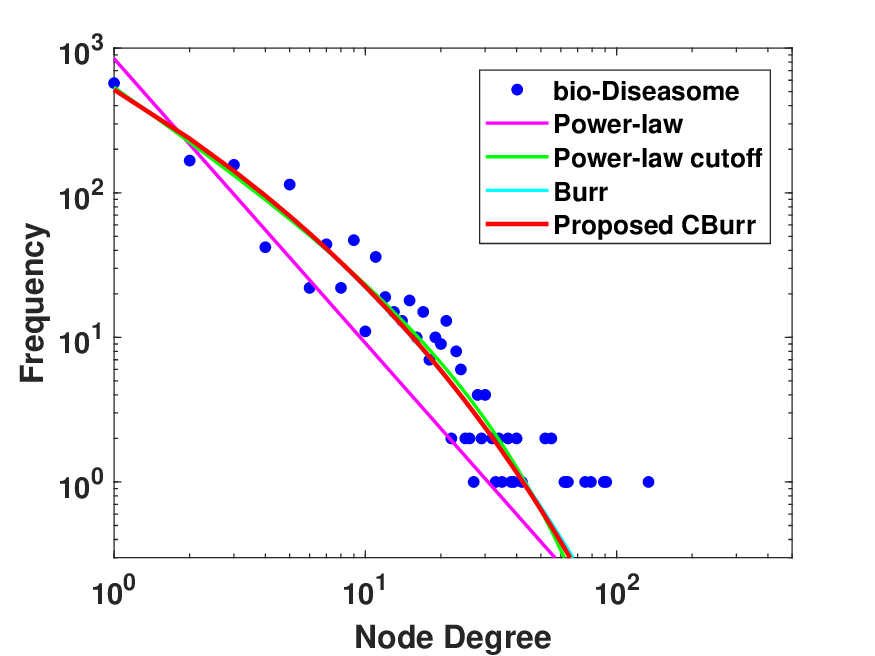}
         \includegraphics[width=8.1cm, height=7.0cm]{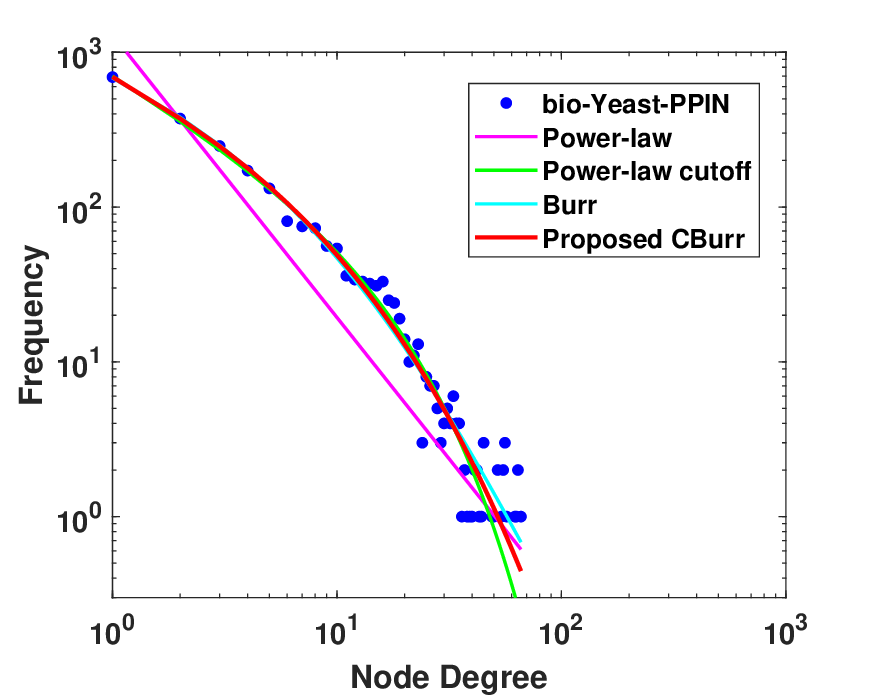}
          \includegraphics[width=8.1cm, height=7.0cm]{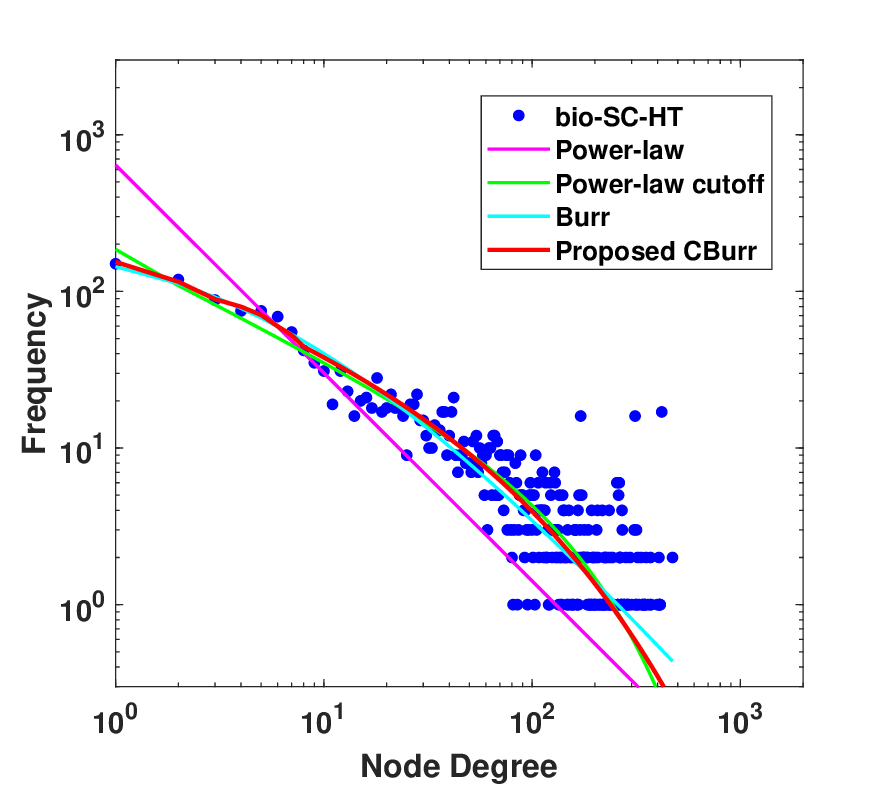}
          \includegraphics[width=8.1cm, height=7.0cm]{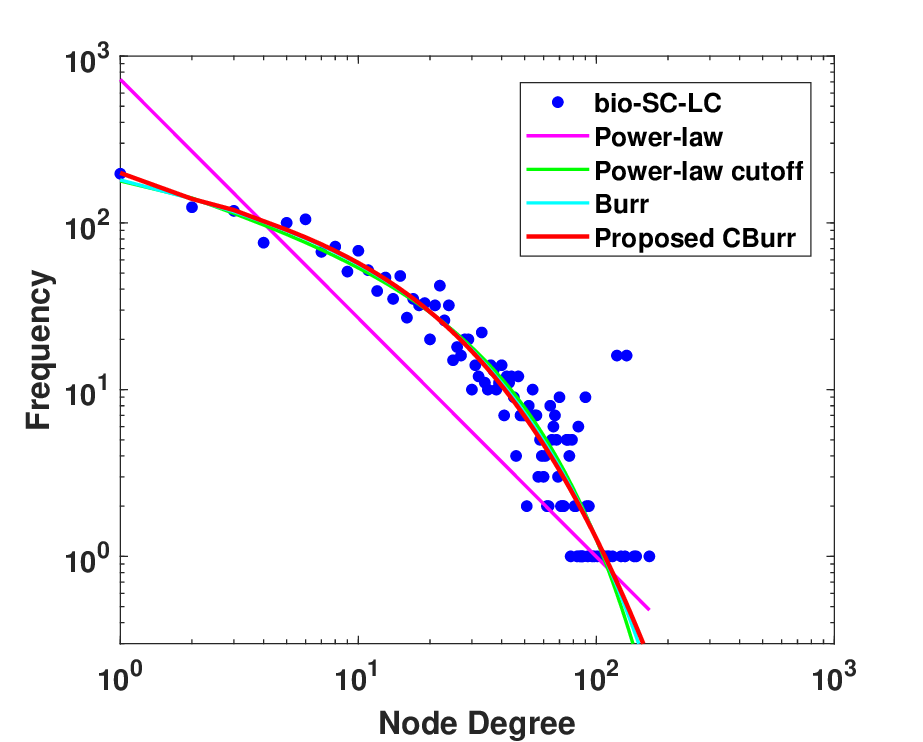}
        \caption{Insights into network connectivity patterns and distribution characteristics are provided by the log-log scale degree distribution of the (top-left) Bio-Dmela, (top-right) Bio-Grid-Fruitfly, (middle-left) Bio-Diseasome, (middle-right) Bio-Yeast-PPIN, (bottom-left) Bio-SC-HT, and (bottom-right) Bio-SC-LC networks.}
        \label{fig:log_log_plots}
\end{figure*}

\begin{figure*}
     \centering
         \includegraphics[width=8.1cm, height=7.0cm]{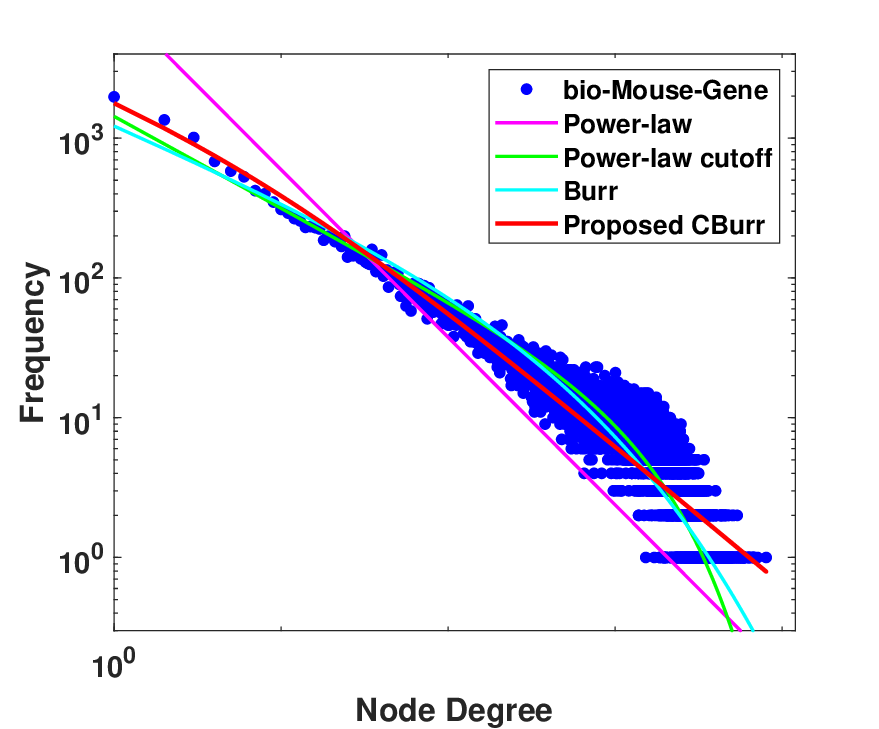}
         \includegraphics[width=8.1cm, height=7.0cm]{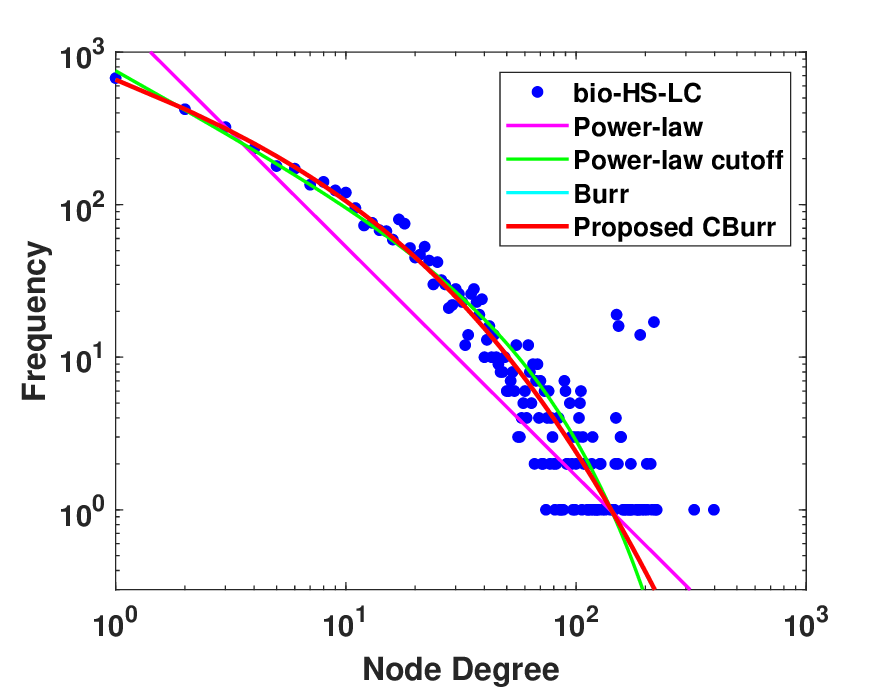}
         \includegraphics[width=8.1cm, height=7.0cm]{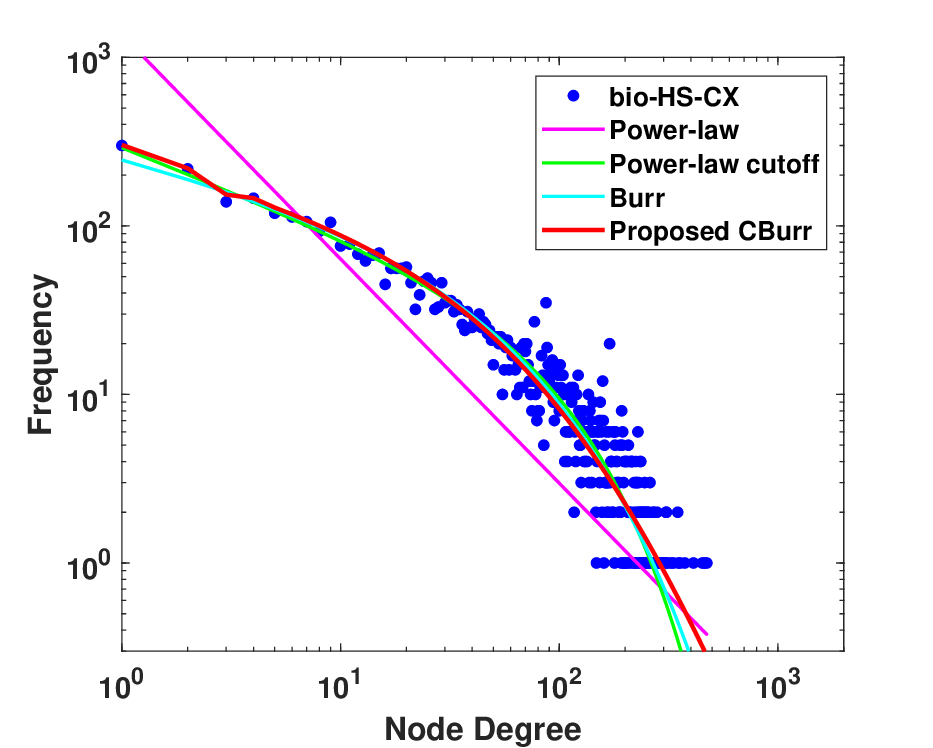}
         \includegraphics[width=8.1cm, height=7.0cm]{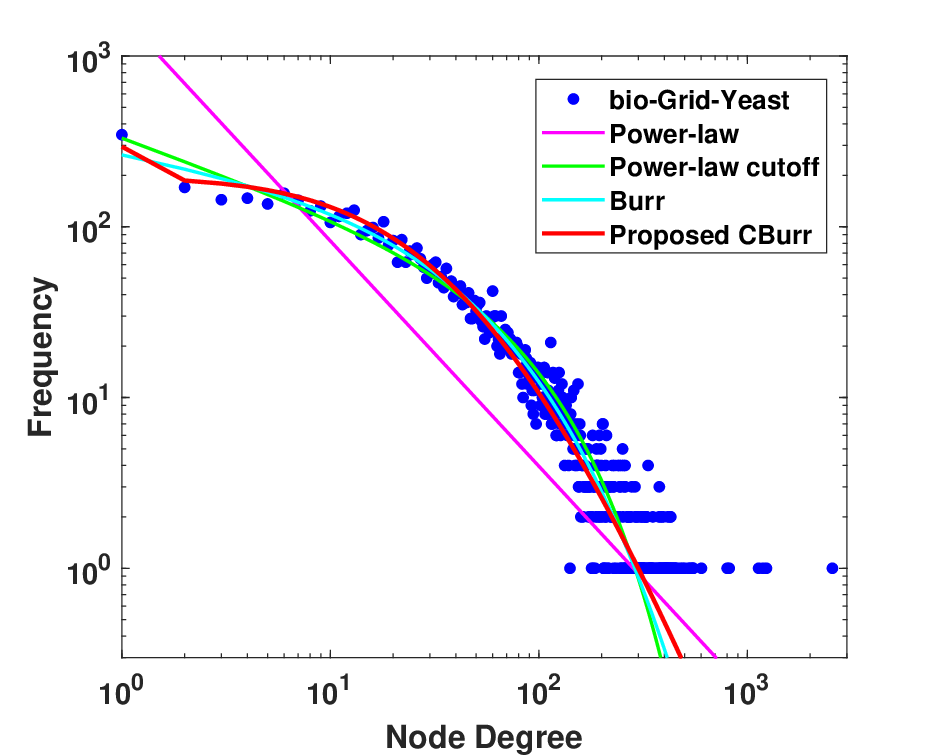}
          \includegraphics[width=8.1cm, height=7.0cm]{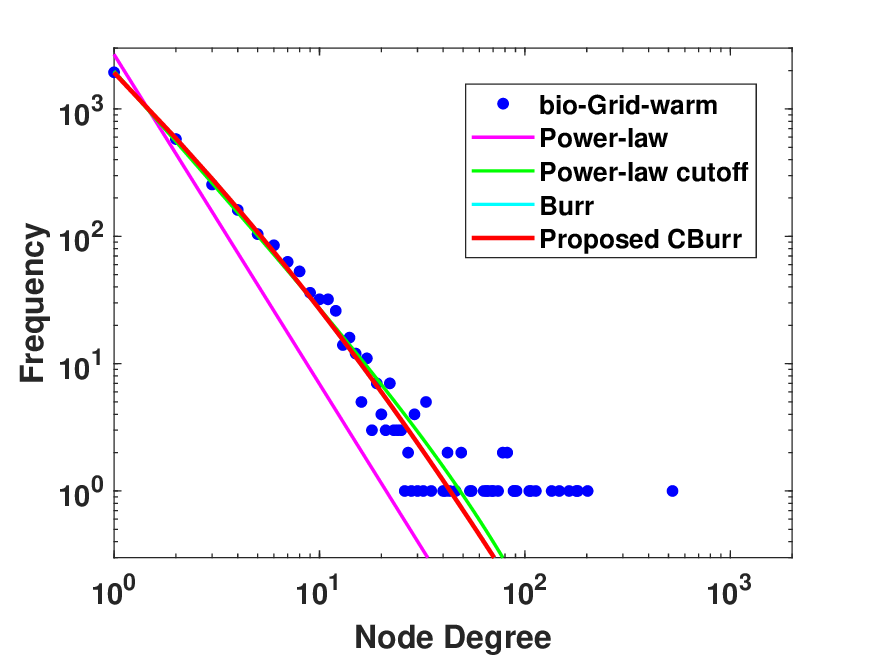}
          \includegraphics[width=8.1cm, height=7.0cm]{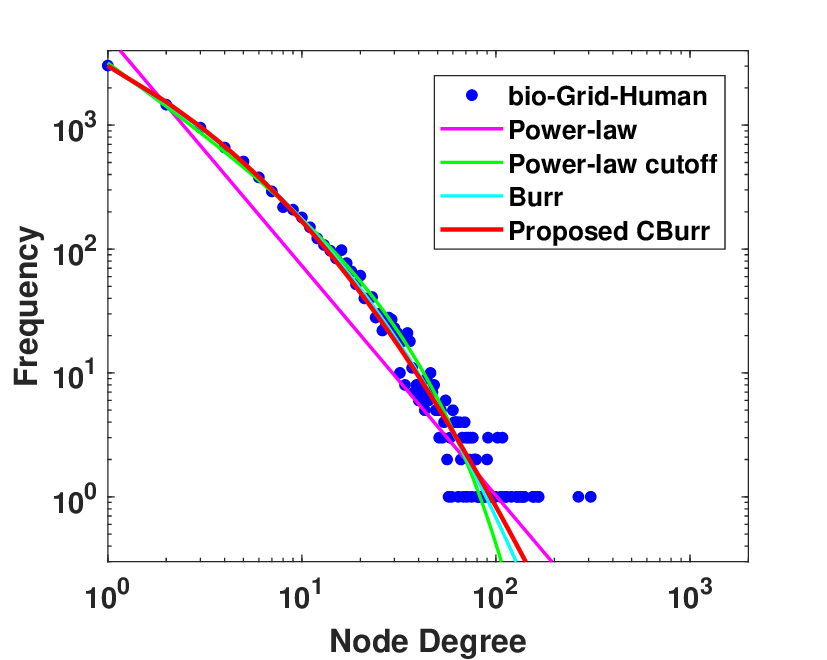}
        \caption{Insights into network connectivity patterns and distribution characteristics are provided by the log-log scale degree distribution of the (top-left) bio-Mouse-Gene, (top-right) Bio HS-LC, (middle-left) Bio-HS-CX, (middle-right) Bio-Grid-Yeast, (bottom-left) Bio-Grid-Warm, and (bottom-right) Bio-Grid-Human networks.}
        \label{fig:log_log_plots2}
\end{figure*}

In the $\chi^2$ test for goodness of fit, the Null Hypothesis $(H_o)$ is that the observed data is consistent with the predicted values of the fitted distribution. If the p-value of the test is lower than 0.05, then the test rejects the Null Hypothesis; otherwise, we fail to reject $(H_o)$ \cite{efron1994introduction}. When the Bootstrap Chi-square test p-value is higher than $0.9$, it suggests the test provides insufficient evidence against the observed data fits well with the proposed CBurr distribution model. This is further supported by the observation that eight out of twelve datasets in Table \ref{table:1estparameter} have p-values higher than 0.9. The high p-values indicate that {there is not a significant deviation between the observed and model-generated degree distributions, confirming the CBurr's strong goodness-of-fit.} The proposed CBurr model {captures both the heavy tail and non-monotonic body structure often seen in many of the tested biological network data. A $p$-value less than $0.9$, while not necessarily indicating a poor fit, suggests that the CBurr may not fully capture all aspects of the observed degree distribution, particularly in the body or tail of the distribution. This could be due to model misspecification, parameter instability, or data heterogeneity, such as the presence of multiple network regimes or domain-specific noise. Nonetheless, if the $p$-value remains above conventional thresholds (e.g., $> 0.05$), the model is still statistically acceptable, albeit with reduced confidence in its robustness.} 

Table \ref{tab:performance} presents statistical measures (RMSE, KLD, and MAE) used to evaluate the performance of the CBurr distribution. The results show that the CBurr distribution generally achieves smaller RMSE and MAE values compared to other distributions, with a few exceptions favoring the power-law cutoff or Burr distribution. Out of twelve datasets, ten of them had the best RMSE performances when using the proposed CBurr distribution model in comparison with others. The Poisson distribution consistently exhibits the poorest performance across all networks. Table \ref{tab:performance} displays the highest values of RMSE, KLD, and MAE, which have been highlighted in bold, across all the datasets among the considered distributions. Regarding KLD values, the CBurr distribution consistently yields smaller values compared to other distributions, except in cases where the power-law cutoff or Burr distribution performs better. This indicates a favorable match between the observed degree distribution and the CBurr distribution in most networks. {As depicted in Table \ref{tab:performance}, CBurr model often behaves similarly to the Burr and cut-off power-law distributions when the Poisson shift parameter $\lambda$ is small, causing the exponential term to approximate $1$ and effectively reducing the model to the standard Burr. Additionally, its survival function structurally resembles the cut-off power-law under certain parameter settings, especially in the tails. Since many biological networks exhibit heavy-tailed but smoothly truncated degree distributions, these models tend to offer similarly good fits. As a result, the added flexibility of the CBurr may not noticeably improve performance unless the data shows more complex, non-monotonic patterns.} In summary, considering the RMSE, KLD, and MAE values, the CBurr distribution outperforms other distributions across all networks. These results strongly support the plausibility of modeling the observed distribution with the CBurr distribution. 
% KL divergence and RMSE are advantageous as performance measures because KL divergence provides a detailed assessment of distributional differences, while RMSE summarizes overall prediction error and enables direct comparisons across models or datasets. Lower RMSE values indicate better model fit and accuracy.

The efficacy of the proposed CBurr distribution can be confirmed by examining the log-log plots of fitted results obtained from various distribution functions. For each network, these plots display the actual frequency distribution, estimated frequency using the CBurr distribution, and estimated frequencies using other distributions such as power-law, power-law cutoff, and Burr distributions. The PDF and CDF, along with supports and parameters, are given in Table \ref{table:pdf cdf}. To make the comparison fair, we implemented them in a similar manner to that done in CBurr. Figs.~\ref{fig:log_log_plots} and \ref{fig:log_log_plots2} showcase these log-log graphs for several biological networks considered in this study. Notably, the CBurr distribution consistently yields a better fit as its curve consistently intersects the midpoint of the observed distribution scatter plot, indicating superior performance compared to the other considered distributions. The proposed CBurr distribution offers an effective and efficient approach for modeling the entire degree distribution of real-world networks, including lower-degree nodes, thereby enhancing our understanding of their structural properties.

\begin{table*}[htbp]
%\footnotesize
\centering
\caption{Parameters, PDF, CDF, and comparison of competing distributions with the proposed CBurr distribution}
\label{table:pdf cdf}
% \resizebox{14cm}{!}{
\begin{tabular}{|c|c|c|c|c|c|}
\hline
Name & pmf/pdf & cdf & support & {No. of} & {Remarks} \\
     &         &     &         & {Parameters} &     \\ \hline
Power-law & $C y^{-\alpha}$ & $\sum_{y=1}^k C y^{-\alpha}$ & $y=1,2,3, \ldots$ & {2} & {Very simple} \\
    &  & & &  & {but inflexible} \\ \hline

Pareto & $\alpha \frac{x_m^\alpha}{y^{\alpha+1}}$ & $\sum_{y=1}^k \alpha \frac{x_m^\alpha}{y^{\alpha+1}}$ & $ y=1,2,3, \ldots$ & {2} & {Heavy-tailed} \\ 
  &  & & &  &  \\ \hline

Log-normal & $\frac{1}{y \sqrt{2 \pi \sigma^2}} \exp \left[-\frac{(\ln y-\mu)^2}{2 \sigma^2}\right]$ & $\frac{1}{2}+\frac{1}{2} \operatorname{erfc}\left(\frac{(\ln y-\mu)}{\sqrt{2} \sigma}\right)$ & $y \in(0, \infty)$ & {2} & {Mean and standard} \\
  &  & & &  & {deviation of log-degrees} \\ \hline

Poisson & $\frac{e^{-\mu} \mu^y}{y!}$ & $\sum_{y=0}^k \frac{e^{-\mu} \mu^y}{y!}$ & $y=0,1,2, \ldots$ & {1} & {Light-tailed} \\
   &  & & &  &  \\ \hline

Power-law Cutoff & $C y^{-\alpha} e^{-\lambda y}$ & $\sum_{y=1}^k C y^{-\alpha} e^{-\lambda y}$ & $y=1,2,3, \ldots $ & {3} & {Adds exponential cutoff} \\
   &  & & &  &  {to Power-law} \\ \hline

Lomax & $\frac{\alpha}{\gamma}\left(1+\frac{y}{\gamma}\right)^{-\alpha-1}$ & $1-\left(1+\frac{y}{\gamma}\right)^{-\alpha}$ & $y>0, \alpha, \gamma > 0 $ & {2} & {Shifted Pareto} \\
   &  & & &  &  \\ \hline
%& $\gamma$ is scale parameter, $\alpha$ is shape parameter\\

Burr & $ c \alpha \gamma^{-c}y^{c-1}\left[1+\left(\frac{y}{\gamma}\right)^{c}\right]^{-\alpha-1}$ & $ 1-\left[1+\left(\frac{y}{\gamma}\right)^{c}\right]^{-\alpha}$ & $y>0, \alpha, \gamma, c>0 $ & {3} & {Captures wide range} \\
   &  & & &  & {of behaviors} \\ \hline
%& $\gamma$ is the scale parameter, and $\alpha, c$ are shape parameters \\

Exponentiated Burr & $\alpha \beta \theta y^{\alpha-1}\frac{\left[1-\left(1+y^\alpha\right)^{-\theta}\right]^{\beta-1}}{\left(1+y^\alpha\right)^{(\theta+1)}}$ & $\left[1-\left(1+y^\alpha\right)^{-\theta}\right]^\beta$ &  $y>0, \alpha, \beta, \theta>0$ & {3} & {Extends Burr with} \\  %& \\
  &  & & &  & {resilience parameter} \\ \hline
  
Burr MO & $\frac{\alpha c k y^{c-1}\left(1+y^c\right)^{-(k+1)}}{\left[1-(1-\alpha)\left(1+y^c\right)^{-k}\right]^2}$ & $\frac{1-\left(1+y^c\right)^{-k}}{1-(1-\alpha)\left(1+y^c\right)^{-k}}$ & $y>0, \alpha, c, k>0$ & {3} & {Extends Burr with} \\  
  &  & & &  & {tilt parameter} \\ \hline

Proposed CBurr & R.H.S. of Eqn. \ref{e12} & $1 -$ R.H.S. of Eqn. \ref{cburrcdf} & $y>0, \lambda, \alpha, \beta, \theta>0$ & {4} & 
{Extends Burr with $\lambda$} \\ 
  &  & & &  & {(Poisson rate), allows skewing,} \\ 
  &  & & &  & {controls body-tail balance} \\ \hline
\end{tabular}
\end{table*}

\section{Discussions} \label{conclusion}
This article introduced the CBurr distribution for modeling the node degree distribution of biological networks. The CBurr distribution is derived from a family of continuous probability distributions, allowing for the inclusion of a Poisson-shift parameter to enhance the flexibility and efficiency of the Burr XII distribution. Our data analysis and experiments have provided strong evidence supporting the superiority of the proposed CBurr distribution over traditional power-law distributions. The CBurr distribution demonstrates better modeling capabilities, capturing the intricate connectivity patterns and characteristics observed in real-world heavy-tailed biological networks. By incorporating the CBurr distribution, we are able to represent the entire node degree distribution accurately, considering both highly connected nodes (hubs) and nodes with lower degrees. This comprehensive modeling approach enables a more realistic representation of the network structure and facilitates a deeper understanding of biological systems. 
%In future research, we will explore the applicability of the CBurr distribution in a broader range of network types and investigate its performance across various statistical properties.

In conclusion, the proposed CBurr distribution offers a valuable tool for the effective and efficient modeling of node degree distributions in biological networks. Its flexibility, accuracy, and ability to capture complex connectivity patterns make it a promising choice for analyzing and interpreting biological network data. By conducting extensive analyses of different network structures and characteristics, we aim to gain a deeper understanding of the versatility and robustness of the proposed distribution. {While the CBurr distribution offers high flexibility, it comes with increased computational cost due to the normalization constant and four-parameter estimation. It risks overfitting, especially on smaller or sparse datasets, and its parameters can be difficult to interpret biologically. While excellent for empirical fit, it lacks a generative mechanism rooted in biological network theory. This can be considered a possible future research avenue for understanding the generative mechanism of biological networks using the proposed CBurr model. As future research, exploring CBurr distribution for analyzing social networks, collaboration networks, temporal networks, and web networks will demonstrate its effectiveness and contribute to the modeling and analysis of diverse real-world networks in other areas.}

\section*{Data Availability Statement}
Datasets are collected from the network repository \url{https://networkrepository.com/bio.php}. Codes and implementation of the models are available at \url{https://github.com/ctanujit/CBurr}.

\begin{acknowledgments}
The authors would like to thank the editor and the two learned reviewers for their constructive discussions and valuable comments. 
\end{acknowledgments}

%\section*{Appendixes}

\bibliography{bibliography}% Produces the bibliography via BibTeX.

\providecommand{\noopsort}[1]{}\providecommand{\singleletter}[1]{#1}%
\begin{thebibliography}{67}%
\makeatletter
\providecommand \@ifxundefined [1]{%
 \@ifx{#1\undefined}
}%
\providecommand \@ifnum [1]{%
 \ifnum #1\expandafter \@firstoftwo
 \else \expandafter \@secondoftwo
 \fi
}%
\providecommand \@ifx [1]{%
 \ifx #1\expandafter \@firstoftwo
 \else \expandafter \@secondoftwo
 \fi
}%
\providecommand \natexlab [1]{#1}%
\providecommand \enquote  [1]{``#1''}%
\providecommand \bibnamefont  [1]{#1}%
\providecommand \bibfnamefont [1]{#1}%
\providecommand \citenamefont [1]{#1}%
\providecommand \href@noop [0]{\@secondoftwo}%
\providecommand \href [0]{\begingroup \@sanitize@url \@href}%
\providecommand \@href[1]{\@@startlink{#1}\@@href}%
\providecommand \@@href[1]{\endgroup#1\@@endlink}%
\providecommand \@sanitize@url [0]{\catcode `\\12\catcode `\$12\catcode `\&12\catcode `\#12\catcode `\^12\catcode `\_12\catcode `\%12\relax}%
\providecommand \@@startlink[1]{}%
\providecommand \@@endlink[0]{}%
\providecommand \url  [0]{\begingroup\@sanitize@url \@url }%
\providecommand \@url [1]{\endgroup\@href {#1}{\urlprefix }}%
\providecommand \urlprefix  [0]{URL }%
\providecommand \Eprint [0]{\href }%
\providecommand \doibase [0]{http://dx.doi.org/}%
\providecommand \selectlanguage [0]{\@gobble}%
\providecommand \bibinfo  [0]{\@secondoftwo}%
\providecommand \bibfield  [0]{\@secondoftwo}%
\providecommand \translation [1]{[#1]}%
\providecommand \BibitemOpen [0]{}%
\providecommand \bibitemStop [0]{}%
\providecommand \bibitemNoStop [0]{.\EOS\space}%
\providecommand \EOS [0]{\spacefactor3000\relax}%
\providecommand \BibitemShut  [1]{\csname bibitem#1\endcsname}%
\let\auto@bib@innerbib\@empty
%</preamble>
\bibitem [{\citenamefont {Barab{\'a}si}\ and\ \citenamefont {Albert}(1999)}]{barabasi1999emergence}%
  \BibitemOpen
  \bibfield  {author} {\bibinfo {author} {\bibfnamefont {A.-L.}\ \bibnamefont {Barab{\'a}si}}\ and\ \bibinfo {author} {\bibfnamefont {R.}~\bibnamefont {Albert}},\ }\bibfield  {title} {\enquote {\bibinfo {title} {Emergence of scaling in random networks},}\ }\href@noop {} {\bibfield  {journal} {\bibinfo  {journal} {science}\ }\textbf {\bibinfo {volume} {286}},\ \bibinfo {pages} {509--512} (\bibinfo {year} {1999})}\BibitemShut {NoStop}%
\bibitem [{\citenamefont {Broido}\ and\ \citenamefont {Clauset}(2019)}]{broido2019scale}%
  \BibitemOpen
  \bibfield  {author} {\bibinfo {author} {\bibfnamefont {A.~D.}\ \bibnamefont {Broido}}\ and\ \bibinfo {author} {\bibfnamefont {A.}~\bibnamefont {Clauset}},\ }\bibfield  {title} {\enquote {\bibinfo {title} {Scale-free networks are rare},}\ }\href@noop {} {\bibfield  {journal} {\bibinfo  {journal} {Nature communications}\ }\textbf {\bibinfo {volume} {10}},\ \bibinfo {pages} {1--10} (\bibinfo {year} {2019})}\BibitemShut {NoStop}%
\bibitem [{\citenamefont {Chattopadhyay}\ \emph {et~al.}(2021{\natexlab{a}})\citenamefont {Chattopadhyay}, \citenamefont {Chakraborty}, \citenamefont {Ghosh},\ and\ \citenamefont {Das}}]{chattopadhyay2021uncovering}%
  \BibitemOpen
  \bibfield  {author} {\bibinfo {author} {\bibfnamefont {S.}~\bibnamefont {Chattopadhyay}}, \bibinfo {author} {\bibfnamefont {T.}~\bibnamefont {Chakraborty}}, \bibinfo {author} {\bibfnamefont {K.}~\bibnamefont {Ghosh}}, \ and\ \bibinfo {author} {\bibfnamefont {A.~K.}\ \bibnamefont {Das}},\ }\bibfield  {title} {\enquote {\bibinfo {title} {Uncovering patterns in heavy-tailed networks: A journey beyond scale-free},}\ }in\ \href@noop {} {\emph {\bibinfo {booktitle} {Proceedings of the 3rd ACM India Joint International Conference on Data Science \& Management of Data (8th ACM IKDD CODS \& 26th COMAD)}}}\ (\bibinfo {year} {2021})\ pp.\ \bibinfo {pages} {136--144}\BibitemShut {NoStop}%
\bibitem [{\citenamefont {Wuchty}(2001)}]{wuchty2001scale}%
  \BibitemOpen
  \bibfield  {author} {\bibinfo {author} {\bibfnamefont {S.}~\bibnamefont {Wuchty}},\ }\bibfield  {title} {\enquote {\bibinfo {title} {Scale-free behavior in protein domain networks},}\ }\href@noop {} {\bibfield  {journal} {\bibinfo  {journal} {Molecular biology and evolution}\ }\textbf {\bibinfo {volume} {18}},\ \bibinfo {pages} {1694--1702} (\bibinfo {year} {2001})}\BibitemShut {NoStop}%
\bibitem [{\citenamefont {Arita}(2005)}]{arita2005scale}%
  \BibitemOpen
  \bibfield  {author} {\bibinfo {author} {\bibfnamefont {M.}~\bibnamefont {Arita}},\ }\bibfield  {title} {\enquote {\bibinfo {title} {Scale-freeness and biological networks},}\ }\href@noop {} {\bibfield  {journal} {\bibinfo  {journal} {Journal of biochemistry}\ }\textbf {\bibinfo {volume} {138}},\ \bibinfo {pages} {1--4} (\bibinfo {year} {2005})}\BibitemShut {NoStop}%
\bibitem [{\citenamefont {Khanin}\ and\ \citenamefont {Wit}(2006)}]{khanin2006scale}%
  \BibitemOpen
  \bibfield  {author} {\bibinfo {author} {\bibfnamefont {R.}~\bibnamefont {Khanin}}\ and\ \bibinfo {author} {\bibfnamefont {E.}~\bibnamefont {Wit}},\ }\bibfield  {title} {\enquote {\bibinfo {title} {How scale-free are biological networks},}\ }\href@noop {} {\bibfield  {journal} {\bibinfo  {journal} {Journal of computational biology}\ }\textbf {\bibinfo {volume} {13}},\ \bibinfo {pages} {810--818} (\bibinfo {year} {2006})}\BibitemShut {NoStop}%
\bibitem [{\citenamefont {Wuchty}, \citenamefont {Ravasz},\ and\ \citenamefont {Barab{\'a}si}(2006)}]{wuchty2006architecture}%
  \BibitemOpen
  \bibfield  {author} {\bibinfo {author} {\bibfnamefont {S.}~\bibnamefont {Wuchty}}, \bibinfo {author} {\bibfnamefont {E.}~\bibnamefont {Ravasz}}, \ and\ \bibinfo {author} {\bibfnamefont {A.-L.}\ \bibnamefont {Barab{\'a}si}},\ }\bibfield  {title} {\enquote {\bibinfo {title} {The architecture of biological networks},}\ }\href@noop {} {\bibfield  {journal} {\bibinfo  {journal} {Complex systems science in biomedicine}\ ,\ \bibinfo {pages} {165--181}} (\bibinfo {year} {2006})}\BibitemShut {NoStop}%
\bibitem [{\citenamefont {Jeong}\ \emph {et~al.}(2000)\citenamefont {Jeong}, \citenamefont {Tombor}, \citenamefont {Albert}, \citenamefont {Oltvai},\ and\ \citenamefont {Barab{\'a}si}}]{jeong2000large}%
  \BibitemOpen
  \bibfield  {author} {\bibinfo {author} {\bibfnamefont {H.}~\bibnamefont {Jeong}}, \bibinfo {author} {\bibfnamefont {B.}~\bibnamefont {Tombor}}, \bibinfo {author} {\bibfnamefont {R.}~\bibnamefont {Albert}}, \bibinfo {author} {\bibfnamefont {Z.~N.}\ \bibnamefont {Oltvai}}, \ and\ \bibinfo {author} {\bibfnamefont {A.-L.}\ \bibnamefont {Barab{\'a}si}},\ }\bibfield  {title} {\enquote {\bibinfo {title} {The large-scale organization of metabolic networks},}\ }\href@noop {} {\bibfield  {journal} {\bibinfo  {journal} {Nature}\ }\textbf {\bibinfo {volume} {407}},\ \bibinfo {pages} {651--654} (\bibinfo {year} {2000})}\BibitemShut {NoStop}%
\bibitem [{\citenamefont {Fell}\ and\ \citenamefont {Wagner}(2000)}]{fell2000small}%
  \BibitemOpen
  \bibfield  {author} {\bibinfo {author} {\bibfnamefont {D.~A.}\ \bibnamefont {Fell}}\ and\ \bibinfo {author} {\bibfnamefont {A.}~\bibnamefont {Wagner}},\ }\bibfield  {title} {\enquote {\bibinfo {title} {The small world of metabolism},}\ }\href@noop {} {\bibfield  {journal} {\bibinfo  {journal} {Nature biotechnology}\ }\textbf {\bibinfo {volume} {18}},\ \bibinfo {pages} {1121--1122} (\bibinfo {year} {2000})}\BibitemShut {NoStop}%
\bibitem [{\citenamefont {Ma}\ and\ \citenamefont {Zeng}(2003)}]{ma2003connectivity}%
  \BibitemOpen
  \bibfield  {author} {\bibinfo {author} {\bibfnamefont {H.-W.}\ \bibnamefont {Ma}}\ and\ \bibinfo {author} {\bibfnamefont {A.-P.}\ \bibnamefont {Zeng}},\ }\bibfield  {title} {\enquote {\bibinfo {title} {The connectivity structure, giant strong component and centrality of metabolic networks},}\ }\href@noop {} {\bibfield  {journal} {\bibinfo  {journal} {Bioinformatics}\ }\textbf {\bibinfo {volume} {19}},\ \bibinfo {pages} {1423--1430} (\bibinfo {year} {2003})}\BibitemShut {NoStop}%
\bibitem [{\citenamefont {Uetz}\ \emph {et~al.}(2000)\citenamefont {Uetz}, \citenamefont {Giot}, \citenamefont {Cagney}, \citenamefont {Mansfield}, \citenamefont {Judson}, \citenamefont {Knight}, \citenamefont {Lockshon}, \citenamefont {Narayan}, \citenamefont {Srinivasan}, \citenamefont {Pochart} \emph {et~al.}}]{uetz2000comprehensive}%
  \BibitemOpen
  \bibfield  {author} {\bibinfo {author} {\bibfnamefont {P.}~\bibnamefont {Uetz}}, \bibinfo {author} {\bibfnamefont {L.}~\bibnamefont {Giot}}, \bibinfo {author} {\bibfnamefont {G.}~\bibnamefont {Cagney}}, \bibinfo {author} {\bibfnamefont {T.~A.}\ \bibnamefont {Mansfield}}, \bibinfo {author} {\bibfnamefont {R.~S.}\ \bibnamefont {Judson}}, \bibinfo {author} {\bibfnamefont {J.~R.}\ \bibnamefont {Knight}}, \bibinfo {author} {\bibfnamefont {D.}~\bibnamefont {Lockshon}}, \bibinfo {author} {\bibfnamefont {V.}~\bibnamefont {Narayan}}, \bibinfo {author} {\bibfnamefont {M.}~\bibnamefont {Srinivasan}}, \bibinfo {author} {\bibfnamefont {P.}~\bibnamefont {Pochart}},  \emph {et~al.},\ }\bibfield  {title} {\enquote {\bibinfo {title} {A comprehensive analysis of protein--protein interactions in saccharomyces cerevisiae},}\ }\href@noop {} {\bibfield  {journal} {\bibinfo  {journal} {Nature}\ }\textbf {\bibinfo {volume} {403}},\ \bibinfo {pages} {623--627} (\bibinfo {year} {2000})}\BibitemShut {NoStop}%
\bibitem [{\citenamefont {Schwikowski}, \citenamefont {Uetz},\ and\ \citenamefont {Fields}(2000)}]{schwikowski2000network}%
  \BibitemOpen
  \bibfield  {author} {\bibinfo {author} {\bibfnamefont {B.}~\bibnamefont {Schwikowski}}, \bibinfo {author} {\bibfnamefont {P.}~\bibnamefont {Uetz}}, \ and\ \bibinfo {author} {\bibfnamefont {S.}~\bibnamefont {Fields}},\ }\bibfield  {title} {\enquote {\bibinfo {title} {A network of protein--protein interactions in yeast},}\ }\href@noop {} {\bibfield  {journal} {\bibinfo  {journal} {Nature biotechnology}\ }\textbf {\bibinfo {volume} {18}},\ \bibinfo {pages} {1257--1261} (\bibinfo {year} {2000})}\BibitemShut {NoStop}%
\bibitem [{\citenamefont {Maslov}\ and\ \citenamefont {Sneppen}(2002)}]{maslov2002specificity}%
  \BibitemOpen
  \bibfield  {author} {\bibinfo {author} {\bibfnamefont {S.}~\bibnamefont {Maslov}}\ and\ \bibinfo {author} {\bibfnamefont {K.}~\bibnamefont {Sneppen}},\ }\bibfield  {title} {\enquote {\bibinfo {title} {Specificity and stability in topology of protein networks},}\ }\href@noop {} {\bibfield  {journal} {\bibinfo  {journal} {Science}\ }\textbf {\bibinfo {volume} {296}},\ \bibinfo {pages} {910--913} (\bibinfo {year} {2002})}\BibitemShut {NoStop}%
\bibitem [{\citenamefont {Rual}\ \emph {et~al.}(2005)\citenamefont {Rual}, \citenamefont {Venkatesan}, \citenamefont {Hao}, \citenamefont {Hirozane-Kishikawa}, \citenamefont {Dricot}, \citenamefont {Li}, \citenamefont {Berriz}, \citenamefont {Gibbons}, \citenamefont {Dreze}, \citenamefont {Ayivi-Guedehoussou} \emph {et~al.}}]{rual2005towards}%
  \BibitemOpen
  \bibfield  {author} {\bibinfo {author} {\bibfnamefont {J.-F.}\ \bibnamefont {Rual}}, \bibinfo {author} {\bibfnamefont {K.}~\bibnamefont {Venkatesan}}, \bibinfo {author} {\bibfnamefont {T.}~\bibnamefont {Hao}}, \bibinfo {author} {\bibfnamefont {T.}~\bibnamefont {Hirozane-Kishikawa}}, \bibinfo {author} {\bibfnamefont {A.}~\bibnamefont {Dricot}}, \bibinfo {author} {\bibfnamefont {N.}~\bibnamefont {Li}}, \bibinfo {author} {\bibfnamefont {G.~F.}\ \bibnamefont {Berriz}}, \bibinfo {author} {\bibfnamefont {F.~D.}\ \bibnamefont {Gibbons}}, \bibinfo {author} {\bibfnamefont {M.}~\bibnamefont {Dreze}}, \bibinfo {author} {\bibfnamefont {N.}~\bibnamefont {Ayivi-Guedehoussou}},  \emph {et~al.},\ }\bibfield  {title} {\enquote {\bibinfo {title} {Towards a proteome-scale map of the human protein--protein interaction network},}\ }\href@noop {} {\bibfield  {journal} {\bibinfo  {journal} {Nature}\ }\textbf {\bibinfo {volume} {437}},\ \bibinfo {pages} {1173--1178} (\bibinfo {year} {2005})}\BibitemShut {NoStop}%
\bibitem [{\citenamefont {Rzhetsky}\ and\ \citenamefont {Gomez}(2001)}]{rzhetsky2001birth}%
  \BibitemOpen
  \bibfield  {author} {\bibinfo {author} {\bibfnamefont {A.}~\bibnamefont {Rzhetsky}}\ and\ \bibinfo {author} {\bibfnamefont {S.~M.}\ \bibnamefont {Gomez}},\ }\bibfield  {title} {\enquote {\bibinfo {title} {Birth of scale-free molecular networks and the number of distinct dna and protein domains per genome},}\ }\href@noop {} {\bibfield  {journal} {\bibinfo  {journal} {Bioinformatics}\ }\textbf {\bibinfo {volume} {17}},\ \bibinfo {pages} {988--996} (\bibinfo {year} {2001})}\BibitemShut {NoStop}%
\bibitem [{\citenamefont {Tong}\ \emph {et~al.}(2004)\citenamefont {Tong}, \citenamefont {Lesage}, \citenamefont {Bader}, \citenamefont {Ding}, \citenamefont {Xu}, \citenamefont {Xin}, \citenamefont {Young}, \citenamefont {Berriz}, \citenamefont {Brost}, \citenamefont {Chang} \emph {et~al.}}]{tong2004global}%
  \BibitemOpen
  \bibfield  {author} {\bibinfo {author} {\bibfnamefont {A.~H.~Y.}\ \bibnamefont {Tong}}, \bibinfo {author} {\bibfnamefont {G.}~\bibnamefont {Lesage}}, \bibinfo {author} {\bibfnamefont {G.~D.}\ \bibnamefont {Bader}}, \bibinfo {author} {\bibfnamefont {H.}~\bibnamefont {Ding}}, \bibinfo {author} {\bibfnamefont {H.}~\bibnamefont {Xu}}, \bibinfo {author} {\bibfnamefont {X.}~\bibnamefont {Xin}}, \bibinfo {author} {\bibfnamefont {J.}~\bibnamefont {Young}}, \bibinfo {author} {\bibfnamefont {G.~F.}\ \bibnamefont {Berriz}}, \bibinfo {author} {\bibfnamefont {R.~L.}\ \bibnamefont {Brost}}, \bibinfo {author} {\bibfnamefont {M.}~\bibnamefont {Chang}},  \emph {et~al.},\ }\bibfield  {title} {\enquote {\bibinfo {title} {Global mapping of the yeast genetic interaction network},}\ }\href@noop {} {\bibfield  {journal} {\bibinfo  {journal} {science}\ }\textbf {\bibinfo {volume} {303}},\ \bibinfo {pages} {808--813} (\bibinfo {year} {2004})}\BibitemShut {NoStop}%
\bibitem [{\citenamefont {Lee}\ \emph {et~al.}(2002)\citenamefont {Lee}, \citenamefont {Rinaldi}, \citenamefont {Robert}, \citenamefont {Odom}, \citenamefont {Bar-Joseph}, \citenamefont {Gerber}, \citenamefont {Hannett}, \citenamefont {Harbison}, \citenamefont {Thompson}, \citenamefont {Simon} \emph {et~al.}}]{lee2002transcriptional}%
  \BibitemOpen
  \bibfield  {author} {\bibinfo {author} {\bibfnamefont {T.~I.}\ \bibnamefont {Lee}}, \bibinfo {author} {\bibfnamefont {N.~J.}\ \bibnamefont {Rinaldi}}, \bibinfo {author} {\bibfnamefont {F.}~\bibnamefont {Robert}}, \bibinfo {author} {\bibfnamefont {D.~T.}\ \bibnamefont {Odom}}, \bibinfo {author} {\bibfnamefont {Z.}~\bibnamefont {Bar-Joseph}}, \bibinfo {author} {\bibfnamefont {G.~K.}\ \bibnamefont {Gerber}}, \bibinfo {author} {\bibfnamefont {N.~M.}\ \bibnamefont {Hannett}}, \bibinfo {author} {\bibfnamefont {C.~T.}\ \bibnamefont {Harbison}}, \bibinfo {author} {\bibfnamefont {C.~M.}\ \bibnamefont {Thompson}}, \bibinfo {author} {\bibfnamefont {I.}~\bibnamefont {Simon}},  \emph {et~al.},\ }\bibfield  {title} {\enquote {\bibinfo {title} {Transcriptional regulatory networks in saccharomyces cerevisiae},}\ }\href@noop {} {\bibfield  {journal} {\bibinfo  {journal} {science}\ }\textbf {\bibinfo {volume} {298}},\ \bibinfo {pages} {799--804} (\bibinfo {year} {2002})}\BibitemShut {NoStop}%
\bibitem [{\citenamefont {Farkas}\ \emph {et~al.}(2003)\citenamefont {Farkas}, \citenamefont {Jeong}, \citenamefont {Vicsek}, \citenamefont {Barab{\'a}si},\ and\ \citenamefont {Oltvai}}]{farkas2003topology}%
  \BibitemOpen
  \bibfield  {author} {\bibinfo {author} {\bibfnamefont {I.}~\bibnamefont {Farkas}}, \bibinfo {author} {\bibfnamefont {H.}~\bibnamefont {Jeong}}, \bibinfo {author} {\bibfnamefont {T.}~\bibnamefont {Vicsek}}, \bibinfo {author} {\bibfnamefont {A.-L.}\ \bibnamefont {Barab{\'a}si}}, \ and\ \bibinfo {author} {\bibfnamefont {Z.~N.}\ \bibnamefont {Oltvai}},\ }\bibfield  {title} {\enquote {\bibinfo {title} {The topology of the transcription regulatory network in the yeast, saccharomyces cerevisiae},}\ }\href@noop {} {\bibfield  {journal} {\bibinfo  {journal} {Physica A: Statistical Mechanics and its Applications}\ }\textbf {\bibinfo {volume} {318}},\ \bibinfo {pages} {601--612} (\bibinfo {year} {2003})}\BibitemShut {NoStop}%
\bibitem [{\citenamefont {Luscombe}\ \emph {et~al.}(2004)\citenamefont {Luscombe}, \citenamefont {Babu}, \citenamefont {Yu}, \citenamefont {Snyder}, \citenamefont {Teichmann},\ and\ \citenamefont {Gerstein}}]{luscombe2004genomic}%
  \BibitemOpen
  \bibfield  {author} {\bibinfo {author} {\bibfnamefont {N.~M.}\ \bibnamefont {Luscombe}}, \bibinfo {author} {\bibfnamefont {M.~M.}\ \bibnamefont {Babu}}, \bibinfo {author} {\bibfnamefont {H.}~\bibnamefont {Yu}}, \bibinfo {author} {\bibfnamefont {M.}~\bibnamefont {Snyder}}, \bibinfo {author} {\bibfnamefont {S.~A.}\ \bibnamefont {Teichmann}}, \ and\ \bibinfo {author} {\bibfnamefont {M.}~\bibnamefont {Gerstein}},\ }\bibfield  {title} {\enquote {\bibinfo {title} {Genomic analysis of regulatory network dynamics reveals large topological changes},}\ }\href@noop {} {\bibfield  {journal} {\bibinfo  {journal} {Nature}\ }\textbf {\bibinfo {volume} {431}},\ \bibinfo {pages} {308--312} (\bibinfo {year} {2004})}\BibitemShut {NoStop}%
\bibitem [{\citenamefont {Haggarty}, \citenamefont {Clemons},\ and\ \citenamefont {Schreiber}(2003)}]{haggarty2003chemical}%
  \BibitemOpen
  \bibfield  {author} {\bibinfo {author} {\bibfnamefont {S.~J.}\ \bibnamefont {Haggarty}}, \bibinfo {author} {\bibfnamefont {P.~A.}\ \bibnamefont {Clemons}}, \ and\ \bibinfo {author} {\bibfnamefont {S.~L.}\ \bibnamefont {Schreiber}},\ }\bibfield  {title} {\enquote {\bibinfo {title} {Chemical genomic profiling of biological networks using graph theory and combinations of small molecule perturbations},}\ }\href@noop {} {\bibfield  {journal} {\bibinfo  {journal} {Journal of the American Chemical Society}\ }\textbf {\bibinfo {volume} {125}},\ \bibinfo {pages} {10543--10545} (\bibinfo {year} {2003})}\BibitemShut {NoStop}%
\bibitem [{\citenamefont {Dorogovtsev}\ and\ \citenamefont {Mendes}(2003)}]{dorogovtsev2003evolution}%
  \BibitemOpen
  \bibfield  {author} {\bibinfo {author} {\bibfnamefont {S.~N.}\ \bibnamefont {Dorogovtsev}}\ and\ \bibinfo {author} {\bibfnamefont {J.~F.}\ \bibnamefont {Mendes}},\ }\href@noop {} {\emph {\bibinfo {title} {Evolution of networks: From biological nets to the Internet and WWW}}}\ (\bibinfo  {publisher} {Oxford university press},\ \bibinfo {year} {2003})\BibitemShut {NoStop}%
\bibitem [{\citenamefont {Lima-Mendez}\ and\ \citenamefont {Van~Helden}(2009)}]{lima2009powerful}%
  \BibitemOpen
  \bibfield  {author} {\bibinfo {author} {\bibfnamefont {G.}~\bibnamefont {Lima-Mendez}}\ and\ \bibinfo {author} {\bibfnamefont {J.}~\bibnamefont {Van~Helden}},\ }\bibfield  {title} {\enquote {\bibinfo {title} {The powerful law of the power law and other myths in network biology},}\ }\href@noop {} {\bibfield  {journal} {\bibinfo  {journal} {Molecular BioSystems}\ }\textbf {\bibinfo {volume} {5}},\ \bibinfo {pages} {1482--1493} (\bibinfo {year} {2009})}\BibitemShut {NoStop}%
\bibitem [{\citenamefont {Miranda}\ \emph {et~al.}(2016)\citenamefont {Miranda}, \citenamefont {Pinto}, \citenamefont {Baptista},\ and\ \citenamefont {La~Guardia}}]{miranda2016theoretical}%
  \BibitemOpen
  \bibfield  {author} {\bibinfo {author} {\bibfnamefont {P.~J.}\ \bibnamefont {Miranda}}, \bibinfo {author} {\bibfnamefont {S.~E. d.~S.}\ \bibnamefont {Pinto}}, \bibinfo {author} {\bibfnamefont {M.~S.}\ \bibnamefont {Baptista}}, \ and\ \bibinfo {author} {\bibfnamefont {G.~G.}\ \bibnamefont {La~Guardia}},\ }\bibfield  {title} {\enquote {\bibinfo {title} {Theoretical knock-outs on biological networks},}\ }\href@noop {} {\bibfield  {journal} {\bibinfo  {journal} {Journal of Theoretical Biology}\ }\textbf {\bibinfo {volume} {403}},\ \bibinfo {pages} {38--44} (\bibinfo {year} {2016})}\BibitemShut {NoStop}%
\bibitem [{\citenamefont {Saucan}, \citenamefont {Samal},\ and\ \citenamefont {Jost}(2021)}]{saucan2021simple}%
  \BibitemOpen
  \bibfield  {author} {\bibinfo {author} {\bibfnamefont {E.}~\bibnamefont {Saucan}}, \bibinfo {author} {\bibfnamefont {A.}~\bibnamefont {Samal}}, \ and\ \bibinfo {author} {\bibfnamefont {J.}~\bibnamefont {Jost}},\ }\bibfield  {title} {\enquote {\bibinfo {title} {A simple differential geometry for complex networks},}\ }\href@noop {} {\bibfield  {journal} {\bibinfo  {journal} {Network Science}\ }\textbf {\bibinfo {volume} {9}},\ \bibinfo {pages} {S106--S133} (\bibinfo {year} {2021})}\BibitemShut {NoStop}%
\bibitem [{\citenamefont {Brede}(2012)}]{newman2010networks}%
  \BibitemOpen
  \bibfield  {author} {\bibinfo {author} {\bibfnamefont {M.}~\bibnamefont {Brede}},\ }\href@noop {} {\enquote {\bibinfo {title} {Networks—an introduction. mark ej newman.(2010, oxford university press.). 772 pages. isbn-978-0-19-920665-0.}}\ } (\bibinfo {year} {2012})\BibitemShut {NoStop}%
\bibitem [{\citenamefont {Barab{\'a}si}(2013)}]{barabasi2013network}%
  \BibitemOpen
  \bibfield  {author} {\bibinfo {author} {\bibfnamefont {A.-L.}\ \bibnamefont {Barab{\'a}si}},\ }\bibfield  {title} {\enquote {\bibinfo {title} {Network science},}\ }\href@noop {} {\bibfield  {journal} {\bibinfo  {journal} {Philosophical Transactions of the Royal Society A: Mathematical, Physical and Engineering Sciences}\ }\textbf {\bibinfo {volume} {371}},\ \bibinfo {pages} {20120375} (\bibinfo {year} {2013})}\BibitemShut {NoStop}%
\bibitem [{\citenamefont {Albert}\ and\ \citenamefont {Barab{\'a}si}(2002)}]{albert2002statistical}%
  \BibitemOpen
  \bibfield  {author} {\bibinfo {author} {\bibfnamefont {R.}~\bibnamefont {Albert}}\ and\ \bibinfo {author} {\bibfnamefont {A.-L.}\ \bibnamefont {Barab{\'a}si}},\ }\bibfield  {title} {\enquote {\bibinfo {title} {Statistical mechanics of complex networks},}\ }\href@noop {} {\bibfield  {journal} {\bibinfo  {journal} {Reviews of modern physics}\ }\textbf {\bibinfo {volume} {74}},\ \bibinfo {pages} {47} (\bibinfo {year} {2002})}\BibitemShut {NoStop}%
\bibitem [{\citenamefont {Artime}\ \emph {et~al.}(2024)\citenamefont {Artime}, \citenamefont {Grassia}, \citenamefont {De~Domenico}, \citenamefont {Gleeson}, \citenamefont {Makse}, \citenamefont {Mangioni}, \citenamefont {Perc},\ and\ \citenamefont {Radicchi}}]{artime2024robustness}%
  \BibitemOpen
  \bibfield  {author} {\bibinfo {author} {\bibfnamefont {O.}~\bibnamefont {Artime}}, \bibinfo {author} {\bibfnamefont {M.}~\bibnamefont {Grassia}}, \bibinfo {author} {\bibfnamefont {M.}~\bibnamefont {De~Domenico}}, \bibinfo {author} {\bibfnamefont {J.~P.}\ \bibnamefont {Gleeson}}, \bibinfo {author} {\bibfnamefont {H.~A.}\ \bibnamefont {Makse}}, \bibinfo {author} {\bibfnamefont {G.}~\bibnamefont {Mangioni}}, \bibinfo {author} {\bibfnamefont {M.}~\bibnamefont {Perc}}, \ and\ \bibinfo {author} {\bibfnamefont {F.}~\bibnamefont {Radicchi}},\ }\bibfield  {title} {\enquote {\bibinfo {title} {Robustness and resilience of complex networks},}\ }\href@noop {} {\bibfield  {journal} {\bibinfo  {journal} {Nature Reviews Physics}\ }\textbf {\bibinfo {volume} {6}},\ \bibinfo {pages} {114--131} (\bibinfo {year} {2024})}\BibitemShut {NoStop}%
\bibitem [{\citenamefont {Newman}(2003)}]{newman2003structure}%
  \BibitemOpen
  \bibfield  {author} {\bibinfo {author} {\bibfnamefont {M.~E.}\ \bibnamefont {Newman}},\ }\bibfield  {title} {\enquote {\bibinfo {title} {The structure and function of complex networks},}\ }\href@noop {} {\bibfield  {journal} {\bibinfo  {journal} {SIAM review}\ }\textbf {\bibinfo {volume} {45}},\ \bibinfo {pages} {167--256} (\bibinfo {year} {2003})}\BibitemShut {NoStop}%
\bibitem [{\citenamefont {Clote}(2020)}]{clote2020rna}%
  \BibitemOpen
  \bibfield  {author} {\bibinfo {author} {\bibfnamefont {P.}~\bibnamefont {Clote}},\ }\bibfield  {title} {\enquote {\bibinfo {title} {Are rna networks scale-free?}}\ }\href@noop {} {\bibfield  {journal} {\bibinfo  {journal} {Journal of mathematical biology}\ }\textbf {\bibinfo {volume} {80}},\ \bibinfo {pages} {1291--1321} (\bibinfo {year} {2020})}\BibitemShut {NoStop}%
\bibitem [{\citenamefont {Muzio}, \citenamefont {O’Bray},\ and\ \citenamefont {Borgwardt}(2021)}]{muzio2021biological}%
  \BibitemOpen
  \bibfield  {author} {\bibinfo {author} {\bibfnamefont {G.}~\bibnamefont {Muzio}}, \bibinfo {author} {\bibfnamefont {L.}~\bibnamefont {O’Bray}}, \ and\ \bibinfo {author} {\bibfnamefont {K.}~\bibnamefont {Borgwardt}},\ }\bibfield  {title} {\enquote {\bibinfo {title} {Biological network analysis with deep learning},}\ }\href@noop {} {\bibfield  {journal} {\bibinfo  {journal} {Briefings in bioinformatics}\ }\textbf {\bibinfo {volume} {22}},\ \bibinfo {pages} {1515--1530} (\bibinfo {year} {2021})}\BibitemShut {NoStop}%
\bibitem [{\citenamefont {Jin}\ \emph {et~al.}(2021)\citenamefont {Jin}, \citenamefont {Zeng}, \citenamefont {Xia}, \citenamefont {Huang},\ and\ \citenamefont {Liu}}]{jin2021application}%
  \BibitemOpen
  \bibfield  {author} {\bibinfo {author} {\bibfnamefont {S.}~\bibnamefont {Jin}}, \bibinfo {author} {\bibfnamefont {X.}~\bibnamefont {Zeng}}, \bibinfo {author} {\bibfnamefont {F.}~\bibnamefont {Xia}}, \bibinfo {author} {\bibfnamefont {W.}~\bibnamefont {Huang}}, \ and\ \bibinfo {author} {\bibfnamefont {X.}~\bibnamefont {Liu}},\ }\bibfield  {title} {\enquote {\bibinfo {title} {Application of deep learning methods in biological networks},}\ }\href@noop {} {\bibfield  {journal} {\bibinfo  {journal} {Briefings in bioinformatics}\ }\textbf {\bibinfo {volume} {22}},\ \bibinfo {pages} {1902--1917} (\bibinfo {year} {2021})}\BibitemShut {NoStop}%
\bibitem [{\citenamefont {Pan}\ \emph {et~al.}(2022)\citenamefont {Pan}, \citenamefont {You}, \citenamefont {Li}, \citenamefont {Huang}, \citenamefont {Guo}, \citenamefont {Yu}, \citenamefont {Wang},\ and\ \citenamefont {Zhao}}]{pan2022dwppi}%
  \BibitemOpen
  \bibfield  {author} {\bibinfo {author} {\bibfnamefont {J.}~\bibnamefont {Pan}}, \bibinfo {author} {\bibfnamefont {Z.-H.}\ \bibnamefont {You}}, \bibinfo {author} {\bibfnamefont {L.-P.}\ \bibnamefont {Li}}, \bibinfo {author} {\bibfnamefont {W.-Z.}\ \bibnamefont {Huang}}, \bibinfo {author} {\bibfnamefont {J.-X.}\ \bibnamefont {Guo}}, \bibinfo {author} {\bibfnamefont {C.-Q.}\ \bibnamefont {Yu}}, \bibinfo {author} {\bibfnamefont {L.-P.}\ \bibnamefont {Wang}}, \ and\ \bibinfo {author} {\bibfnamefont {Z.-Y.}\ \bibnamefont {Zhao}},\ }\bibfield  {title} {\enquote {\bibinfo {title} {Dwppi: A deep learning approach for predicting protein--protein interactions in plants based on multi-source information with a large-scale biological network},}\ }\href@noop {} {\bibfield  {journal} {\bibinfo  {journal} {Frontiers in Bioengineering and Biotechnology}\ }\textbf {\bibinfo {volume} {10}},\ \bibinfo {pages} {807522} (\bibinfo {year} {2022})}\BibitemShut {NoStop}%
\bibitem [{\citenamefont {Yang}\ \emph {et~al.}(2022)\citenamefont {Yang}, \citenamefont {Li}, \citenamefont {Wu}, \citenamefont {Yu}, \citenamefont {Xu}, \citenamefont {Chu},\ and\ \citenamefont {Zhang}}]{yang2022deep}%
  \BibitemOpen
  \bibfield  {author} {\bibinfo {author} {\bibfnamefont {J.}~\bibnamefont {Yang}}, \bibinfo {author} {\bibfnamefont {Z.}~\bibnamefont {Li}}, \bibinfo {author} {\bibfnamefont {W.~K.~K.}\ \bibnamefont {Wu}}, \bibinfo {author} {\bibfnamefont {S.}~\bibnamefont {Yu}}, \bibinfo {author} {\bibfnamefont {Z.}~\bibnamefont {Xu}}, \bibinfo {author} {\bibfnamefont {Q.}~\bibnamefont {Chu}}, \ and\ \bibinfo {author} {\bibfnamefont {Q.}~\bibnamefont {Zhang}},\ }\bibfield  {title} {\enquote {\bibinfo {title} {Deep learning identifies explainable reasoning paths of mechanism of action for drug repurposing from multilayer biological network},}\ }\href@noop {} {\bibfield  {journal} {\bibinfo  {journal} {Briefings in Bioinformatics}\ }\textbf {\bibinfo {volume} {23}},\ \bibinfo {pages} {bbac469} (\bibinfo {year} {2022})}\BibitemShut {NoStop}%
\bibitem [{\citenamefont {Ma}\ \emph {et~al.}(2023)\citenamefont {Ma}, \citenamefont {Zhang}, \citenamefont {Li}, \citenamefont {Jiang}, \citenamefont {Wang}, \citenamefont {Guo}, \citenamefont {Li}, \citenamefont {Bi}, \citenamefont {Jiang},\ and\ \citenamefont {Wei}}]{ma2023predicting}%
  \BibitemOpen
  \bibfield  {author} {\bibinfo {author} {\bibfnamefont {W.}~\bibnamefont {Ma}}, \bibinfo {author} {\bibfnamefont {S.}~\bibnamefont {Zhang}}, \bibinfo {author} {\bibfnamefont {Z.}~\bibnamefont {Li}}, \bibinfo {author} {\bibfnamefont {M.}~\bibnamefont {Jiang}}, \bibinfo {author} {\bibfnamefont {S.}~\bibnamefont {Wang}}, \bibinfo {author} {\bibfnamefont {N.}~\bibnamefont {Guo}}, \bibinfo {author} {\bibfnamefont {Y.}~\bibnamefont {Li}}, \bibinfo {author} {\bibfnamefont {X.}~\bibnamefont {Bi}}, \bibinfo {author} {\bibfnamefont {H.}~\bibnamefont {Jiang}}, \ and\ \bibinfo {author} {\bibfnamefont {Z.}~\bibnamefont {Wei}},\ }\bibfield  {title} {\enquote {\bibinfo {title} {Predicting drug-target affinity by learning protein knowledge from biological networks},}\ }\href@noop {} {\bibfield  {journal} {\bibinfo  {journal} {IEEE Journal of Biomedical and Health Informatics}\ }\textbf {\bibinfo {volume} {27}},\ \bibinfo {pages} {2128--2137} (\bibinfo {year} {2023})}\BibitemShut {NoStop}%
\bibitem [{\citenamefont {Ito}\ \emph {et~al.}(2000)\citenamefont {Ito}, \citenamefont {Tashiro}, \citenamefont {Muta}, \citenamefont {Ozawa}, \citenamefont {Chiba}, \citenamefont {Nishizawa}, \citenamefont {Yamamoto}, \citenamefont {Kuhara},\ and\ \citenamefont {Sakaki}}]{ito2000toward}%
  \BibitemOpen
  \bibfield  {author} {\bibinfo {author} {\bibfnamefont {T.}~\bibnamefont {Ito}}, \bibinfo {author} {\bibfnamefont {K.}~\bibnamefont {Tashiro}}, \bibinfo {author} {\bibfnamefont {S.}~\bibnamefont {Muta}}, \bibinfo {author} {\bibfnamefont {R.}~\bibnamefont {Ozawa}}, \bibinfo {author} {\bibfnamefont {T.}~\bibnamefont {Chiba}}, \bibinfo {author} {\bibfnamefont {M.}~\bibnamefont {Nishizawa}}, \bibinfo {author} {\bibfnamefont {K.}~\bibnamefont {Yamamoto}}, \bibinfo {author} {\bibfnamefont {S.}~\bibnamefont {Kuhara}}, \ and\ \bibinfo {author} {\bibfnamefont {Y.}~\bibnamefont {Sakaki}},\ }\bibfield  {title} {\enquote {\bibinfo {title} {Toward a protein--protein interaction map of the budding yeast: a comprehensive system to examine two-hybrid interactions in all possible combinations between the yeast proteins},}\ }\href@noop {} {\bibfield  {journal} {\bibinfo  {journal} {Proceedings of the National Academy of Sciences}\ }\textbf {\bibinfo {volume} {97}},\ \bibinfo {pages} {1143--1147} (\bibinfo {year}
  {2000})}\BibitemShut {NoStop}%
\bibitem [{\citenamefont {Clauset}, \citenamefont {Shalizi},\ and\ \citenamefont {Newman}(2009)}]{clauset2009power}%
  \BibitemOpen
  \bibfield  {author} {\bibinfo {author} {\bibfnamefont {A.}~\bibnamefont {Clauset}}, \bibinfo {author} {\bibfnamefont {C.~R.}\ \bibnamefont {Shalizi}}, \ and\ \bibinfo {author} {\bibfnamefont {M.~E.}\ \bibnamefont {Newman}},\ }\bibfield  {title} {\enquote {\bibinfo {title} {Power-law distributions in empirical data},}\ }\href@noop {} {\bibfield  {journal} {\bibinfo  {journal} {SIAM review}\ }\textbf {\bibinfo {volume} {51}},\ \bibinfo {pages} {661--703} (\bibinfo {year} {2009})}\BibitemShut {NoStop}%
\bibitem [{\citenamefont {Goldstein}, \citenamefont {Morris},\ and\ \citenamefont {Yen}(2004)}]{goldstein2004problems}%
  \BibitemOpen
  \bibfield  {author} {\bibinfo {author} {\bibfnamefont {M.~L.}\ \bibnamefont {Goldstein}}, \bibinfo {author} {\bibfnamefont {S.~A.}\ \bibnamefont {Morris}}, \ and\ \bibinfo {author} {\bibfnamefont {G.~G.}\ \bibnamefont {Yen}},\ }\bibfield  {title} {\enquote {\bibinfo {title} {Problems with fitting to the power-law distribution},}\ }\href@noop {} {\bibfield  {journal} {\bibinfo  {journal} {The European Physical Journal B-Condensed Matter and Complex Systems}\ }\textbf {\bibinfo {volume} {41}},\ \bibinfo {pages} {255--258} (\bibinfo {year} {2004})}\BibitemShut {NoStop}%
\bibitem [{\citenamefont {Salem}(2018)}]{salem2018biological}%
  \BibitemOpen
  \bibfield  {author} {\bibinfo {author} {\bibfnamefont {M.~S.~Z.}\ \bibnamefont {Salem}},\ }\bibfield  {title} {\enquote {\bibinfo {title} {Biological networks: An introductory review},}\ }\href@noop {} {\bibfield  {journal} {\bibinfo  {journal} {Journal of Proteomics and Genomics Research}\ }\textbf {\bibinfo {volume} {2}},\ \bibinfo {pages} {41} (\bibinfo {year} {2018})}\BibitemShut {NoStop}%
\bibitem [{\citenamefont {Voitalov}\ \emph {et~al.}(2019)\citenamefont {Voitalov}, \citenamefont {van~der Hoorn}, \citenamefont {van~der Hofstad},\ and\ \citenamefont {Krioukov}}]{voitalov2019scale}%
  \BibitemOpen
  \bibfield  {author} {\bibinfo {author} {\bibfnamefont {I.}~\bibnamefont {Voitalov}}, \bibinfo {author} {\bibfnamefont {P.}~\bibnamefont {van~der Hoorn}}, \bibinfo {author} {\bibfnamefont {R.}~\bibnamefont {van~der Hofstad}}, \ and\ \bibinfo {author} {\bibfnamefont {D.}~\bibnamefont {Krioukov}},\ }\bibfield  {title} {\enquote {\bibinfo {title} {Scale-free networks well done},}\ }\href@noop {} {\bibfield  {journal} {\bibinfo  {journal} {Physical Review Research}\ }\textbf {\bibinfo {volume} {1}},\ \bibinfo {pages} {033034} (\bibinfo {year} {2019})}\BibitemShut {NoStop}%
\bibitem [{\citenamefont {Chattopadhyay}\ \emph {et~al.}(2021{\natexlab{b}})\citenamefont {Chattopadhyay}, \citenamefont {Chakraborty}, \citenamefont {Ghosh},\ and\ \citenamefont {Das}}]{chattopadhyay2021modified}%
  \BibitemOpen
  \bibfield  {author} {\bibinfo {author} {\bibfnamefont {S.}~\bibnamefont {Chattopadhyay}}, \bibinfo {author} {\bibfnamefont {T.}~\bibnamefont {Chakraborty}}, \bibinfo {author} {\bibfnamefont {K.}~\bibnamefont {Ghosh}}, \ and\ \bibinfo {author} {\bibfnamefont {A.~K.}\ \bibnamefont {Das}},\ }\bibfield  {title} {\enquote {\bibinfo {title} {Modified lomax model: A heavy-tailed distribution for fitting large-scale real-world complex networks},}\ }\href@noop {} {\bibfield  {journal} {\bibinfo  {journal} {Social Network Analysis and Mining}\ }\textbf {\bibinfo {volume} {11}},\ \bibinfo {pages} {43} (\bibinfo {year} {2021}{\natexlab{b}})}\BibitemShut {NoStop}%
\bibitem [{\citenamefont {Chakraborty}\ \emph {et~al.}(2022)\citenamefont {Chakraborty}, \citenamefont {Chattopadhyay}, \citenamefont {Das}, \citenamefont {Kumar},\ and\ \citenamefont {Senthilnath}}]{chakraborty2022searching}%
  \BibitemOpen
  \bibfield  {author} {\bibinfo {author} {\bibfnamefont {T.}~\bibnamefont {Chakraborty}}, \bibinfo {author} {\bibfnamefont {S.}~\bibnamefont {Chattopadhyay}}, \bibinfo {author} {\bibfnamefont {S.}~\bibnamefont {Das}}, \bibinfo {author} {\bibfnamefont {U.}~\bibnamefont {Kumar}}, \ and\ \bibinfo {author} {\bibfnamefont {J.}~\bibnamefont {Senthilnath}},\ }\bibfield  {title} {\enquote {\bibinfo {title} {Searching for heavy-tailed probability distributions for modeling real-world complex networks},}\ }\href@noop {} {\bibfield  {journal} {\bibinfo  {journal} {IEEE Access}\ }\textbf {\bibinfo {volume} {10}},\ \bibinfo {pages} {115092--115107} (\bibinfo {year} {2022})}\BibitemShut {NoStop}%
\bibitem [{\citenamefont {Chakraborty}, \citenamefont {Das},\ and\ \citenamefont {Chattopadhyay}(2022)}]{chakraborty2022new}%
  \BibitemOpen
  \bibfield  {author} {\bibinfo {author} {\bibfnamefont {T.}~\bibnamefont {Chakraborty}}, \bibinfo {author} {\bibfnamefont {S.}~\bibnamefont {Das}}, \ and\ \bibinfo {author} {\bibfnamefont {S.}~\bibnamefont {Chattopadhyay}},\ }\bibfield  {title} {\enquote {\bibinfo {title} {A new method for generalizing burr and related distributions},}\ }\href@noop {} {\bibfield  {journal} {\bibinfo  {journal} {Mathematica Slovaca}\ }\textbf {\bibinfo {volume} {72}},\ \bibinfo {pages} {241--264} (\bibinfo {year} {2022})}\BibitemShut {NoStop}%
\bibitem [{\citenamefont {Stumpf}, \citenamefont {Wiuf},\ and\ \citenamefont {May}(2005)}]{stumpf2005subnets}%
  \BibitemOpen
  \bibfield  {author} {\bibinfo {author} {\bibfnamefont {M.~P.}\ \bibnamefont {Stumpf}}, \bibinfo {author} {\bibfnamefont {C.}~\bibnamefont {Wiuf}}, \ and\ \bibinfo {author} {\bibfnamefont {R.~M.}\ \bibnamefont {May}},\ }\bibfield  {title} {\enquote {\bibinfo {title} {Subnets of scale-free networks are not scale-free: sampling properties of networks},}\ }\href@noop {} {\bibfield  {journal} {\bibinfo  {journal} {Proceedings of the National Academy of Sciences}\ }\textbf {\bibinfo {volume} {102}},\ \bibinfo {pages} {4221--4224} (\bibinfo {year} {2005})}\BibitemShut {NoStop}%
\bibitem [{\citenamefont {Burr}(1942)}]{burr1942cumulative}%
  \BibitemOpen
  \bibfield  {author} {\bibinfo {author} {\bibfnamefont {I.~W.}\ \bibnamefont {Burr}},\ }\bibfield  {title} {\enquote {\bibinfo {title} {Cumulative frequency functions},}\ }\href@noop {} {\bibfield  {journal} {\bibinfo  {journal} {The Annals of mathematical statistics}\ }\textbf {\bibinfo {volume} {13}},\ \bibinfo {pages} {215--232} (\bibinfo {year} {1942})}\BibitemShut {NoStop}%
\bibitem [{\citenamefont {Ahmed}\ \emph {et~al.}(2021)\citenamefont {Ahmed}, \citenamefont {Khaleel}, \citenamefont {Oguntunde},\ and\ \citenamefont {Abdal-Hammed}}]{ahmed2021new}%
  \BibitemOpen
  \bibfield  {author} {\bibinfo {author} {\bibfnamefont {M.~T.}\ \bibnamefont {Ahmed}}, \bibinfo {author} {\bibfnamefont {M.~A.}\ \bibnamefont {Khaleel}}, \bibinfo {author} {\bibfnamefont {P.~E.}\ \bibnamefont {Oguntunde}}, \ and\ \bibinfo {author} {\bibfnamefont {M.~K.}\ \bibnamefont {Abdal-Hammed}},\ }\bibfield  {title} {\enquote {\bibinfo {title} {A new version of the exponentiated burr x distribution},}\ }in\ \href@noop {} {\emph {\bibinfo {booktitle} {Journal of Physics: Conference Series}}},\ Vol.\ \bibinfo {volume} {1818}\ (\bibinfo {organization} {IOP Publishing},\ \bibinfo {year} {2021})\ p.\ \bibinfo {pages} {012116}\BibitemShut {NoStop}%
\bibitem [{\citenamefont {Jayakumar}\ and\ \citenamefont {Mathew}(2008)}]{jayakumar2008generalization}%
  \BibitemOpen
  \bibfield  {author} {\bibinfo {author} {\bibfnamefont {K.}~\bibnamefont {Jayakumar}}\ and\ \bibinfo {author} {\bibfnamefont {T.}~\bibnamefont {Mathew}},\ }\bibfield  {title} {\enquote {\bibinfo {title} {On a generalization to marshall-olkin scheme and its application to burr type xii distribution},}\ }\href@noop {} {\bibfield  {journal} {\bibinfo  {journal} {Statistical Papers}\ }\textbf {\bibinfo {volume} {49}},\ \bibinfo {pages} {421} (\bibinfo {year} {2008})}\BibitemShut {NoStop}%
\bibitem [{\citenamefont {Wang}, \citenamefont {Ma},\ and\ \citenamefont {Yao}(2021)}]{wang2021arbitrary}%
  \BibitemOpen
  \bibfield  {author} {\bibinfo {author} {\bibfnamefont {X.}~\bibnamefont {Wang}}, \bibinfo {author} {\bibfnamefont {F.}~\bibnamefont {Ma}}, \ and\ \bibinfo {author} {\bibfnamefont {B.}~\bibnamefont {Yao}},\ }\bibfield  {title} {\enquote {\bibinfo {title} {Arbitrary degree distribution networks with perturbations},}\ }\href@noop {} {\bibfield  {journal} {\bibinfo  {journal} {AIP Advances}\ }\textbf {\bibinfo {volume} {11}} (\bibinfo {year} {2021})}\BibitemShut {NoStop}%
\bibitem [{\citenamefont {Zhang}\ and\ \citenamefont {Zhang}(2019)}]{zhang2019protein}%
  \BibitemOpen
  \bibfield  {author} {\bibinfo {author} {\bibfnamefont {G.}~\bibnamefont {Zhang}}\ and\ \bibinfo {author} {\bibfnamefont {W.}~\bibnamefont {Zhang}},\ }\bibfield  {title} {\enquote {\bibinfo {title} {Protein--protein interaction network analysis of insecticide resistance molecular mechanism in drosophila melanogaster},}\ }\href@noop {} {\bibfield  {journal} {\bibinfo  {journal} {Archives of Insect Biochemistry and Physiology}\ }\textbf {\bibinfo {volume} {100}},\ \bibinfo {pages} {e21523} (\bibinfo {year} {2019})}\BibitemShut {NoStop}%
\bibitem [{\citenamefont {Singh}, \citenamefont {Xu},\ and\ \citenamefont {Berger}(2008{\natexlab{a}})}]{singh2008-isorank-multi}%
  \BibitemOpen
  \bibfield  {author} {\bibinfo {author} {\bibfnamefont {R.}~\bibnamefont {Singh}}, \bibinfo {author} {\bibfnamefont {J.}~\bibnamefont {Xu}}, \ and\ \bibinfo {author} {\bibfnamefont {B.}~\bibnamefont {Berger}},\ }\bibfield  {title} {\enquote {\bibinfo {title} {Global alignment of multiple protein interaction networks with application to functional orthology detection},}\ }\href@noop {} {\bibfield  {journal} {\bibinfo  {journal} {PNAS}\ }\textbf {\bibinfo {volume} {105}},\ \bibinfo {pages} {12763--12768} (\bibinfo {year} {2008}{\natexlab{a}})}\BibitemShut {NoStop}%
\bibitem [{\citenamefont {Marshall}\ and\ \citenamefont {Olkin}(1997)}]{marshall1997new}%
  \BibitemOpen
  \bibfield  {author} {\bibinfo {author} {\bibfnamefont {A.~W.}\ \bibnamefont {Marshall}}\ and\ \bibinfo {author} {\bibfnamefont {I.}~\bibnamefont {Olkin}},\ }\bibfield  {title} {\enquote {\bibinfo {title} {A new method for adding a parameter to a family of distributions with application to the exponential and weibull families},}\ }\href@noop {} {\bibfield  {journal} {\bibinfo  {journal} {Biometrika}\ }\textbf {\bibinfo {volume} {84}},\ \bibinfo {pages} {641--652} (\bibinfo {year} {1997})}\BibitemShut {NoStop}%
\bibitem [{\citenamefont {Cox}(1972)}]{cox1972regression}%
  \BibitemOpen
  \bibfield  {author} {\bibinfo {author} {\bibfnamefont {D.~R.}\ \bibnamefont {Cox}},\ }\bibfield  {title} {\enquote {\bibinfo {title} {Regression models and life-tables},}\ }\href@noop {} {\bibfield  {journal} {\bibinfo  {journal} {Journal of the Royal Statistical Society: Series B (Methodological)}\ }\textbf {\bibinfo {volume} {34}},\ \bibinfo {pages} {187--202} (\bibinfo {year} {1972})}\BibitemShut {NoStop}%
\bibitem [{\citenamefont {Panja}\ \emph {et~al.}(2023)\citenamefont {Panja}, \citenamefont {Chakraborty}, \citenamefont {Kumar},\ and\ \citenamefont {Liu}}]{panja2023epicasting}%
  \BibitemOpen
  \bibfield  {author} {\bibinfo {author} {\bibfnamefont {M.}~\bibnamefont {Panja}}, \bibinfo {author} {\bibfnamefont {T.}~\bibnamefont {Chakraborty}}, \bibinfo {author} {\bibfnamefont {U.}~\bibnamefont {Kumar}}, \ and\ \bibinfo {author} {\bibfnamefont {N.}~\bibnamefont {Liu}},\ }\bibfield  {title} {\enquote {\bibinfo {title} {Epicasting: an ensemble wavelet neural network for forecasting epidemics},}\ }\href@noop {} {\bibfield  {journal} {\bibinfo  {journal} {Neural Networks}\ }\textbf {\bibinfo {volume} {165}},\ \bibinfo {pages} {185--212} (\bibinfo {year} {2023})}\BibitemShut {NoStop}%
\bibitem [{\citenamefont {Greenwood}\ \emph {et~al.}(1979)\citenamefont {Greenwood}, \citenamefont {Landwehr}, \citenamefont {Matalas},\ and\ \citenamefont {Wallis}}]{greenwood1979probability}%
  \BibitemOpen
  \bibfield  {author} {\bibinfo {author} {\bibfnamefont {J.~A.}\ \bibnamefont {Greenwood}}, \bibinfo {author} {\bibfnamefont {J.~M.}\ \bibnamefont {Landwehr}}, \bibinfo {author} {\bibfnamefont {N.~C.}\ \bibnamefont {Matalas}}, \ and\ \bibinfo {author} {\bibfnamefont {J.~R.}\ \bibnamefont {Wallis}},\ }\bibfield  {title} {\enquote {\bibinfo {title} {Probability weighted moments: definition and relation to parameters of several distributions expressable in inverse form},}\ }\href@noop {} {\bibfield  {journal} {\bibinfo  {journal} {Water resources research}\ }\textbf {\bibinfo {volume} {15}},\ \bibinfo {pages} {1049--1054} (\bibinfo {year} {1979})}\BibitemShut {NoStop}%
\bibitem [{\citenamefont {Tadikamalla}(1980)}]{tadikamalla1980look}%
  \BibitemOpen
  \bibfield  {author} {\bibinfo {author} {\bibfnamefont {P.~R.}\ \bibnamefont {Tadikamalla}},\ }\bibfield  {title} {\enquote {\bibinfo {title} {A look at the burr and related distributions},}\ }\href@noop {} {\bibfield  {journal} {\bibinfo  {journal} {International Statistical Review/Revue Internationale de Statistique}\ ,\ \bibinfo {pages} {337--344}} (\bibinfo {year} {1980})}\BibitemShut {NoStop}%
\bibitem [{\citenamefont {Johnson}, \citenamefont {Kotz},\ and\ \citenamefont {Balakrishnan}(1995)}]{johnson1995continuous}%
  \BibitemOpen
  \bibfield  {author} {\bibinfo {author} {\bibfnamefont {N.~L.}\ \bibnamefont {Johnson}}, \bibinfo {author} {\bibfnamefont {S.}~\bibnamefont {Kotz}}, \ and\ \bibinfo {author} {\bibfnamefont {N.}~\bibnamefont {Balakrishnan}},\ }\href@noop {} {\emph {\bibinfo {title} {Continuous univariate distributions, volume 2}}},\ Vol.\ \bibinfo {volume} {289}\ (\bibinfo  {publisher} {John wiley \& sons},\ \bibinfo {year} {1995})\BibitemShut {NoStop}%
\bibitem [{\citenamefont {Singh}, \citenamefont {Xu},\ and\ \citenamefont {Berger}(2008{\natexlab{b}})}]{singh2008global}%
  \BibitemOpen
  \bibfield  {author} {\bibinfo {author} {\bibfnamefont {R.}~\bibnamefont {Singh}}, \bibinfo {author} {\bibfnamefont {J.}~\bibnamefont {Xu}}, \ and\ \bibinfo {author} {\bibfnamefont {B.}~\bibnamefont {Berger}},\ }\bibfield  {title} {\enquote {\bibinfo {title} {Global alignment of multiple protein interaction networks with application to functional orthology detection},}\ }\href@noop {} {\bibfield  {journal} {\bibinfo  {journal} {Proceedings of the National Academy of Sciences}\ }\textbf {\bibinfo {volume} {105}},\ \bibinfo {pages} {12763--12768} (\bibinfo {year} {2008}{\natexlab{b}})}\BibitemShut {NoStop}%
\bibitem [{\citenamefont {Liu}\ \emph {et~al.}(2015)\citenamefont {Liu}, \citenamefont {Wu}, \citenamefont {Miao},\ and\ \citenamefont {Wu}}]{liu2015regnetwork}%
  \BibitemOpen
  \bibfield  {author} {\bibinfo {author} {\bibfnamefont {Z.-P.}\ \bibnamefont {Liu}}, \bibinfo {author} {\bibfnamefont {C.}~\bibnamefont {Wu}}, \bibinfo {author} {\bibfnamefont {H.}~\bibnamefont {Miao}}, \ and\ \bibinfo {author} {\bibfnamefont {H.}~\bibnamefont {Wu}},\ }\bibfield  {title} {\enquote {\bibinfo {title} {Regnetwork: an integrated database of transcriptional and post-transcriptional regulatory networks in human and mouse},}\ }\href@noop {} {\bibfield  {journal} {\bibinfo  {journal} {Database}\ }\textbf {\bibinfo {volume} {2015}} (\bibinfo {year} {2015})}\BibitemShut {NoStop}%
\bibitem [{\citenamefont {Albert}(2005)}]{albert2005scale}%
  \BibitemOpen
  \bibfield  {author} {\bibinfo {author} {\bibfnamefont {R.}~\bibnamefont {Albert}},\ }\bibfield  {title} {\enquote {\bibinfo {title} {Scale-free networks in cell biology},}\ }\href@noop {} {\bibfield  {journal} {\bibinfo  {journal} {Journal of cell science}\ }\textbf {\bibinfo {volume} {118}},\ \bibinfo {pages} {4947--4957} (\bibinfo {year} {2005})}\BibitemShut {NoStop}%
\bibitem [{\citenamefont {Goh}\ \emph {et~al.}(2007)\citenamefont {Goh}, \citenamefont {Cusick}, \citenamefont {Valle}, \citenamefont {Childs}, \citenamefont {Vidal},\ and\ \citenamefont {Barab{\'a}si}}]{goh2007human}%
  \BibitemOpen
  \bibfield  {author} {\bibinfo {author} {\bibfnamefont {K.-I.}\ \bibnamefont {Goh}}, \bibinfo {author} {\bibfnamefont {M.~E.}\ \bibnamefont {Cusick}}, \bibinfo {author} {\bibfnamefont {D.}~\bibnamefont {Valle}}, \bibinfo {author} {\bibfnamefont {B.}~\bibnamefont {Childs}}, \bibinfo {author} {\bibfnamefont {M.}~\bibnamefont {Vidal}}, \ and\ \bibinfo {author} {\bibfnamefont {A.-L.}\ \bibnamefont {Barab{\'a}si}},\ }\bibfield  {title} {\enquote {\bibinfo {title} {The human disease network},}\ }\href@noop {} {\bibfield  {journal} {\bibinfo  {journal} {Proceedings of the National Academy of Sciences}\ }\textbf {\bibinfo {volume} {104}},\ \bibinfo {pages} {8685--8690} (\bibinfo {year} {2007})}\BibitemShut {NoStop}%
\bibitem [{\citenamefont {Goh}\ and\ \citenamefont {Choi}(2012)}]{goh2012exploring}%
  \BibitemOpen
  \bibfield  {author} {\bibinfo {author} {\bibfnamefont {K.-I.}\ \bibnamefont {Goh}}\ and\ \bibinfo {author} {\bibfnamefont {I.-G.}\ \bibnamefont {Choi}},\ }\bibfield  {title} {\enquote {\bibinfo {title} {Exploring the human diseasome: the human disease network},}\ }\href@noop {} {\bibfield  {journal} {\bibinfo  {journal} {Briefings in functional genomics}\ }\textbf {\bibinfo {volume} {11}},\ \bibinfo {pages} {533--542} (\bibinfo {year} {2012})}\BibitemShut {NoStop}%
\bibitem [{\citenamefont {Bu}\ \emph {et~al.}(2003)\citenamefont {Bu}, \citenamefont {Zhao}, \citenamefont {Cai}, \citenamefont {Xue}, \citenamefont {Zhu}, \citenamefont {Lu}, \citenamefont {Zhang}, \citenamefont {Sun}, \citenamefont {Ling}, \citenamefont {Zhang} \emph {et~al.}}]{bu2003topological}%
  \BibitemOpen
  \bibfield  {author} {\bibinfo {author} {\bibfnamefont {D.}~\bibnamefont {Bu}}, \bibinfo {author} {\bibfnamefont {Y.}~\bibnamefont {Zhao}}, \bibinfo {author} {\bibfnamefont {L.}~\bibnamefont {Cai}}, \bibinfo {author} {\bibfnamefont {H.}~\bibnamefont {Xue}}, \bibinfo {author} {\bibfnamefont {X.}~\bibnamefont {Zhu}}, \bibinfo {author} {\bibfnamefont {H.}~\bibnamefont {Lu}}, \bibinfo {author} {\bibfnamefont {J.}~\bibnamefont {Zhang}}, \bibinfo {author} {\bibfnamefont {S.}~\bibnamefont {Sun}}, \bibinfo {author} {\bibfnamefont {L.}~\bibnamefont {Ling}}, \bibinfo {author} {\bibfnamefont {N.}~\bibnamefont {Zhang}},  \emph {et~al.},\ }\bibfield  {title} {\enquote {\bibinfo {title} {Topological structure analysis of the protein--protein interaction network in budding yeast},}\ }\href@noop {} {\bibfield  {journal} {\bibinfo  {journal} {Nucleic acids research}\ }\textbf {\bibinfo {volume} {31}},\ \bibinfo {pages} {2443--2450} (\bibinfo {year} {2003})}\BibitemShut {NoStop}%
\bibitem [{\citenamefont {Stark}\ \emph {et~al.}(2006)\citenamefont {Stark}, \citenamefont {Breitkreutz}, \citenamefont {Reguly}, \citenamefont {Boucher}, \citenamefont {Breitkreutz},\ and\ \citenamefont {Tyers}}]{stark2006biogrid}%
  \BibitemOpen
  \bibfield  {author} {\bibinfo {author} {\bibfnamefont {C.}~\bibnamefont {Stark}}, \bibinfo {author} {\bibfnamefont {B.-J.}\ \bibnamefont {Breitkreutz}}, \bibinfo {author} {\bibfnamefont {T.}~\bibnamefont {Reguly}}, \bibinfo {author} {\bibfnamefont {L.}~\bibnamefont {Boucher}}, \bibinfo {author} {\bibfnamefont {A.}~\bibnamefont {Breitkreutz}}, \ and\ \bibinfo {author} {\bibfnamefont {M.}~\bibnamefont {Tyers}},\ }\bibfield  {title} {\enquote {\bibinfo {title} {Biogrid: a general repository for interaction datasets},}\ }\href@noop {} {\bibfield  {journal} {\bibinfo  {journal} {Nucleic acids research}\ }\textbf {\bibinfo {volume} {34}},\ \bibinfo {pages} {D535--D539} (\bibinfo {year} {2006})}\BibitemShut {NoStop}%
\bibitem [{\citenamefont {Oughtred}\ \emph {et~al.}(2019)\citenamefont {Oughtred}, \citenamefont {Stark}, \citenamefont {Breitkreutz}, \citenamefont {Rust}, \citenamefont {Boucher}, \citenamefont {Chang}, \citenamefont {Kolas}, \citenamefont {O’Donnell}, \citenamefont {Leung}, \citenamefont {McAdam} \emph {et~al.}}]{oughtred2019biogrid}%
  \BibitemOpen
  \bibfield  {author} {\bibinfo {author} {\bibfnamefont {R.}~\bibnamefont {Oughtred}}, \bibinfo {author} {\bibfnamefont {C.}~\bibnamefont {Stark}}, \bibinfo {author} {\bibfnamefont {B.-J.}\ \bibnamefont {Breitkreutz}}, \bibinfo {author} {\bibfnamefont {J.}~\bibnamefont {Rust}}, \bibinfo {author} {\bibfnamefont {L.}~\bibnamefont {Boucher}}, \bibinfo {author} {\bibfnamefont {C.}~\bibnamefont {Chang}}, \bibinfo {author} {\bibfnamefont {N.}~\bibnamefont {Kolas}}, \bibinfo {author} {\bibfnamefont {L.}~\bibnamefont {O’Donnell}}, \bibinfo {author} {\bibfnamefont {G.}~\bibnamefont {Leung}}, \bibinfo {author} {\bibfnamefont {R.}~\bibnamefont {McAdam}},  \emph {et~al.},\ }\bibfield  {title} {\enquote {\bibinfo {title} {The biogrid interaction database: 2019 update},}\ }\href@noop {} {\bibfield  {journal} {\bibinfo  {journal} {Nucleic acids research}\ }\textbf {\bibinfo {volume} {47}},\ \bibinfo {pages} {D529--D541} (\bibinfo {year} {2019})}\BibitemShut {NoStop}%
\bibitem [{\citenamefont {Rossi}\ and\ \citenamefont {Ahmed}(2015)}]{nr-aaai15}%
  \BibitemOpen
  \bibfield  {author} {\bibinfo {author} {\bibfnamefont {R.~A.}\ \bibnamefont {Rossi}}\ and\ \bibinfo {author} {\bibfnamefont {N.~K.}\ \bibnamefont {Ahmed}},\ }\bibfield  {title} {\enquote {\bibinfo {title} {The network data repository with interactive graph analytics and visualization},}\ }in\ \href {http://networkrepository.com} {\emph {\bibinfo {booktitle} {Proceedings of the Twenty-Ninth AAAI Conference on Artificial Intelligence}}}\ (\bibinfo {year} {2015})\BibitemShut {NoStop}%
\bibitem [{\citenamefont {Rossi}\ \emph {et~al.}(2018)\citenamefont {Rossi}, \citenamefont {Ahmed}, \citenamefont {Zhou},\ and\ \citenamefont {Eldardiry}}]{rossi2018interactive}%
  \BibitemOpen
  \bibfield  {author} {\bibinfo {author} {\bibfnamefont {R.~A.}\ \bibnamefont {Rossi}}, \bibinfo {author} {\bibfnamefont {N.~K.}\ \bibnamefont {Ahmed}}, \bibinfo {author} {\bibfnamefont {R.}~\bibnamefont {Zhou}}, \ and\ \bibinfo {author} {\bibfnamefont {H.}~\bibnamefont {Eldardiry}},\ }\bibfield  {title} {\enquote {\bibinfo {title} {Interactive visual graph mining and learning},}\ }\href@noop {} {\bibfield  {journal} {\bibinfo  {journal} {ACM Transactions on Intelligent Systems and Technology (TIST)}\ }\textbf {\bibinfo {volume} {9}},\ \bibinfo {pages} {1--25} (\bibinfo {year} {2018})}\BibitemShut {NoStop}%
\bibitem [{\citenamefont {Efron}\ and\ \citenamefont {Tibshirani}(1994)}]{efron1994introduction}%
  \BibitemOpen
  \bibfield  {author} {\bibinfo {author} {\bibfnamefont {B.}~\bibnamefont {Efron}}\ and\ \bibinfo {author} {\bibfnamefont {R.~J.}\ \bibnamefont {Tibshirani}},\ }\href@noop {} {\emph {\bibinfo {title} {An introduction to the bootstrap}}}\ (\bibinfo  {publisher} {Chapman and Hall/CRC},\ \bibinfo {year} {1994})\BibitemShut {NoStop}%
\end{thebibliography}%

\end{document}